\documentclass{aa}  

\usepackage{graphicx}
\usepackage{txfonts}
\usepackage{mathrsfs}
\usepackage{graphicx}   
\usepackage{amsmath}    
\usepackage{todonotes}
\usepackage{mathtools}
\usepackage{bm}
\usepackage{multicol}
\usepackage{graphicx}  
\usepackage{soul}
\usepackage[utf8]{inputenc}
\usepackage{lmodern}
\usepackage{diagbox}
\usepackage{dsfont}
\usepackage{changepage}
\usepackage{ulem}
\usepackage{natbib}
\bibpunct{(}{)}{;}{a}{}{,}
\usepackage{longtable}
\usepackage{multirow}
\usepackage{amssymb}
\usepackage{nicematrix}
\usepackage{listings}
\usepackage[hidelinks,citecolor=blue,linkcolor=blue]{hyperref}
\hypersetup{
    colorlinks=true,
    linkcolor=blue,
    urlcolor=green,
    citecolor=blue
} 

\begin{document} 

\definecolor{forestgreen}{RGB}{24, 129, 24}
\newcommand\luba[1]{\emph{{\color{forestgreen}#1}}}
\newcommand\sm[1]{\emph{{\color{red}#1}}}
\newcommand\libu[1]{\emph{{\color{blue}#1}}}
\newcommand\orc[1]{\href{https://orcid.org/#1}{\includegraphics[width=3mm]{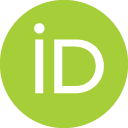}}}

\newenvironment{redtext}{\color{purple}}{}

        \title{{Constraining differential rotation in $\gamma$ Doradus stars \\ from inertial dips properties}}
   
\titlerunning{Constraining differential rotation in $\gamma$ Doradus stars from inertial dips properties}
   \author{L. Barrault
          \inst{1}\orc{0009-0007-7748-900X}
          \and
          S. Mathis\inst{2}\orc{0000-0001-9491-8012}
          \and
          L. Bugnet\inst{1}\orc{0000-0003-0142-4000}
          }

   \institute{Institute of Science and Technology Austria (ISTA), Am Campus 1, 3400 Klosterneuburg, Austria \\
            \email{lucas.barrault@ist.ac.at}
         \and
             Université Paris-Saclay, Université Paris Cité, CEA, CNRS, AIM, 91191, Gif-sur-Yvette, France\\
             }

   \date{Received XXX}

 
  \abstract
   {The presence of dips in the gravity modes period spacing versus period diagram of $\gamma$ Doradus stars is now well established by recent asteroseismic studies. Such Lorentzian-shaped inertial dips arise from the interaction of gravito-inertial modes in the radiative envelope of intermediate-mass main sequence stars with pure inertial modes in their convective core. They allow to study stellar internal properties. This window on stellar internal dynamics is extremely valuable in the context of the understanding of angular momentum transport inside stars, since it allows us to probe rotation in their core.}
   {We investigate the signature and the detectability of a differential rotation between the convective core and the near-core region inside $\gamma$ Doradus stars from the inertial dip properties.}
   {We study the coupling between gravito-inertial modes in the radiative zone and pure inertial modes in the convective core in the sub-inertial regime, allowing for a two-zones differential rotation from the two sides of the core-to-envelope boundary. 
   We solve the coupling equation numerically and match the result to an analytical derivation of the Lorentzian dip properties. We then use typical values of measured near-core rotation and buoyancy travel time to infer ranges of parameters for which differential core to near-core rotation would be detectable in current \textsl{Kepler} data.}
   {We show that increasing the convective core rotation with respect to the near-core rotation leads to a shift of the period of the observed dip to lower periods. In addition, the dip gets deeper and thinner as the convective core rotation increases. We demonstrate that such a signature is detectable in \textsl{Kepler} data, given appropriate dip parameter ranges and near-core structural properties.}
   {Studying the dip properties in asteroseismic data thus allows to access core to near-core radial differential rotation and better understand the transport of angular momentum at convective-radiative interfaces in intermediate-mass main sequence stars.}


   \maketitle
%

\section{Introduction}
\par The efficiency of angular momentum transport inside stars inferred from standard hydrodynamic evolution modelling is underestimated compared to observations. Cores are rotating slower than they are expected during a significant portion of the stellar evolution: on the Main Sequence \citep[MS][]{Aerts2017TheOscillations, Eggenberger2019RotationInteriors}, as well as in subgiants \citep[SG,][]{Deheuvels2014SeismicGiants,Gehan2018CoreRanges,Eggenberger2019AsteroseismologyPhase}, Red Giant Branch stars \citep[RGB,][]{Deheuvels2012SeismicKepler,Mosser2012SpinGiants} and Red Clump stars \citep[RC,][]{Deheuvels2015SeismicStars, Mosser2024LockedGiants}. As theoretical insights on the dynamical processes at play in stellar interiors are being developed, with candidates such as internal gravity waves \citep{Talon2005HydrodynamicalDwarfs,Belkacem2015AngularModes,Pincon2017CanStars}, meridional circulation  \citep{Rieutord2006TheModel,Decressin2009DiagnosesStars}, instabilities-driven turbulence: \citep{Zahn1992CirculationStars,Mathis2018AnisotropicZones}, and internal magnetism \citep{Spruit1999DifferentialInteriors,Spruit2002DynamoInterior,Mathis2005TransportField,Fuller2019SlowingCores,Petitdemange2023Spin-downLayers,Barrere2023NumericalProto-magnetars}, tools needed to accurately probe the rotation rate at different locations inside the star are also needed to better identify transport mechanisms.
\par During the MS, gravity oscillation modes (g-modes) are unique probes of the stellar radiative zone's properties. However, the rotation rate of the convective core remains unconstrained. Yet, measuring convective-core rotation holds valuable insights on stellar dynamics and chemical mixing since the properties of convection themselves are affected by rotation \citep{Augustson2019APenetration}, and rotation is a key parameter for core dynamo \citep{Brun2005SimulationsAction,Augustson2016THESTARS}, which is a progenitor of internal magnetism in evolved phases, the key to model compact object properties \citep[e.g.][for white dwarfs]{Bagnulo2022MultipleDwarfs}. Measuring intermediate-mass and massive stars' core rotation would thus allow us to better understand angular momentum redistribution and magnetic field generation in convective zones, happening over dynamical timescales \citep{Brun2004Global-scaleEnvelope, Brun2022PoweringStars}. 
\par Two families of stars on the MS have gained significant interest, exhibiting clear g-mode pulsations in \textsl{Kepler} \citep{Li2020Gravity-modeKepler,Papics2017SignaturesStars,Pedersen2022InternalTransport} and TESS \citep{Antoci2019TheMission,Garcia2022InternalDays} data: the massive Slowly-Pulsating B stars (SPB) and the intermediate-mass $\gamma$ Doradus stars. In the case of these g-mode pulsators, rotation alters the character of gravity modes propagating in the envelope. The Coriolis force acts as a restoring force alongside with the buoyancy and modifies the structure of the modes, becoming gravito-inertial modes. Rotation changes the spatial structure of such modes \citep{Lee1997Low-frequencyDependence,Dintrans2000OscillationsTheory,Mathis2008AngularWaves}, as well as their spacing in period \citep{Ballot20122DStars,Bouabid2013EffectsStars}, which would be constant in the asymptotic, non-rotating case \citep{Tassoul1980AsymptoticPulsations}.
\par Following this observation, the period-spacing between gravity modes of consecutive radial orders and of the same horizontal structure holds valuable information about the radiative interior of the stars. Any deviation from this constant pattern can arise from dynamical processes and mixing. Modulations in the period-spacing were proven to arise from sharp discontinuities of the mean molecular weight \citep{Miglio2008ProbingStars} and can be used as a test for profiles of convective penetration and entropy thermal stratification \citep{Michielsen2019ProbingAsteroseismology}. The slope of the period spacings in an inertial frame has been used to measure near-core rotation rates, a location at which g-modes reach the highest sensitivity \citep{VanReeth2015DetectingStudies,Ouazzani2017AStars,Christophe2018DecipheringStars}. With the surface rotation rate inferred from near-surface acoustic modes (p-modes) in the case of mixed $\gamma$ Doradus - $\delta$ Scuti pulsators \citep{Kurtz2014Asteroseismic11145123,Saio2015Asteroseismic9244992}, or inferred from rotational spot modulation \citep{VanReeth2018SensitivityStars}, the differential near-core to surface rotation has been measured in $\gamma$ Doradus stars, and proven to be limited, the ratio of surface rotation to core rotation ranging between 0.97 and 1.02 in the latter study. The sample of $\gamma$ Doradus stars exhibiting gravito-inertial modes is now large, with 611 stars in \textsl{Kepler} data showing prograde Kelvin g-modes, which have the highest visibility, and retrograde r-modes, corresponding to global Rossby waves \citep[see][for gravito-inertial modes typologies]{Townsend2003AsymptoticStars,Mathis2008AngularWaves}. Among those, 58 have measured near-core to surface differential rotation \citep{Li2020Gravity-modeKepler}.
\par Recently, \citet{Ouazzani2020FirstRevealed} made a breakthrough in the analysis of the period-spacing pattern, proving that pure inertial modes propagating in the convective core of the star, couple through the convective-radiative boundary with gravito-inertial modes. This process results in a characteristic dip in the period-spacing pattern, which was later confirmed by \citet{Saio2021RotationModes}, and observed in Kelvin g-modes series of 16 $\gamma$ Doradus stars. The location of the dip in period, as well as its width and depth, were proven to depend on stellar parameters and evolution stage, and the prescription of convective penetration in the radiative zone.
As a detailed understanding of the formation of the dip was lacking, \cite{Tokuno2022AsteroseismologyOscillations} described theoretically the coupling mechanism and found a Lorentzian shape for the dip, specific to this coupling mechanism compared to the dips created by a strong gradient in molecular weight \citep{Kurtz2014Asteroseismic11145123,Saio2015Asteroseismic9244992,Schmid2016Asteroseismic10080943B,Murphy2016Near-uniform7661054,Pedersen2018TheSpacings,Michielsen2019ProbingAsteroseismology,Li2019PeriodStars,Wu2020AsteroseismicCore}. This study derives a key coupling parameter controlling the formation of the dip, $\epsilon$, depending on the density stratification as well as the near-core rotation rate of the star. \citet{Galoy2024PropertiesStars} further conducted a thorough numerical study of the inertial dip formation using the spectral 2D oscillation code TOP \citep{Reese2009PulsationMethod, Ballot20122DStars}, deriving empirical relations for the width and the location of the dips as a function of the frequencies of the pure inertial modes in an isolated convective core and the near-core stratification gradient. This opens the possibility of future studies allowing structure inversions from the inertial dip structure observed in rapid g-mode pulsators from asteroseismic data. This study derives as well an improvement of the model of \cite{Tokuno2022AsteroseismologyOscillations}, taking into account a multi-mode interaction from both sides of the boundary.
\par The $\epsilon$ coupling parameter was estimated by forward modelling methods in \cite{Aerts2023ModeStars}, in a sample of 37 $\gamma$ Doradus and 26 SPB stars, and its dependency on the stellar age and the near-core rotation was established. The SPB sample currently used is not as rife as the $\gamma$ Doradus one and contains slower rotators, hence partly explaining the non-confirmed detection of inertial dips in SPBs. However, the $\gamma$ Doradus stars with confirmed inertial dips detection do not stand out in terms of their coupling parameter compared to the overall sample. A sole reasoning on the coupling parameter appears not fully satisfying in explaining the relatively low proportion of inertial dips findings in observed $\gamma$ Doradus period-spacing patterns (3 \%). This is mitigated by the fact that no ensemble analysis of inertial dip has been made so far, \citet{Saio2021RotationModes} relying on a visual inspection for their study.

\par 
While \citet{Ouazzani2020FirstRevealed}, \citet{Tokuno2022AsteroseismologyOscillations} and \citet{Galoy2024PropertiesStars} remained in the framework of solid-body rotation, \citet{Saio2021RotationModes} explored the possibility of allowing a steep differential rotation profile from the convective core to the near-core regions. Their study reveals a minimum of 0.85 for the ratio of the core rotation rate over the near-core rotation rate in stars exhibiting inertial dips. \citet{Moyano2024AngularStars} further used these results to put constraints on the angular momentum redistribution processes at play in the radiative interior of $\gamma$ Doradus stars. For that matter, it is key to further explore the effect of differential rotation, as we expect it to Doppler-shift the location of the dip in the period-spacing pattern and it might lead to a modification in its morphology.
We thus aim at extending the theoretical explanations of the signature of rotation on the g-modes period spacing given by \citet{Tokuno2022AsteroseismologyOscillations} to allow for differential rotation, and investigate the effects of differentiality on the coupling mechanism, the subsequent dip formation, and the shape of the dip. We propose an analytical two-zone model, with both the convective core and the radiative envelope rotating as solid bodies with different rotation rates, serving as a first laboratory toward the comprehension of the effect of differential rotation on the coupling. We aim to provide predictions for the measurement of such two-zone differential rotation, and investigate the detectability of such differentiality in perspective with other processes shifting the location of the dip: the density stratification in the core and the gradient of stratification at the interface of the two zones \citep{Ouazzani2020FirstRevealed}.
\par The outline of the paper is as follows: in Section 2, we describe the theoretical framework we are using, and summarize the results of \citet{Tokuno2022AsteroseismologyOscillations}. In Section 3, we derive the coupling equation in the presence of differential rotation and solve it both numerically and analytically. We describe the characteristics of the modified lorentzian profile obtained with differential rotation, in the case of Kelvin g-modes. In Section 4, we summarize the effects of differential rotation and put our findings in perspective with other processes influencing the position of the dip, tackling the detectability of such differential rotation in the analysis of inertial dips. We then conclude in Section 5 on the potentiality of differential rotation measurements and expose as well the limitations of our model and the leads that could be pursued by future studies on inertial dips.

\section{Framework and derivation of the structure of the modes from both sides of the convective-radiative boundary}

\subsection{Selection of a $\gamma$ Doradus star}

We are working on a model of g-mode pulsating main sequence star displaying a convective core surrounded by a radiative envelope. Two different stellar pulsators are classically comprised in this analysis: $\gamma$ Doradus stars, of 1.3 to 2.0 $\mathrm{M}_{\odot}$, and SPB stars, of 2 to 7 $\mathrm{M}_{\odot}$. In this work, we will focus on the case of $\gamma$ Doradus stars, as (1) more oscillation modes are detected in this stellar class, (2) the structural discontinuities exhibited by SPB stars would make the period-spacings modulations due to prominent molecular weight or temperature gradient, making the detection of a dip more difficult and (3) the sample of $\gamma$ Doradus stars observed period spacings is more abundant than the one of SPB stars and inertial dips in SPB stars were searched for in \cite{Saio2021RotationModes} and none were found in the sample studied by \cite{Papics2017SignaturesStars}.

The classical instability strip of $\gamma$ Doradus covers in the low-mass range stars close to their zero-age main sequence (ZAMS), and in the high-mass range stars close to their terminal-age main-sequence (TAMS). However, this picture is complicated by the detection of $\gamma$ Doradus at locations in the color-magnitude diagram far from the classical instability strip \citep[see e.g. Fig. 12 of ][]{Li2020Gravity-modeKepler}.

\subsection{Physical framework and rotational set-up}
We now present the relevant assumptions and physical approximations used to study core-to-envelope mode coupling, and discuss them with regard to the state of the art on gravito-inertial modes in the envelope and pure inertial modes in the core.

\subsubsection{Gravito-inertial modes typology}

Gravito-inertial modes form the predominant features in the spectrum of $\gamma$ Doradus stars. Their restoring forces are both buoyancy and the Coriolis force and are strongly affected by the latter both in their spatial structure and frequencies, compared to their pure gravity mode counterparts in non-rotating stars. Gravito-inertial modes can be classically separated into four classes based on their geometry and their restoring mechanism, as described for instance in \citet{Mathis2008AngularWaves}: Poincaré modes, existing in the non-rotating case, that are internal gravity waves modified by the Coriolis acceleration in rotating stars. Their radial wavenumber rapidly increases with rotation \citep{Townsend2003AsymptoticStars}, as their damping; r-modes, which are purely retrograde waves existing only for rapid rotators in the sub-inertial regime, driven by the conservation of the vorticity and curvature effect; Yanai modes, mix of the aforementionned modes with smaller radial nodes, hence less damped than the Poincaré modes; and Kelvin g-modes, also driven by the conservation of vorticity, which are purely prograde modes trapped around the equator of the star.

We stay within the framework of Kelvin g-modes, for two reasons:
First, Kelvin g-modes have the highest visibility in $\gamma$ Doradus stars and were by far the most observed in the sample of 611 stars analyzed in \cite{Li2020Gravity-modeKepler}.
Second, r-modes that were also observed in this sample would not display dips arising from the interacting mechanism we are describing, as \cite{Saio2021RotationModes} pointed out. Poincaré modes remain unobserved, while only 7 stars show Yanai modes to this date \citep{VanReeth2018SensitivityStars, Li2019PeriodStars}.


\subsubsection{Frequency regimes and relevant approximations}
\label{subsubsec:Freq_reg}

These gravito-inertial modes have different propagation regions based on the frequency regime at which they appear. As studied by \citet{Prat2016AsymptoticDynamics}, noting $\sigma_{\rm env}$ and $\sigma_{\rm core}$ the frequencies of the wave in the frames co-rotating with the envelope and the core, respectively, gravito-inertial waves are evanescent in the core in the super-inertial regime ($\sigma_{\text{env}} > 2\Omega_{\text{env}}$ and $\sigma_{\text{core}} > 2\Omega_{\text{core}})$, whereas they can penetrate in the core as pure inertial modes in the sub-inertial regime ($\sigma_{\text{core}} < 2\Omega_{\text{core}}$ and $\sigma_{\text{env}} < 2\Omega_{\text{env}}$). We will thus stay within the sub-inertial regime in both regions in our study, ignoring a trans-inertial regime in which we have ($\sigma_{\rm env} <  2\Omega_{\text{env}}$ and $\sigma_{\rm core} >  2\Omega_{\text{core}}$ or $\sigma_{\rm env} >  2\Omega_{\text{env}}$ and $\sigma_{\rm core} <  2\Omega_{\text{core}}$), which results in a chaotic behavior of the rays and reduced coupling between the core and the envelope \citep{Prat2018AsymptoticRotation}. Gravito-inertial modes are described in solid lines in Fig.~\ref{fig:sketch_gamma}. They are propagating between two turning points for which their frequency in the co-rotating frame equates the near-core Brunt-Väisälä ($N$) frequency ($r_{\text{a}}$) and the minimum between the near-surface Brunt-Väisälä and Lamb ($L$) frequencies ($r_{\text{b}}$).

In the sub-inertial regime, waves thus propagate in the core and gain a purely inertial character, because buoyancy can no longer act as a restoring force in an unstratified region. While these waves have been long studied and observed in geophysics, \cite{Wu2005ORIGINMODES} made a comprehensive study of the properties of pure inertial modes in an astrophysical context.

The equation governing the structure of pure inertial modes is only separable for solid-body rotating spheres with constant density. We thus choose for the density profile a uniform averaged density in the core, for the study to remain analytical. This assumption will be further discussed in subsection \ref{subsec:dens_strat}.  Pure inertial waves are symbolized as dashed oscillations in Fig.~\ref{fig:sketch_gamma} and propagate within the convective core $r< R_{\text{core}}$.

We adopt the Traditional Approximation of Rotation (TAR) within the envelope \citep[e.g.][]{Gerkema2008GeophysicalApproximation}. The TAR consists in neglecting the latitudinal component of the rotation vector in the Coriolis acceleration. The TAR is known to hold in highly stratified layers of stellar interiors. For the sub-inertial Kelvin g-modes in which we are interested, we must stay within the regime $N >(\gg) 2\Omega$, for which the TAR solution approaches the solution obtained with a full treatment of the Coriolis force \citep{Prat2016AsymptoticDynamics}. This cannot be satisfied in the convective core, with the Brunt-Väisälä frequency reaching non-positive real values in this zone. Hence the inertial dips could only be seen in numerical studies for which the TAR was lifted \citep{Saio2018TheoryStars,Ouazzani2020FirstRevealed}. As for the envelope, the strong molecular weight gradient near the core ensures high values of the Brunt-Väisälä frequency even in the near-core region. This sharp stratification gradient at the core-envelope boundary ensures that the TAR holds for gravito-inertial modes, and is thus part of our studied cases, as well as in \cite{Tokuno2022AsteroseismologyOscillations}. This allows the region of radial coordinates $[R_{\text{core}}; r_{\text{a}}]$ in which the TAR fails and gravito-inertial modes become evanescent to remain small so that the structure of such modes remain close to the form predicted by the TAR at $r_{\text{a}}$.
Throughout this study, we neglect as well the wave's gravific potential fluctuation \citep{Cowling1941TheStars}, approximation justified by the strongly radially oscillating character of the modes.

\subsubsection{Differential rotation framework}

\begin{figure}
    \centering
    \includegraphics[width=0.8\linewidth]{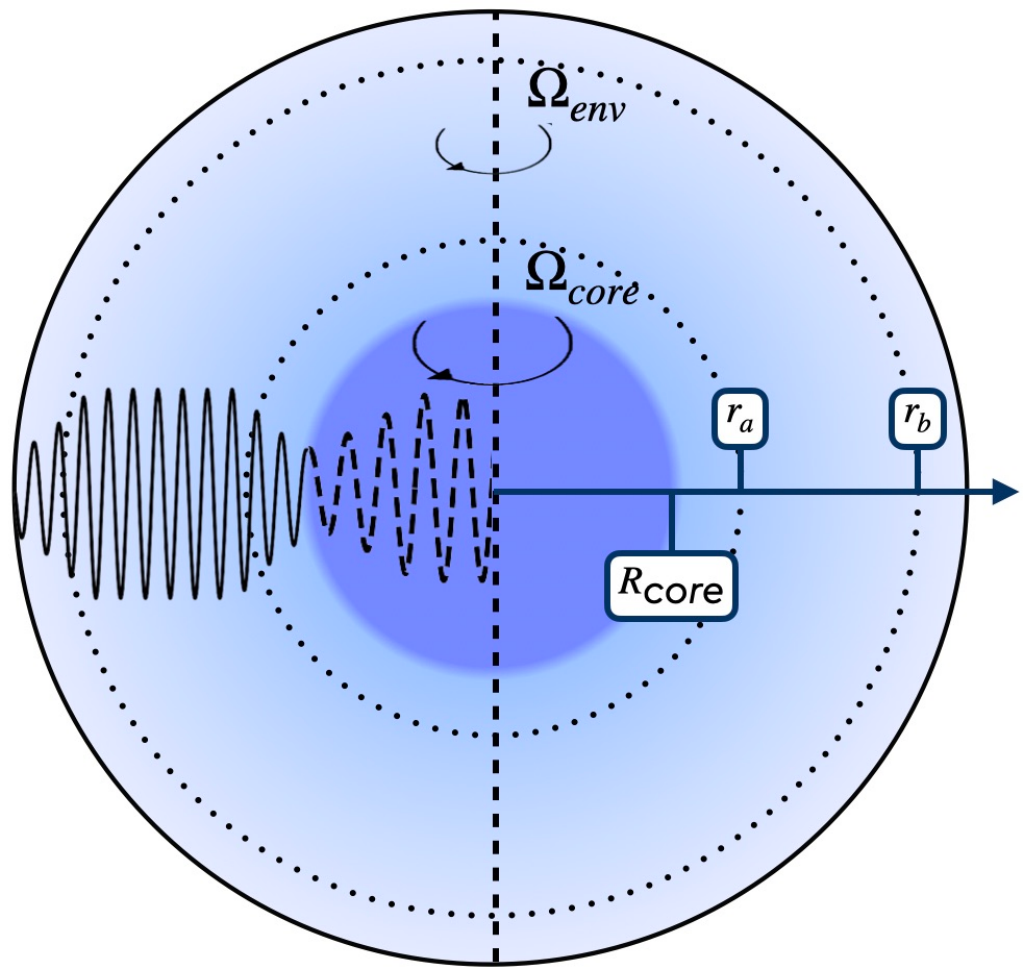}
    \caption{Sketch of the rotation profile described: $\Omega_{\text{core}}$ for $r<R_{\text{core}}$, $\Omega_{\text{env}}$ for $r>R_{\text{core}}$ gravito-inertial modes are propagating in the zone $r\in[r_{\text{a}};r_{\text{b}}]$, being evanescent in the region $r\in[R_{\text{core}};r_{\text{a}}]$. Pure-inertial modes are trapped in the convective core $r<R_{\text{core}}$. The matching of the relevant quantities is done at the location $r=R_{\text{core}}$.}
    \label{fig:sketch_gamma}
\end{figure}

We allow a differential rotation between the convective core and the radiative surrounding envelope. This takes the form of 
a bi-layer rotation rate, with $\Omega_{\text{env}}$ the rotation rate of the radiative envelope and $\Omega_{\text{core}}$ the rotation rate of the convective core (see Fig.~\ref{fig:sketch_gamma}), parametrized as:

\begin{equation}
\Omega_{\text{env}} = \alpha_{\text{rot}} \Omega_{\text{core}}
\label{eq:rel_freq}
\end{equation}
with $\alpha_{\text{rot}}$ accounting for the degree of differential rotation between the core and the envelope.

In the convective core, the rotation is known to approach cylindricity from the application of the Taylor-Proudmann theorem. The core rotation profile was treated by hydrodynamical \citep{Browning2004SimulationsOvershooting} and magneto-hydrodynamical \citep{Brun2005SimulationsAction,Featherstone2009EffectsStars,Augustson2016THESTARS} simulations, showing an impact of the magnetic field on the limitation of cylindrical differential rotation in the core (5 \% of radial differential rotation at the equator for model M4 of \citet{Augustson2016THESTARS}). We decide to stay in the framework of solid-body rotation to maintain the analyticity of the inertial mode solutions. Indeed, the introduction of a convective core differential rotation breaks the regularity of the solutions: the wave structure is of an attractor type \citep{Baruteau2013InertialShell} and critical layers can form, that damp the inertial waves \citep{Astoul2021TheWaves}. Additionally, seismology allows probing an averaged rotation rate of a spherical zone, thus advocating for a solid-body treatment of the core rotation rate in our study, while a two-zone analytical treatment is useful to get a precise physical understanding of the dip formation in the context of differential rotation.

As for the radiative zone, transport processes can lead to differential rotation \citep{Maeder2000THESTARS,Rogers2015ONSTARS}. The topic was treated in \cite{VanReeth2018SensitivityStars}, where it was shown that differential rotation within the envelope was hard to constrain from period-spacings, unless multiple mode series are exhibited, and was less than $5\%$ between the surface and the near-core region in the stars in which a measurement was possible from starspot modulations.
This approximation of a solid-body rotating radiative zone is thus done because differential rotation is limited and the key quantity for the coupling is only the near-core rotation. We hereby emphasize that the uniform rotation rate of the envelope $\Omega_{\rm env}$ we take would correspond to the near-core rotation rate in a differentially rotating radiative zone. Differential rotation in our study is then probed between the convective core and the inner radiative envelope near the core and does not correspond to the differential rotation probed by the curvature gravito-inertial modes and rotational spot modulations \citep[see e.g.][]{VanReeth2018SensitivityStars}.

\subsection{Formalism and description of the relevant reference frames}

We here describe the expansion of the quantities in both the two co-rotating frames relevant to this study and the inertial one.
In the rotational set-up previously described, each scalar and vectorial field will be expanded in Fourier's series in the azimuthal angle $\varphi$ and time t:
\begin{equation}
X(r,\theta,\varphi,t) = \sum_{\sigma_{\text{zone}}, m}X'(r,\theta) e^{i(m\varphi + \sigma_{\text{zone}} t)}
\end{equation}
\begin{equation}
\boldsymbol{x}(r,\theta,\varphi,t) = \sum_{\sigma_{\text{zone}}, m}\boldsymbol{x}'(r,\theta) e^{i(m\varphi + \sigma_{\text{zone}} t)} \, .
\end{equation}
We emphasize that with the chosen sign convention, modes with negative $m$ are prograde, and modes with positive $m$ are retrograde.

$\sigma_{\text{zone}}$ being either $\sigma_{\text{env}}$ or 
$\sigma_{\text{core}}$, the two local wave angular frequencies related to the frames rotating at respectively $\Omega_{\text{env}}$ and $\Omega_{\text{core}}$, linked to the frequency in the inertial frame $\sigma_{\rm in}$ as such:
\begin{equation}
    \sigma_{\text{env}} = \sigma_{\text{in}} + m\Omega_{\text{env}}  \, ,
\end{equation}

\begin{equation}
    \sigma_{\text{core}} = \sigma_{\text{in}} + m\Omega_{\text{core}}  \, .
\end{equation}
The frequency decomposition in the two co-rotating frames is thus dependent on the zone in which the waves propagate. This is a differing key point from the solid-body rotation case treated by \cite{Tokuno2022AsteroseismologyOscillations}.
Since we are considering a differentially-rotating case, we derive the expressions for the quantities in the co-rotating frames, and further ensure the matching between the two zones in the inertial frame. Taking into account a Doppler shift related to the two zones on either side of the boundary in the phase term, the relevant quantities to match in the decomposition are thus:
\begin{equation}
    X'(r, \theta)e^{i(m \varphi + \sigma_{\text{zone}}t)}e^{-im\Omega_{\text{zone}}t} = X'(r, \theta)e^{i(m \varphi + \sigma_{\text{in}}t)}
\end{equation}
and
\begin{equation}
    \boldsymbol{x}'(r, \theta)e^{i(m \varphi + \sigma_{\text{zone}}t)}e^{-im\Omega_{\text{zone}}t} = \boldsymbol{x}'(r, \theta)e^{i(m \varphi + \sigma_{\text{in}}t)}
\end{equation}


We further define the two spin parameters, which are the relevant parameters in the following analysis, as:
\begin{equation}
    s_{\text{env}} = \frac{2\Omega_{\text{env}}}{\sigma_{\text{env}}}  \, ,
\end{equation}
\begin{equation}
    s_{\text{core}} = \frac{2\Omega_{\text{core}}}{\sigma_{\text{core}}}  \, .
\end{equation}
We place ourselves in the sub-inertial regime, for which $\sigma_{\text{env}} < 2\Omega_{\text{env}}$ and $\sigma_{\text{core}} < 2\Omega_{\text{core}}$, thus $s_{\text{env}} > 1$ and $s_{\text{env}} > 1$. As argued in paragraph \ref{subsubsec:Freq_reg}, this is the regime where inertial modes can propagate in the convective core, and we are thus considering low-frequency waves.

\subsection{Expression of mode structures from both sides of the boundary}
With this bi-layer differentially rotating framework defined, the derivation of the structure of modes from both sides of the boundary is very similar to subsection 3.1.1 and 3.1.2 of \citet{Tokuno2022AsteroseismologyOscillations}, respectively for gravito-inertial modes propagating in the radiative envelope and pure inertial modes propagating in the convective core. The spin parameter the functions depend on in those cases is no longer fixed at a single value $s$, but rather at $s_{\rm env}$ and $s_{\rm core}$. The full calculations are given respectively in Appendix \ref{Appendix:g-i} and \ref{Appendix:pure_inertial} for readability. \\
We obtain the expressions for the radial displacement and the Eulerian perturbation of pressure at $R_{\rm core}$.\\
For gravito-inertial modes, corresponding to Hough solutions \citep[see e.g.][]{Lee1997Low-frequencyDependence}, the structure at the boundary depends on the hypothesis made on the continuity or discontinuity of the near-core Brunt-Väisälä profile. In the former continuous case:
\begin{equation}
\frac{\xi_{r}'}{r}\Bigg|_{R_{\text{core}}} = \sum_{k}a_{k}\epsilon X_{k}^{m}(s_{\text{env}})\Theta_{k}^{m}(\mu; s_{\text{env}})
\label{eq:displacement_env_in}
\end{equation}

\noindent and
\begin{equation}
\frac{p'}{\bar{p}\Gamma_{1}}\Bigg|_{R_{\text{core}}} = 
\frac{\sigma_{\text{env}}^{2} R_{\text{core}}^{2}}{c_{S}^{2}}\sum_{k}a_{k}Y_{k}^{m}(s_{\text{env}})\Theta_{k}^{m}(\mu ; s_{\text{env}}) \, .
\label{eq:pressure_env_in}
\end{equation}
In the latter discontinuous case, the functions $X_{k}^{m}$ and $Y_{k}^{m}$ are replaced by respectively $\tilde{X}_{k}^{m}$ and $\tilde{Y}_{k}^{m}$ described in Appendix \ref{Appendix:g-i}.

\noindent We have introduced $\mu = \cos \theta$, $\Theta_{k}^{m}$ the Hough functions described in the Appendix \ref{Appendix:g-i} and $c_{S}$ the local sound speed. For pure inertial modes, we obtain Bryan solutions \citep{Bryan1889TheEllipticity}:
\begin{equation}
\frac{\xi_{r}'}{r}\Big{|}_{R_{\text{core}}}= \sum_{l}b_{l}C_{l}^{m}(1/s_{\text{core}})\tilde{P}_{l}^{m}(\mu) \, ,
\label{eq:displacement_core_in}
\end{equation}
and

\begin{equation}
\frac{p'}{\bar{p}\Gamma_{1}} \Big{|}_{R_{\text{core}}} 
=\frac{\sigma_{\text{core}}^{2}R_{\text{core}}^{2}}{c_{S}^{2}}\sum_{l}b_{l}P_{l}^{m}(1/s_{\text{core}})\tilde{P}_{l}^{m}(\mu) \, ,
\label{eq:pressure_core_in}
\end{equation}
Where $P_{l}^{m}$ are the Legendre polynomials and $\tilde{P}_{l}^{m}$ their normalized counterparts (see Appendix \ref{Appendix:pure_inertial}). These solutions take into account the Doppler shift induced by the differential rotation.

\section{Coupling of the core and envelope modes in a differentially rotating context}

With the expressions previously derived at both sides of the convective core boundary, we investigate the coupling of pure inertial modes in the convective core and gravito-inertial modes in the radiative envelope considering differential two-zone rotation. We describe the relevant assumptions allowing us to simplify the coupling equation. We solve the equation numerically and derive an analytical shape for the resulting inertial dip in the period-spacing pattern.

\subsection{Matrix formulation of the problem}
\label{subsec:matrix}
In this section, we investigate the problem of the continuity of the relevant quantities in the specific context of a continuous Brunt-Väisälä profile, and a two-zone differential rotation described in Fig.~\ref{fig:sketch_gamma}. The matching must be achieved at fixed azimuthal number $m$ and frequency $\sigma_{\rm in}$ in the inertial frame. Thus, compared to a one-zone rotation model, the problem can no longer be equivalent to working at a fixed spin parameter in the two-zone case. The discontinuous case is treated in subsection \ref{subsec:disc}. 

The two quantities that must stay continuous at the two sides of the interface are the radial Lagrangian displacement $\xi_{r}$ and the Lagrangian perturbation of pressure $\delta p = p' + \xi_{r}(\frac{\partial \bar{p}}{\partial r})$. Except for the rotation frequency, background quantities are continuous at the interface in the case of the continuous Brunt-Väisälä profile.

Given the density concentration in the core of $\gamma$ Doradus stars, and the critical rotation frequency scaling as $\Omega_{\rm crit} \propto \sqrt{\mathcal{G}\rho}$, even for fast rotators close to $\Omega_{\text{surface}} = 0.5 \Omega_{\rm crit,\text{surface}}$, the rotation frequency at the edge of the convective core will be negligible compared to the critical rotation rate at this location. Thus the contribution of the centrifugal acceleration is considered to be negligible with respect to the local gravity at the core/envelope boundary in this study \citep[e.g.][]{Ballot2010GravityStars}.

The continuity of the radial Lagrangian displacement being already ensured, the continuity of the Lagrangian perturbation of pressure is equivalent to the one of the Eulerian pressure $p'$. Using Eqs.~ \eqref{eq:displacement_env_in} and \eqref{eq:displacement_core_in}, the matching equations for the Lagrangian displacement writes:
\begin{align}
    \sum_{k}&a_{k}\epsilon X_{k}^{m}(s_{\text{env}})\Theta_{k}^{m}(\mu;s_{\text{env}})\nonumber \\ & =\sum_{l}b_{l}C_{l}^{m}(1/s_{\text{core}})\tilde{P}_{l}^{m}(\mu) \, .
    \label{eq:match_displacement_dif}
\end{align}
The matching of the Eulerian perturbation of the pressure writes, using Eq.~ \eqref{eq:pressure_env_in} and Eq.~ \eqref{eq:pressure_core_in}:
\begin{align}
    \sigma_{\text{env}}^{2}\sum_{k}&a_{k}Y_{k}^{m}(s_{\text{env}})\Theta_{k}^{m}(\mu;s_{\text{env}})\nonumber \\ &= \sigma_{\text{core}}^{2}\sum_{l}b_{l}P_{l}^{m}(1/s_{\text{core}})\tilde{P}_{l}^{m}(\mu) \, .
    \label{eq:match_pressure_dif}
\end{align}

Projecting each of the last two equations $(\boldsymbol{\cdot})$ on Hough functions, by taking
 $\int_{-1}^{1}(\boldsymbol{\cdot})\Theta_{k}^{m}(\mu;s_{\text{env}})\mathrm{d}\mu$, we reduce the dimensionality of the equations, taking profit of the orthogonality of Hough functions.

\begin{equation}
    a_{k}\epsilon X_{k}^{m}(s_{\rm env})=\sum_{l}b_{l}C_{l}^{m}(1/s_{\text{core}})c_{k,l}\, ,
    \label{eq:proj_dis}
\end{equation}

\noindent and 

\begin{equation}
    \sigma_{\text{env}}^{2}a_{k}Y_{k}^{m}(s_{\text{env}})=\sigma_{ \text{core}}^{2}\sum_{l}b_{l}P_{l}^{m}(1/s_{\text{core}})c_{k,l} \, ,
    \label{eq:Y}
\end{equation}

\noindent the coefficient $c_{k,l}$ being defined as: 
\begin{equation}
    c_{k,l} \equiv \int_{-1}^{1}\Theta_{k}^{m}(\mu, s_{\text{env}})\tilde{P}_{l}^{m}(\mu)\mathrm{d}\mu\, .
    \label{eq:def_c}
\end{equation}


Injecting the expressions for the $a_{k}$ given in Eq. \eqref{eq:proj_dis} in Eq. \eqref{eq:Y}, one can recast the equations in a matrix form:

\begin{equation}
[\mathcal{M}(s_{\text{core}},s_{\text{env}}) - \epsilon \mathcal{N}(s_{\text{core}},s_{\text{env}})]\vec{b} = \vec{0} \, ,
\label{eq:matrix_eq}
\end{equation}
$\vec{b}$ being the column vector of the terms $b_{l}$. The matrices are defined as:
\begin{equation}
    [\mathcal{M}]_{k,l} = c_{k,l}\sigma_{\text{env}}^{2}Y_{k}^{m}(s_{\text{env}})C_{l}^{m}(1/s_{\text{core}})
\end{equation}
and
\begin{equation}
    [\mathcal{N}]_{k,l} = c_{k,l}\sigma_{\text{core}}^{2}X_{k}^{m}(s_{\text{env}})P_{l}^{m}(1/s_{\text{core}}) \, .
\end{equation}
For $\vec{b}$ to be non trivial, we have the following condition:
\begin{equation}
    \text{det}[\mathcal{M}(s_{\text{core}},s_{\text{env}})-\epsilon \mathcal{N}(s_{\text{core}},s_{\text{env}})] = 0 \, .
\end{equation}

To determine the frequencies of the modes, the system has to be closed by an equation linking $s_{\text{core}}$ and $s_{\text{env}}$. Such an equation arises from the fact that the match must occur at fixed $\sigma_{\text{in}}$, the frequency in the inertial frame. We have in each zone:
\begin{equation}
    s_{\text{zone}} = \frac{2\Omega_{\text{zone}}}{\sigma_{\text{in}} + m\Omega_{\text{zone}}}
\end{equation}
Thus, with Eq.~ \eqref{eq:rel_freq}:
\begin{equation}
    s_{\text{env}} = \frac{\alpha_{\text{rot}}s_{\text{core}}}{1+\frac{m}{2}(\alpha_{\text{rot}}-1)s_{\text{core}}} \, .
    \label{eq:link_s}
\end{equation}
For concision, we define $G$ as:
\begin{equation}
    G(s) = \frac{\alpha_{\text{rot}}s}{1+\frac{m}{2}(\alpha_{\text{rot}}-1)s}
\end{equation}
so that $s_{\text{env}} = G(s_{\text{core}}) \Longleftrightarrow s_{\text{core}} = G^{-1}(s_{\text{env}})$.

Due to the choice of the bi-layer rotation profile, the function $G^{-1}$ contains a pole at $1+\frac{m}{2}(1/\alpha_{\rm rot} - 1)s_{\rm env}$, which corresponds to the case of infinitely low frequencies of the core modes: $\sigma_{\rm core} = 0$. The dip model using this function $G^{-1}$ thus must be used for envelope spin parameters below this limit.

\subsection{Synthesized coupling equation}
\label{subsec:determinant}

First, putting $\epsilon = 0$, case of $\frac{dN^{2}}{dr} \rightarrow \infty$ at $R_{\text{core}}$, and no avoided crossing, we see that we have either $Y_{k}^{m} = 0$ (pure inertial mode in the core) or $C_{l}^{m} = 0$ (pure gravito-inertial mode in the envelope). 
We then suppose $Y_{k}^{m}(s_{\text{env}}) = O(\epsilon)$ and $C_{l}^{m}(s_{\text{core}}) = O(\epsilon)$.
In a situation of $\epsilon \neq 0 \ll 1$, the mode extends now over all radial coordinates of the star and is no longer confined either in the convective core or the radiative envelope. Considering a fixed energy input from the source of excitation (flux-blocking at the bottom of the outermost convective zone in the case of $\gamma$ Doradus stars), the amplitude of an interacting mode on the two sides of the boundary must be underdominant compared to non-interacting modes. This is why we can consider that both $Y_{k}^{m}$ and $C_{l}^{m}$ are of the order $O(\epsilon)$.

The spectrum of the pure inertial modes in frequency is sparser than the one of the gravito-inertial modes. We then consider the interaction of one mode of the core with several modes of the envelope with a single value of $k$. The determinant thus reads:

\begin{align}
&\text{det}[\mathcal{M}(s_{\text{core}},s_{\text{env}})-\epsilon \mathcal{N}(s_{\text{core}},s_{\text{env}})] \nonumber\\ &=\text{det}\left(\begin{NiceArray}{ccc:c:ccc}
 &  &  & / &  & & \\
 & O(1) &  & / & & O(1)& \\
 &  &  & / &  & & \\
\hdottedline
 /& / & / & / & / & / & / \\
\hdottedline
 &  &  &  /&  & & \\
 &  O(1)&  & / &  & O(1) & \\
 &  &  &  /&  & & 
\end{NiceArray}\right)\nonumber \\ &=
Y_{k}^{m}(s_{\text{env}})C_{l}^{m}(1/s_{\text{core}}) \nonumber\\
& \times \text{det}
\left(\begin{NiceArray}{ccc:c:ccc}
 &  &  &  &  & & \\
 & O(1) &  & O(1) & & O(1)& \\
 &  &  &  &  & & \\
\hdottedline
 & O(1) &  & O(1/\epsilon) & & O(1) &  \\
\hdottedline
 &  &  &  &  & & \\
 &  O(1)&  &O(1)  &  & O(1) & \\
 &  &  &  &  & & 
\end{NiceArray}\right)
\, .
\end{align}

The matrix at the right-hand side of this equation has one dominant term, at the crossing of the $l$th line and the $k$th column, the coupled modes.
Putting this determinant null, each term in its development with $\epsilon$ must be null. We have for the leading term of order $O(1/\epsilon)$:
\begin{equation}
    c_{k,l} - \epsilon\frac{P_{l}^{m}(1/s_{\text{core}})\sigma_{\text{core}}^{2}X_{k}^{m}(s_{\text{env}})}{C_{l}^{m}(1/s_{\text{core}})\sigma_{\text{env}}^{2}Y_{k}^{m}(s_{\text{env}})} c_{k,l}= 0 \, .
\end{equation}
We thus obtain:
\begin{equation}
    \frac{C_{l}^{m}(1/s_{\text{core}})\sigma_{\text{env}}^{2}Y_{k}^{m}(s_{\text{env}})}{P_{l}^{m}(1/s_{\text{core}})\sigma_{\text{core}}^{2}X_{k}^{m}(s_{\text{env}})} \simeq \epsilon \, .
    \label{eq:coupling}
\end{equation}
This reasoning holds only if $c_{k,l}$ is not negligibly small, of the order of $\epsilon$. This geometric consideration was first used as a criterion for the interaction of modes by \cite{Ouazzani2020FirstRevealed} and lies on top of the criterion on the small value of $\epsilon$ developed in \cite{Tokuno2022AsteroseismologyOscillations}. \citet{Galoy2024PropertiesStars} go further in the development of the determinant by not only taking the most dominant term but by reducing this infinite system of equations to a finite one, truncating it to the mode interactions with appreciable $c_{k,l}$ values. The complexity is increased by the fact that the number of considered interactions is not known a priori and can vary from one interaction to another. Therefore, using \citet{Tokuno2022AsteroseismologyOscillations}'s model can be considered as less precise but more general. We let this promising lead of using the \citet{Galoy2024PropertiesStars} model in the context of differential rotation for further studies.
Defining: 
\begin{equation}
    F_{l}^{m}(s_{\text{core}}) \equiv -\frac{C_{l}^{m}(1/s_{\text{core}})}{P_{l}^{m}(1/s_{\text{core}})} \, ,
\end{equation}
the coupling equation between the Lagrangian displacement and the Lagrangian pressure perturbation in case of differential rotation thus reads, with the expressions of $X_{k}^{m}$ and $Y_{k}^{m}$ given in Appendix~\ref{Appendix:g-i}:
\begin{eqnarray}
\lefteqn{F_{l}^{m}(s_{\text{core}})\frac{\sqrt{3}}{2} \frac{\alpha}{\Lambda_{k}^{m}(s_{\text{env}})}s_{\text{env}}^{2/3}\frac{\sigma_{\text{env}}^{2}}{\sigma_{\text{core}}^{2}}} \nonumber \\
& & \times \left[\cot\left(\frac{\pi^2s_{\text{env}}}{\Omega_{\text{env}}\Pi_{0}}-\frac{\pi}{6}\right)+\frac{1}{\sqrt{3}}\right] \simeq \epsilon \, .
\label{eq:final_coupling}
\end{eqnarray}

\subsection{Solving the coupling system}
\label{subsec:Num_section}

\begin{figure}[ht]
        \includegraphics[width = \linewidth]{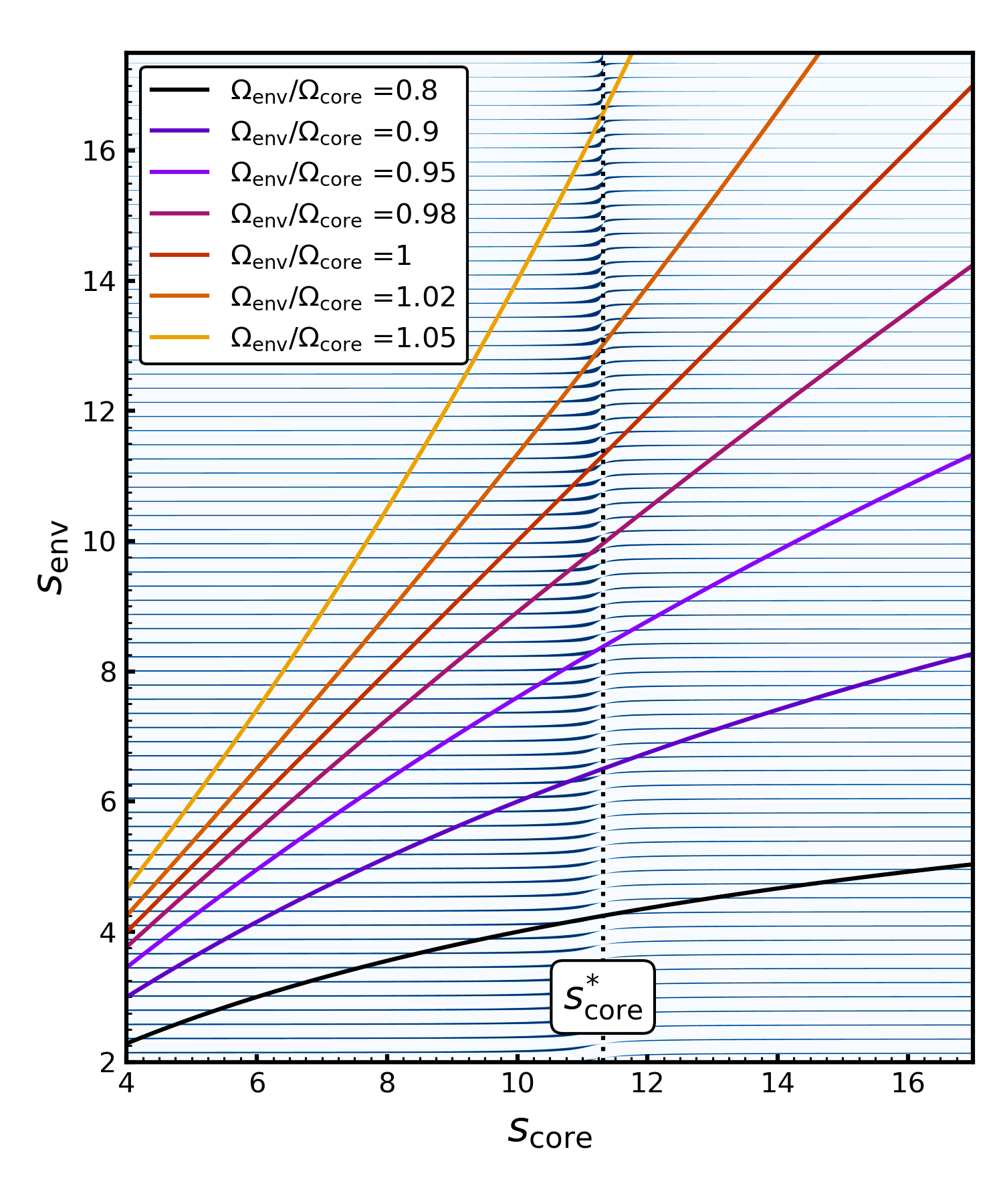}
        \caption{Numerical solutions of Eq.~\eqref{eq:final_coupling}, for the interaction between the $(k=0,m=-1)$ Kelvin g-modes and a $(l=3,m=-1)$ pure inertial mode in the core, as described in Section \ref{sec:Application}, with $\epsilon = 6.19\times10^{-3}$. The location of the inertial modes in the $(s_{\text{core}};s_{\text{env}})$ coordinates are to be found at the intersection between the zeros of the coupling equation (in blue) and the characteristics $s_{\text{env}} = G(s_{\text{core}})$, denoted by full lines for different ratios $\Omega_{\text{env}}/\Omega_{\text{core}}$. The core spin parameter of the pure inertial mode ($s_{\rm core}^{*}$) is denoted by a dashed vertical line.}
        \label{fig:numerics}
\end{figure}

The solving method differs from the solid-body rotation case. Instead of finding the zeros of a 1D function, we have to find the zeros of a 2D function: $g_{\epsilon}(s_{\text{core}}, s_{\text{env}}) = 0$, $g_{\epsilon}$ being defined as the LHS of Eq.~\eqref{eq:final_coupling} minus the coupling parameter. We compute numerically such zeros, for various $\epsilon$ values, and their location in the $(s_{\text{core}},s_{\text{env}})$ plane is given by the blue lines in Fig.~\ref{fig:numerics} for a specific $\epsilon$ value. Far from the spin parameter of the pure inertial mode $s^{*}_{\text{core}}$ (vertical dashed line in Fig.~\ref{fig:numerics}), the gravito-inertial modes in the envelope are not influenced by the interaction and the zeros are regularly spaced in envelope spin parameters (y-axis). On the other hand, the interaction with the pure inertial mode shifts the envelope spin parameters of the zeros close to $s^{*}_{\text{core}}$.

In addition, we are evolving on specific coordinates of the surface $g_{\epsilon}(s_{\text{core}}, s_{\text{env}}) = 0$. Indeed, $s_{\text{core}}$ and $s_{\text{env}}$ are linked \textit{via} Eq.~\eqref{eq:link_s}. The solution of the coupling rotation is thus to be found on a characteristic unique of each $\alpha_{\text{rot}}$. For differential rotations ranging from $\alpha_{\rm rot} = {\Omega_{\rm env}}/{\Omega_{\rm core}} = 0.80$ to $\alpha_{\rm rot} = 1.05$, the characteristics are marked as full lines on Fig.~\ref{fig:numerics}. The solid-body rotating case is shown by the red line. Depending on the characteristic, hence the differential rotation value, we can see that the location of the spin parameter of the envelope at which the interaction occurs is shifted. The local slope of the curve at $s_{\text{core}} = s^{*}_{\text{core}}$ also varies, which shows that with an increasing differential rotation from the core to the envelope, less gravito-inertial modes will be influenced by the pure inertial mode, thus the dip will be thinner with increasing differentiality. The curvature of the characteristics introduces a non-symmetricity of the dip relative to its minimum, which however stays small for this range of differential rotation. We further interpret this effect in subsection ~\ref{subsec:effect_diff_rot}.

\subsection{Frequency-dependence of the dips}
\label{subsec:freq_dep}
We recall from subsection \ref{subsec:matrix} that $s_{\text{env}} = G(s_{\text{core}})$. The coupling equation is then:
\begin{eqnarray}
\lefteqn{F_{l}^{m}(G^{-1}(s_{\text{env}}))\frac{\sqrt{3}}{2}\frac{\alpha}{\Lambda_{k}^{m}(s_{\text{env}})}s_{\text{env}}^{2/3}\alpha_{\text{rot}}^{2}\frac{G^{-1}(s_{\text{env}})^{2}}{s_{\text{env}}^{2}}} \nonumber \\ 
& & \times \left[\cot\Big{(}\frac{\pi^2s_{\text{env}}}{\Omega_{\text{env}}\Pi_{0,\text{env}}}-\frac{\pi}{6}\Big{)}+\frac{1}{\sqrt{3}}\right] \simeq \epsilon
\end{eqnarray}
We will now expand the front function near its zero, $s_{\text{core}}^{*}$, considering the interaction to happen at close spin parameter values. The values of the zeros are the ones given by Eq.~ \eqref{eq:def_C}. Those values are presented in Table 1 of \cite{Ouazzani2020FirstRevealed}. We obtain:
\begin{align}
    &\frac{F_{l}^{m}(s_{\text{core}})G(s_{\text{core}})^{2/3}}{\Lambda_{k}^{m}(G(s_{\text{core}}))}\left(\frac{s_{\text{core}}}{G(s_{\text{core}})}\right)^{2} \nonumber\\
    &=(s_{\text{core}}-s_{\text{core}}^{*})\frac{\mathrm{d}F_{l}^{m}(s)}{\mathrm{d}s}\Big|_{s_{\text{core}}^{*}}\frac{G(s_{\text{core}}^{*})^{2/3}}{\Lambda_{k}^{m}(G(s_{\text{core}}^{*}))}\left(\frac{s_{\text{core}}^{*}}{G(s_{\text{core}}^{*})}\right)^{2} \, .
\end{align}
As in \citet{Tokuno2022AsteroseismologyOscillations}, we rewrite the coupling equation using $V_{\text{diff}}$:
\begin{equation}
    V_{\text{diff}}=-\left[\frac{\mathrm{d}F_{l}^{m}(s)}{\mathrm{d}s}\frac{\sqrt{3}}{2}\frac{\alpha G(s)^{2/3}}{\Lambda_{k}^{m}(G(s))}\alpha_{\text{rot}}^{2}\left(\frac{s}{G(s)}\right)^{2}\right]_{s_{\text{core}}^{*}} \, .
\end{equation}
The correction of V between the differentially and solid-body rotating cases is then:
\begin{equation}
    \frac{V_{\text{diff}}}{V}=\left[\alpha_{\text{rot}}^{2}\Big(\frac{G(s)}{s}\Big)^{-4/3}\Big(\frac{\Lambda_{k}^{m}(G(s))}{\Lambda_{k}^{m}(s)}\Big)^{-1}\right]_{s_{\text{core}}^{*}} \, .
\end{equation}
With that definition, we retrieve, dropping the $\rm env$ subscript for envelope spin parameters for readability:
\begin{equation}
    \cot\left(\frac{\pi^{2}s}{\Omega_{\text{env}}\Pi_{0}}-\frac{\pi}{6}\right)+\frac{1}{\sqrt{3}}\simeq -\frac{\epsilon/V_{\text{diff}}}{G^{-1}(s)-G^{-1}(s)^{*}} \, .
    \label{eq:coupling_simplified}
\end{equation}
To compute the period spacing expression, we write Eq.~ \eqref{eq:coupling_simplified} for two neighbouring solutions $s_{1}$ and $s_{2}$ with $s_{2} > s_{1}$. We assume that the frequency spectrum of the gravito-inertial modes is dense. The details of our computation are given in Appendix \ref{App:Mod_lorentz}. Using the vicinity in period of the solutions, we show that the Taylor expansion gives:

\begin{align}
    &\left[1+\Big(\frac{\epsilon /V_{\text{diff}}}{\bar{s}-s^{*}}\left(\frac{\mathrm{d}G^{-1}}{\mathrm{d}s}\Big{|}_{\frac{\bar{s}+s^{*}}{2}}\right)^{-1}+\frac{1}{\sqrt{3}}\Big)^{2}\right] \nonumber \\
    & \times \Big[\frac{\pi^{2}(s_{1}-s_{2})}{\Omega_{\text{env}} \Pi_{0}}-\pi\Big]\simeq \nonumber \\
    & -\frac{\epsilon}{V_{\text{diff}}}\frac{s_{1}-s_{2}}{(\bar{s}-s^{*})^{2}}\left(\frac{\mathrm{d}G^{-1}}{\mathrm{d}s}\Big{|}_{\frac{\bar{s}+s^{*}}{2}}\right)^{-2}
    \left(\frac{\mathrm{d}G^{-1}}{\mathrm{d}s}\Big{|}_{\bar{s}}\right) \, ,
    \label{eq:Taylor_exp}
\end{align}
with $s^{*} = G(s_{core}^{*})$ and $\bar{s} = (s_1 + s_2)/2$. This equation can thus be easily compared with equation (63) of \citet{Tokuno2022AsteroseismologyOscillations}. The effect of the differential rotation is comprised in the function $G$, through the derivative of $G^{-1}$ and $V_{\rm diff}$. To get further insights we derive the approximate expression of $\Delta \rm P$ as function of $\rm P$. We define $\mathrm{P} = \pi \bar{s}/\Omega_{\text{env}}$, $\Delta \mathrm{P} = \pi(s_{1}-s_{2})/\Omega_{\text{env}}$, $\mathrm{P}^{*} = \pi s^{*}/\Omega_{\text{env}}$ and $\Gamma_{\text{diff}} = \frac{3 \pi \epsilon}{4 \Omega_{\text{env}} V_{\text{diff}}}$. This $P^{*}$ holds the frequency dependence of the location of the dip in the frame co-rotating with the frequency $\Omega_{\text{env}}$.
We have:

\begin{equation}
    \frac{1}{\Delta \mathrm{P}} - \frac{1}{\Pi_0} \simeq \frac{\Gamma_{\text{diff}}/\pi\frac{\mathrm{d}G^{-1}}{\mathrm{d}s}\big|_{\bar{s}}}{\left((\mathrm{P}-\mathrm{P}_{*})\frac{\mathrm{d}G^{-1}}{\mathrm{d}s}\big|_{\frac{\bar{s}+s_{*}}{2}} + \frac{\Gamma_{\text{diff}}}{\sqrt{3}}\right)^{2}+\Gamma_{\text{diff}}^{2}} \, .
\end{equation}
The dip in the period-spacing is thus also approximatively a Lorentzian\footnote{Formally, this is not the profile of a Lorentzian, as the factors $\frac{\mathrm{d}G^{-1}}{\mathrm{d}s}$ hold a frequency dependence.} near the period of the pure inertial mode in the co-rotating frame corresponding to the envelope Then:

\begin{align}
    & \Delta P \simeq \Pi_0 \nonumber \\
    & \left(1 -\frac{\Pi_0 \Gamma_{\text{diff}}/\pi \frac{\mathrm{d}G^{-1}}{\mathrm{d}s}\big|_{\bar{s}}}{\left((P-P_{*})\frac{\mathrm{d}G^{-1}}{\mathrm{d}s}\big|_{\frac{\bar{s}+s_{*}}{2}} + \frac{\Gamma_{\text{diff}}}{\sqrt{3}}\right)^{2}+\Gamma_{\text{diff}}^{2}+ \Pi_0 \frac{\Gamma_{\text{diff}}}{\pi} \frac{\mathrm{d}G^{-1}}{\mathrm{d}s}\big|_{\bar{s}}}\right) \, .
    \label{eqref:Lorentzian_cont}
\end{align}
Naturally, by putting $\alpha_{\text{rot}} = 1$, we retrieve the canonical inverted Lorentzian profile derived in \citet{Tokuno2022AsteroseismologyOscillations}.

\subsection{Case of discontinuous rotation and density profiles}
\label{subsec:disc}
We take the same two-zone rotation profile as previously, and we assume the density at the interface core-radiative envelope to be discontinuous, which produces as well a discontinuous Brunt-Väisälä profile at the core-envelope boundary. This particular case is the one that should be used if the variation of the Brunt-Väisälä frequency near the core happens on a lengthscale that is less extended than the radial wavelength of the mode. First, this is seen in \cite{Ouazzani2020FirstRevealed} for models near the Terminal Age Main Sequence at the upper range of mass of the $\gamma$ Doradus classical instability strip. Second, as the extra-core mixing mechanisms suffer from uncertainties, it is relevant to encompass both continuous and discontinuous cases in our analysis. 
This problem is thus parametrized as such:
\begin{equation}
\lim_{r \to R_{\text{core}}^{-}} \rho = \rho_{\text{core}}
\end{equation}
and 
\begin{equation}
\lim_{r \to R_{\text{core}}^{+}} \rho = \rho_{\text{core}} + \Delta \rho  = \rho_{\text{env}} \, ,
\end{equation}
as for the Brunt-Väisälä frequency:
\begin{equation}
\lim_{r \to R_{\text{core}}^{-}} N = 0 \, ,
\end{equation}
\begin{equation}
N|_{R_{\text{core}}} = +\infty
\end{equation}
and 
\begin{equation}
\lim_{r \to R_{\text{core}}^{+}} N = N_{0} \, .
\end{equation}
Two cases are comprised in this analysis: (1) only the derivative of the density is discontinuous, hence the Brunt-Väisälä frequency. This corresponds to $\Delta \rho = 0$; (2) both the density and its derivative are discontinuous: $\Delta \rho \neq 0$. We emphasize that in this framework, the lower boundary of the gravito-inertial modes cavity and the radial coordinate of the limit of the convective core are assumed to be the same: $r_{a}= R_{core}$.


The derivation of the radial displacement and the Eulerian pressure perturbation from the upper side of the boundary is made in Appendix~\ref{Appendix:g-i}.

Core solutions are the same as previously described in the case of constant density in the core.
The matching equation for Lagrangian displacement thus becomes:
\begin{align}
    \sum_{k}a_{k}&\tilde{\epsilon}\tilde{X}_{k}^{m}(s_{\text{env}})\Theta_{k}^{m}(\mu;s_{\text{env}})\nonumber \\& =\sum_{l}b_{l}C_{l}^{m}(1/s_{\text{core}})\tilde{P}_{l}^{m}(\mu) \, .
    \label{eq:match_displacement_dif_disc}
\end{align}
The pressure equation has to be precisely taken care of. As \cite{Tokuno2022AsteroseismologyOscillations} highlighted, the match of the Lagrangian variation of pressure does not equate to the match of the Eulerian pressure anymore, since background quantities are no longer continuous near the interface. We thus derive the matching equation for $\delta p = p' + \xi_{r}(\frac{\partial \bar{p}}{\partial r})$. With the same reasoning as the one held in subsection \ref{subsec:matrix}, we neglect the contribution of the centrifugal force with respect to the self-gravity.
We rewrite the expression of the Lagrangian pressure perturbation as:
\begin{equation}
    \delta p = \bar{\rho}c_{S}^{2}\left(\frac{p'}{\Gamma_{1}\bar{p}}\right) - r\bar{\rho}\bar{g}\left(\frac{\xi'_{r}}{r}\right) \, ,
\end{equation}
using the hydrostatic equilibrium and the definition of the sound speed, $\bar{g}$ being the local background gravity acceleration. The expressions of the Lagrangian pressure perturbation in the two zones are thus respectively, in the inertial frame:
\begin{align}
\rho_{\text{env}}&\sum_{k}a_{k}\Bigg{[}\sigma_{\text{env}}^{2}R_{\text{core}}^{2}
    \tilde{Y}_{k}^{m}(s_{\text{env}})\nonumber \\ & - \left( \frac{\mathcal{G} M_{\text{core}}}{R_{\text{core}}}\right)\tilde{\epsilon}\tilde{X}_{k}^{m}(s_{\text{env}}) \Bigg{]}
     \times \Theta_{k}^{m}(\mu, s_{\text{env}})
    \label{eq:delta_press_env}
\end{align}
in the envelope, and:
\begin{align}
    \rho_{\text{core}}&\sum_{l}b_{l}\Bigg{[}\sigma_{\text{core}}^{2}R_{\text{core}}^{2}
     P_{l}^{m}(1/s_{\text{core}})\nonumber \\ & - \left( \frac{\mathcal{G} M_{\text{core}}}{R_{\text{core}}}\right)C_{l}^{m}(1/s_{\text{core}}) \Bigg{]}
    \times \tilde{P}_{l}^{m}(\mu)
    \label{eq:delta_press_core}
\end{align}
in the core.
The approximated coupling equation for the pressure reads:
\begin{align}
    &\rho_{\text{env}}\sum_{k}a_{k}\Bigg{[}\sigma_{\text{env}}^{2}
    \tilde{Y}_{k}^{m}(s_{\text{env}})\nonumber  - \frac{\mathcal{G} M_{\text{core}}}{R_{\text{core}}^{3}}\tilde{\epsilon}\tilde{X}_{k}^{m}(s_{\text{env}}) \Bigg{]} \nonumber
    \\ & \times \Theta_{k}^{m}(\mu, s_{\text{env}})\nonumber \\ = &\rho_{\text{core}}\sum_{l}b_{l}\Bigg{[}\sigma_{\text{core}}^{2}
    P_{l}^{m}(1/s_{\text{core}})\nonumber - \frac{\mathcal{G} M_{\text{core}}}{R_{\text{core}}^{3}}C_{l}^{m}(1/s_{\text{core}}) \Bigg{]} \nonumber
    \\ & \times \tilde{P}_{l}^{m}(\mu) \, .
    \label{eq:coupling_wall}
\end{align}

We follow the same method as previously, projecting both coupling equations on Hough functions. The matrix equation has the same form as Eq.~ \eqref{eq:matrix_eq}:
\begin{equation}
[\mathcal{\tilde{M}}(s_{\text{core}},s_{\text{env}}) - \tilde{\epsilon} \mathcal{\tilde{N}}(s_{\text{core}},s_{\text{env}})]\vec{b} = \vec{0} \, .
\label{eq:matrix_eq_disc}
\end{equation}
With:
\begin{eqnarray}
    \mathcal{\tilde{M}} = c_{k,l}\sigma_{\text{env}}^{2}\rho_{\text{env}}\tilde{Y}_{k}^{m}(s_{\text{env}})C_{l}^{m}(1/s_{\text{core}})
\end{eqnarray}
and
\begin{align}
    \mathcal{\tilde{N}} = & c_{k,l}\Big(\sigma_{\text{core}}^{2}\rho_{\text{core}}\tilde{X}_{k}^{m}(s_{\text{env}})P_{l}^{m}(1/s_{\text{core}}) \nonumber\\
    &  + \Delta \rho\frac{\mathcal{G}M_{\text{core}}}{R_{\text{core}}^{3}}\tilde{X}_{k}^{m}(s_{\text{env}})C_{l}^{m}(1/s_{\text{core}})\Big) \, .
\end{align}

The leading term analysis conducted as before leads to the synthetic coupling equation:
\begin{align}
    & \frac{\sigma_{\text{env}}^{2}}{\sigma_{\text{core}}^{2}}\Big(1+\frac{\Delta \rho}{\rho_{\text{core}}}\Big)\cot\left(\frac{\pi^{2}s_{\text{env}}}{\Omega_{\text{env}}\Pi_0}-\frac{\pi}{4}\right) \nonumber \\ &
    \times\Big[\frac{s_{\text{env}}}{2\Lambda_{k}^{m}(s_{\text{env}})^{1/2}}\Big]F_{l}^{m}(s_{\text{core}}) \nonumber
    \\ & \simeq \tilde{\epsilon}\left(1
    - \frac{\Delta \rho}{\rho_{\text{core}}}\frac{\mathcal{G}M_{\text{core}}}{\sigma_{\text{core}}^{2}R_{\text{core}}^{3}}F_{l}^{m}(s_{\text{core}})\right) \, .
\end{align}

This equation, as before, can be computed either numerically or analytically, by an expansion of $F_{l}^{m}$ around its zero $s_{\text{core}}^{*}$. For the analytical expansion, we rely on the function $G$ previously described. The coupling equation reduces to:
\begin{align}
    & \cot\left(\frac{\pi^{2}s}{\Omega_{\text{env}}\Pi_0}-\frac{\pi}{4}\right) \nonumber \\
    & + \tilde{\epsilon}\left(\frac{\alpha_{\text{rot}}G^{-1}(s)}{2\Omega_{\text{env}}}\right)^{2}\frac{\mathcal{G}M_{\text{core}}}{R_{\text{core}}^{3}}\frac{\Delta \rho}{\rho_{\text{env}}} \nonumber\\ 
    & \phantom{+ \tilde{\epsilon}\left(\frac{\alpha_{\text{rot}}G^{-1}(s)}{2\Omega_{\text{env}}}\right)^{2}}\times \Bigg[\frac{2\Lambda_{k}^{m}(G(s))^{1/2}}{\alpha_{\text{rot}}^{2}G(s)}\Big(\frac{G(s)}{s}\Big)^{2}\Bigg]_{s_{\text{core}}^{*}} \nonumber \\
    & \simeq -\frac{\tilde{\epsilon}/\tilde{V}_{\text{diff}}}{G^{-1}(s)-G^{-1}(s)^{*}} \, ,
\end{align}

with:

\begin{align}
    \tilde{V}_{\text{diff}} = & -\left(1+\frac{\Delta \rho}{\rho_{\text{core}}}\right) \nonumber \\
    & \times \Bigg[\frac{\mathrm{d}F_{l}^{m}(s)}{\mathrm{d}s}\frac{\alpha_{\text{rot}}^{2}G(s)}{2\Lambda_{k}^{m}(G(s))^{1/2}}\Big(\frac{s}{G(s)}\Big)^{2}\Bigg]_{s_{\text{core}}^{*}} \, .
\end{align}
We verify that putting $\alpha_{\text{rot}} = 1$, thus $G = \mathds{1}$ reduces the coupling equation to equations (88) and (89) of \cite{Tokuno2022AsteroseismologyOscillations}.
The correction from the solid-body rotating case is thus:
\begin{equation}
    \frac{\tilde{V}_{\text{diff}}}{\tilde{V}} =
    \Bigg[\alpha_{\text{rot}}^{2}\left(\frac{G(s)}{s}\right)^{-1}\left(\frac{\Lambda_{k}^{m}(G(s))}{\Lambda_{k}^{m}(s)}\right)^{-1/2}\Bigg]_{s_{\text{core}}^{*}} \, .
\end{equation}
With the same expansion of the cotangent as before, and $\tilde{\Gamma}_{\text{diff}} = \frac{\pi\tilde{\epsilon}}{\Omega_{\text{env}}\tilde{V}_{\text{diff}}}$, we get:
\begin{equation}
    \frac{1}{\Delta \mathrm{P}} - \frac{1}{\Pi_0} \simeq \frac{\tilde{\Gamma}_{\text{diff}}/\pi \frac{\mathrm{d}G^{-1}}{\mathrm{d}s}\big|_{\bar{s}}}{\left((\mathrm{P}-\mathrm{P}_{*})\frac{\mathrm{d}G^{-1}}{\mathrm{d}s}\big|_{({\bar{s}+s_{*}})/{2}}\right)^{2}+\tilde{\Gamma}_{\text{diff}}^{2}} \, .
\end{equation}

\section{Results \& Discussion}\label{sec:Application}

We apply in this section our theoretical results to a particular Kelvin g-mode for a typical fast-rotating $\gamma$ Doradus star and investigate the detectability of the core-near-core differential rotation from asteroseismic data. To do so, we first analyze the effect of differential rotation on the structure and location of the inertial dips in the period-spacing pattern, in the formalism of a continuous near-core Brunt-Väisälä frequency. We then aim to assess the potential uncertainties on the differential rotation rate for different processes: (1) the profile of the near-core Brunt-Väisälä frequency (2) the intrinsic uncertainty coming from the limited time duration of observations (3) the influence of density stratification in the core. We conclude on the detectability of the differential rotation at the core-near-core interface and on the convective core rotation rate from the dip properties. Caveats of the formalism and perspectives are ultimately discussed.

\subsection{Case star: typical values of buoyancy travel time and rotation rate}
\label{subsec:case_star}

We select a typical fast-rotating $\gamma$ Doradus star, with an envelope rotation rate of $\Omega_{\rm env}/2\pi = 2.16 \, \mathrm{d}^{-1}$, a buoyancy travel time of $\Pi_{0} = 4175 \, \mathrm{s}$. Those values are close to the ones observed in \cite{Saio2021RotationModes} for KIC12066947: ($2.159 \pm 0.002 \, \mathrm{d}^{-1}$,  $4175 \pm 28 \, \mathrm{s}$ respectively), which is a good representative of the $\gamma$ Doradus in their study. These parameters are compatible with stellar modelling prescriptions, such as the ones presented in \cite{Ouazzani2019Models}. As we are taking an assumption of uniform density in the core to maintain the analyticity of the results, the star best fitting our model is a ZAMS star, as the density gradient in the core is increased during evolution \citep{Ouazzani2020FirstRevealed}. KIC12066947 was not found to be a ZAMS star by \cite{Saio2021RotationModes}, having a core hydrogen abundance of 0.53 in the model computed without overshooting, yet we keep the values of $\Pi_0$ and $\Omega_{\text{env}}$ as they remain close to the modelling prescriptions for ZAMS stars. This age dependence will be further discussed in Section~\ref{subsec:dens_strat}.

To further explore the coupling problem in the presence of differential rotation, we study the interaction between a pure inertial mode $(l = 3, m=-1)$ and a Kelvin g-mode $(k = 0, m=-1)$ in the envelope. In this case:
\begin{equation}
    F_{l}^{m}(s) = -\frac{s^2-10 s-15}{(s+1) \left(s^2-5\right)}
\end{equation}
and the spin parameter of the pure inertial mode is $s^*_{\text{core}} = 11.3245$.
The choice of this particular mode is motivated by the fact that: (1) patterns of Kelvin g-mode $(k = 0, m=-1)$ are the most commonly detected in the 611 $\gamma$ Doradus sample of \cite{Li2020Gravity-modeKepler}, forming 62\% of the detected period-spacing patterns; (2) as shown by \cite{Ouazzani2020FirstRevealed}, the pure inertial $(l = 3, m=-1)$ mode is likely to couple with this Kelvin g-mode, having a high geometrical factor of $c_{0,3} = 0.5$ in the absence of differential rotation, remaining non-negligible with differential rotation, as inverstigated in Appendix \ref{App:Mod_lorentz}. Dips arising from this interaction have previously been observed in \cite{Saio2021RotationModes} in the range $s_{\text{env}} \in [8;11]$. The high spin parameter of the pure inertial mode ensures our formalism holds in that case, which will be later discussed in subsection \ref{subsec:limit}.

For this analysis, we rely on \citet{Aerts2023ModeStars} to assess the relevant range of coupling parameters allowed in our study. Among the 37 $\gamma$ Doradus analyzed, the maximum value of coupling parameter is $\epsilon \approx 0.25$, and $\tilde{\epsilon} \approx 0.1$, which corresponds respectively to $\Gamma \approx 40$h and $\tilde{\Gamma} \approx 12$h. Given the reduced amount of analyzed stars and the potentiality of further detection of coupling parameters superior to this threshold, we allow $\Gamma$ and $\tilde{\Gamma}$ parameters to vary in the range $[1;60]$ hours. 

As for the amount of differential rotation allowed in this study, we base ourselves on the detection of core-near-core differential rotation made by \cite{Saio2021RotationModes}. In their sample, the highest core-near-core differentiality is reached for KIC 05985441, with $\alpha_{\rm rot} \approx 0.85$. Conversely, in the opposite regime of the core rotating slower than the envelope, the highest value of differentiality is $\alpha_{\rm rot} \approx 1.01$ (KIC 8330056). The regime of $\alpha_{\rm rot} < 1$ is favored by realistic 2D models able to reproduce the meridional circulation in the radiative envelope of early-type fast rotating stars and to predict the gravity-darkening observed in such stars \citep{EspinosaLara2013Self-consistentStars}. We thus extend the range of analyzed differential rotation to $\alpha_{\rm rot} \in [0.80;1.05]$. The different ranges of the parameters considered are summarized in Table \ref{tab:params}.

\begin{table}[t]
    \centering
    \begin{tabular}{c|c|c|c|c|c}
        & $\epsilon$ & $\Gamma$ (h) & $\Omega_{\text{env}}/2\pi \, (\text{d}^{-1})$ & $\alpha_{\rm rot}$ & $\Pi_{0}$ (s)\\
        \hline
        & & & & & \\
        & & & $2.16$&  & $4175$ \\ 
        
        min & $6.10^{-3}$ & 1 & & 0.9 & \\
        max & 0.37 & 60 & & 1.05 &\\
    \end{tabular}
    \vspace{10pt}
    \caption{Values (for $\Omega_{\rm env}, \Pi_{0}$)) or ranges of values (for $\epsilon, \Gamma, \alpha_{\rm rot}$) of the parameters considered in our analysis.}
    \label{tab:params}
\end{table}

\begin{figure}[h]
    \centering
    \includegraphics[width=\linewidth]{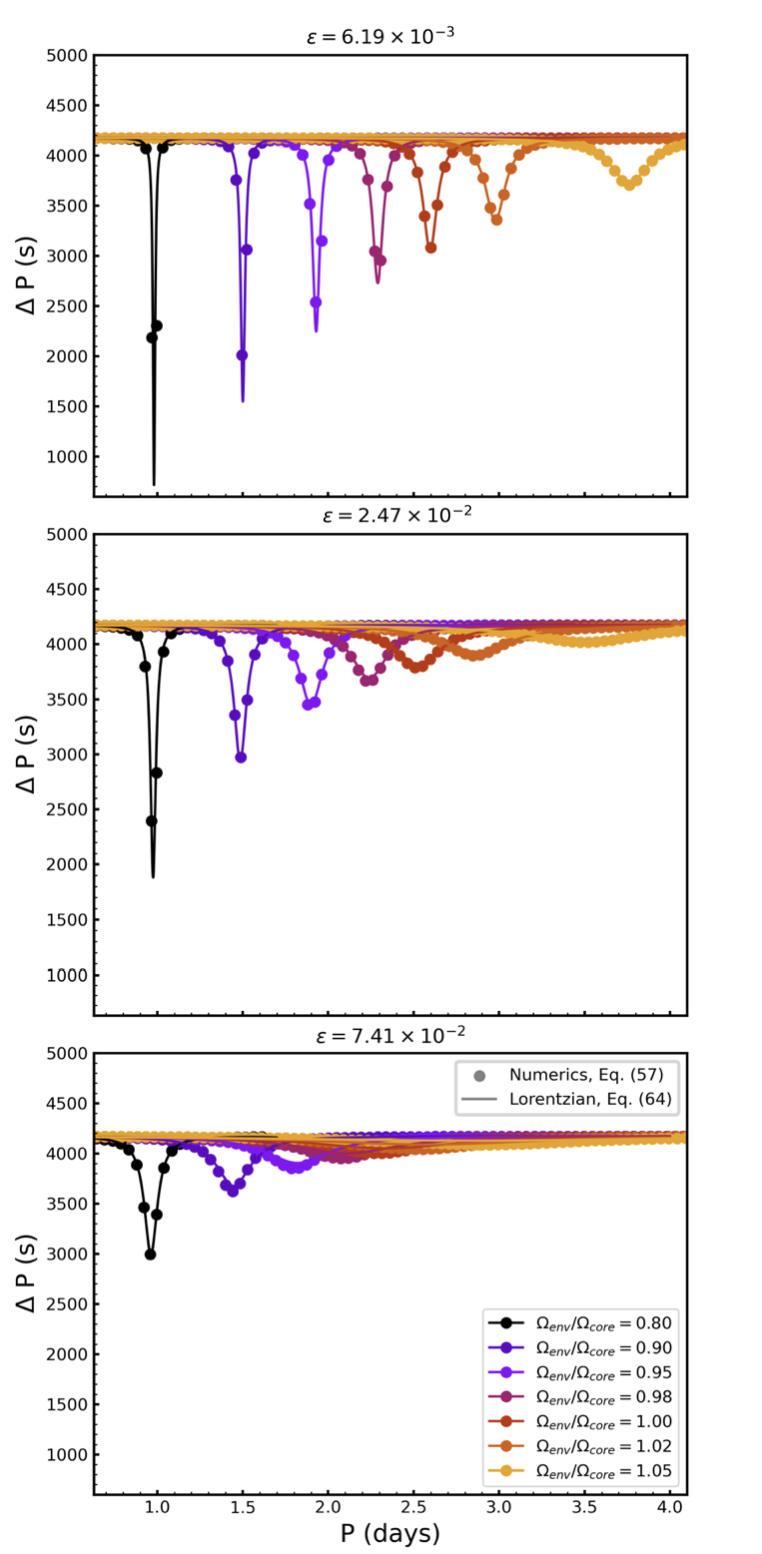}
    \caption{Dips formed by the coupling between $(k = 0, m = -1)$ gravito-inertial modes in the envelope and $(l = 3, m = -1)$ pure inertial modes in the core, in the co-rotating frame related to the envelope. Different panels correspond to coupling parameters of respectively $\epsilon = 6.19\times 10^{-3}$, $1.24\times 10^{-2}$, $7.43\times 10^{-2}$, which give $\Gamma$ = 1.0 (top), 4.0 (middle), and 12.0h (bottom) in the case of no differential rotation.
    For each panel, period-spacing patterns are superimposed on one another, taking values of $\alpha_{\rm rot} = \Omega_{\text{env}}/\Omega_{\text{core}}$ in $[0.80, 0.90, 0.95, 0.98, 1 , 1.02, 1.05]$. The period-spacing series in red shows the case of no differential rotation.}
    \label{fig:dips}
\end{figure}

\subsection{Effects of differential rotation on the morphology of the inertial dip}\label{subsec:effect_diff_rot}


In Fig.~\ref{fig:dips}, the numerical results described in subsection \ref{subsec:Num_section}, obtained by solving Eq. \eqref{eq:coupling} are overplotted on top of the analytical Lorentzian profile described in Eq. \eqref{eqref:Lorentzian_cont}. Three different values for the coupling parameter $\epsilon$ are used, corresponding to $\Gamma$ values in the solid-body rotating case of 1, 4, and 12 hours. These values are both taken for the sake of comparison with the results of \citet{Tokuno2022AsteroseismologyOscillations} in the case of the lower coupling parameter values, as well as to illustrate the case of a moderately high coupling parameter comprised in this study. To follow the evolution of the morphology of the dip with differential rotation, cases from $\alpha_{\rm rot} = 0.8$ to $\alpha_{\rm rot} = 1.05$ are considered in the same figure.

From Fig. \ref{fig:dips}, we first confirm the consistency of the numerical results (dots) with the approximated modified Lorentzian profile (lines), as already predicted by the difference in $O(\epsilon^{2})$ arising from the fact that we take into account the first term in the development of $F_{l}^{m}$ near its zero $s_{\rm core}^{*}$ in Section \ref{subsec:freq_dep}. The dips, as described in the numerical study held in subsection \ref{subsec:Num_section}, arise at an envelope spin parameter varying with the amount of core to near-core differential rotation. From no differential rotation and a spin parameter of $s_{\text{env}}^{*} = s_{\text{core}}^{*}= 11.32$, a differential rotation of $\alpha_{\rm rot} = 0.8$ gives $s_{\text{env}}^{*} = 4.25$. Conversely, a differential rotation of $\alpha_{\rm rot} = 1.05$ gives $s_{\text{env}}^{*} = 16.59$. This corresponds to a shift in the period at which the dip arises in Fig.~\ref{fig:dips}. High core-near-core differentiality corresponds to a low period of the dip, while a faster-rotating envelope increases the period of the dip when compared to the case of uniform rotation. At fixed coupling parameter, with increasing core rotation compared to the envelope, the dip gets also thinner and deeper, being formed of fewer modes. The value of $\Gamma_{\rm diff}$ decreases from $\Gamma$. On the contrary, in regimes where the core rotates slower than the envelope, the dip gets wider and shallower, with many modes composing the dip. The value of $\Gamma_{\rm diff}$ increases from $\Gamma$.

The middle and bottom panels of Fig.~\ref{fig:dips} show the evolution of the dip structure with increasing coupling parameter $\epsilon$ and $\Gamma$ parameter. The results obtained by \cite{Tokuno2022AsteroseismologyOscillations} in the case of no differential rotation still hold in differentially rotating regimes: at fixed $\alpha_{\rm rot}$ parameter, with increasing coupling parameter the dip gets wider, and is shifted to lower periods. We see that in the case of $\Gamma = 12$ hours, dips in the regime of core rotation inferior to envelope rotation are barely distinguishable from the baseline, which in turn questions the detectability of such feature in data.

The effect of an increasing core-to-envelope differential rotation on the dip shape is thus comparable to a decrease of the coupling parameter. This effect can be physically interpreted using Fig.~\ref{fig:numerics}, considering the mode interaction in the reference frame co-rotating with the core. Seen from this frame, the spectrum of gravito-inertial modes gets denser with increasing envelope rotation, sparser with decreasing one. Taking for instance the case $\alpha_{\rm rot} = 1.05$, the intersection of the yellow continuous line with blue thin ones forms modes that are separated by a low spacing in terms of core spin parameter compared to the solid-body rotating one (red line). Conversely in the case $\alpha_{\rm rot} = 0.80$, the lower inclination of the black line with respect to the blue ones forms modes spaced by an increased spacing in core spin parameter. Thus, in the regime of $\alpha_{\rm rot} > 1$, the pure inertial mode is close in period to an increased number of modes, with which it can interact.

\par We quantify this effect in Table~\ref{tab:close_modes}. If $s_{\rm core}^{\rm min}$ and $s_{\rm core}^{\rm max}$ are respectively the lowest and highest value of core spin parameter for which the resulting period spacing is inferior to 98\% of the baseline value, $\Delta n$ is the number of modes in the interval $[s_{\rm core}^{\rm min}, s_{\rm core}^{\rm max}]$. $\Delta s_{\rm core}$=$s_{\rm core}^{\rm max}- s_{\rm core}^{\rm min}$ is the gap in core spin parameter, and $\Delta n/\Delta s_{\rm core}$ a measurement of the density of gravito-inertial modes close to the period of the core pure inertial mode in the frame co-rotating with the core. This density, at fixed $\epsilon$ parameter, increases with decreasing core-to-envelope differential rotation, hence increasing $\alpha_{\rm rot}$ parameter.

This increased coupling efficiency explains the similar effect between a high $\alpha_{\rm rot}$ and a high coupling parameter $\epsilon$ on the dip shape. Conversely, at $\alpha_{\rm rot} < 1$ the pure inertial mode is surrounded by fewer modes and the coupling efficiency is less important, resulting in a thin dip.

\par
This effect is captured in Eq.~\ref{eq:coupling_simplified} by the parameter $\epsilon/V_{\rm diff}$, controlling the depth and the width of the inertial dip structure. The coupling parameter $\epsilon$ is left unchanged by the existence of differential rotation, but $V_{\rm diff}$ increases with increasing core-to-envelope differential rotation (or decreasing $\alpha_{\rm rot}$). This evolution is due to the different angular structure of the gravito-inertial modes, as at fixed core spin parameter the pure inertial mode keeps the same structure as in the solid-body rotating case while the Hough function $\Theta_{k}^{m}(\mu, s_{\rm env})$ varies with the considered envelope spin parameter. Overall, we retrieve the influence of increasing core-to-envelope differential rotation, decreasing the coupling efficiency measured by $\epsilon/V_{\rm diff}$.

\begin{table}[]
    \centering
    \begin{tabular}{c|c|c|c|c|c|c|c}
        $\alpha_{\rm rot}$ & 0.80 & 0.90 & 0.95 & 0.98 & 1.00 & 1.02 & 1.05 \\
        \hline
        $\Delta n$ & 5 & 7 & 10 & 12 & 13 & 15 & 16 \\
        \hline
        \hspace{-1em} $\Delta n/\Delta s_{\rm core}$ \hspace{-0.9em} & 1.4 & 2.4 & 3.4 & 4.5 & 5.5 & 6.9 & 10.2 \\
    \end{tabular}
    \vspace{0.1em}
    \caption{Density of envelope gravito-inertial modes influenced by the core pure inertial mode, for a fixed value of $\epsilon = 2.47\times 10^{-2}$, and varying $\alpha_{\rm rot}$. $\Delta n$ is the number of modes in the interval of length $\Delta s_{\rm core}$=$s_{\rm core}^{\rm max}- s_{\rm core}^{\rm min}$. $\Delta n/\Delta s_{\rm core}$ is the corresponding density of interacting gravito-inertial modes close to the period of the core pure inertial mode in the frame co-rotating with the core.}
    \label{tab:close_modes}
\end{table}

\subsection{Detectability of the core-near-core differential rotation}

\begin{figure*}[t!]
    \centering
    \includegraphics[width=1\linewidth]{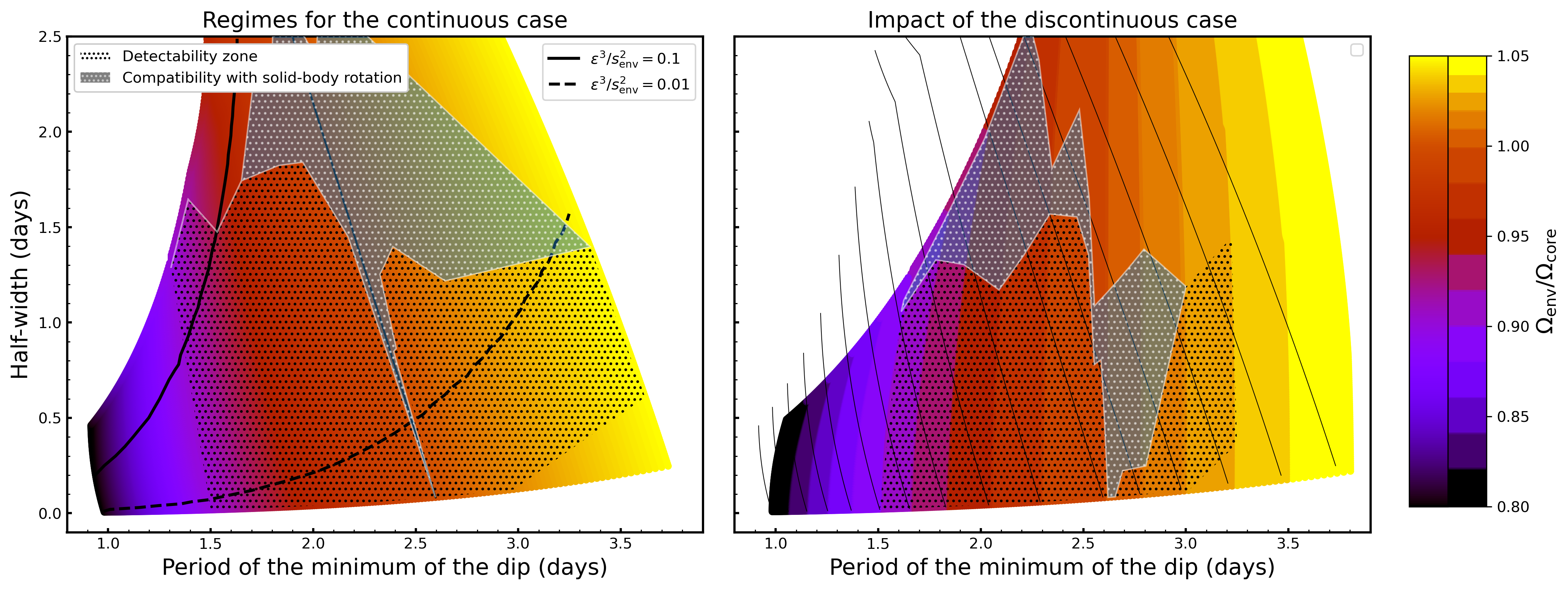}
    \caption{Differential rotation as a function of the period of the minimum of the inertial dip and its half-width in the co-rotating frame of the envelope, for the model star considered with $\Pi_{0} = 4175$s and $\Omega_{\text{env}}/2\pi = 2.16 \, \mathrm{d}^{-1}$. \textsl{Left:} Case of a continuous Brunt-Väisälä frequency near the core, with the solid line separating the regions where the core is rotating faster or slower than the envelope. \textsl{Right:} Case of a discontinuous Brunt-Väisälä frequency near the core. The superimposed contour corresponds to the same levels of differential rotation as highlighted on the sequential colormap, for the continuous Brunt-Väisälä case. On each panel, dashed lines correspond to the limit for which $\epsilon^{3}/s_{\text{env}}^2 < 0.1$ and $\epsilon^{3}/s_{\text{env}}^2 < 0.01$. In each panel, the dotted regions designate the zones for which differential rotation was found to be detectable in our analysis. Black dots designate zones in which we found dips significantly differing from a solid-body rotating case, whereas the white region shows dips in which the solid-body rotating case cannot be excluded.}

    \label{fig:two_cases}
\end{figure*}

In the left panel of Fig.~\ref{fig:two_cases}, we summarize the results by plotting the period of the inertial dip minimum with regards to its half-width in the co-rotating frame related to the envelope, for the extent of parameter $\alpha_{\rm rot}$ and $\Gamma$ described in Table~\ref{tab:params}. The amount of differential rotation is given by the color scale. The computations for this left panel are made in the framework of a continuous Brunt-Väisälä profile near-core.

\subsubsection{Validity of the theoretical expansion}
\label{sec:validity}

Two dotted lines are plotted in the left panel of Fig. ~\ref{fig:two_cases}. Below those lines, the ratio $\epsilon^{3}/s_{\text{env}}^{2}$ is respectively less than $0.1$ and $0.01$. This ratio controls the accuracy of the developments made for the expansion of the gravito-inertial modes to the edge of the convective core, as described by Eq. \eqref{eq:dev}, in the framework of a continuous near-core Brunt-Väisälä frequency. We see that this ratio, for the range of analyzed parameters, reaches high values in the regime in which the core rotates faster than the envelope, due to the decrease in the envelope spin parameter at which the interaction occurs, $s_{\text{env}}^{*}$. We advise against the use of ours and \citet{Tokuno2022AsteroseismologyOscillations}'s formalisms in the regime in which $\epsilon^{3}/s_{\text{env}}^{2} > 0.1$, as the extent of the intermediate region between $R_{\text{core}}$ and $r_{a}$ would make the structure of the gravito-inertial modes significantly altered from Hough functions. Inertial dips would still be detected but further theoretical developments of the wave propagation in this intermediate-region, or numerical studies \citep[e.g.][]{Dintrans2000OscillationsTheory} would be necessary in this case. However, this ratio remains small in the majority of the regions studied, ensuring good validity of our model throughout our analysis.

\subsubsection{Effect of noise in the asteroseismic data on the detectability of the dip}\label{subsubsec:detect}

To address sources of difficult dip structure detection, and thus uncertainty on the detection of differential rotation from the inertial dip structure study, we aim to trace a zone in which the analysis of an inertial dip in data will lead to an accurate estimate of the core to near-core differential rotation. \\
For this, we first construct perturbed period-spacing patterns representative of the number of modes usually found in a period-spacing pattern and their periods, taking into account a realistic estimate of the intrinsic observational uncertainty. We artificially introduce a dip in the period-spacing pattern. The generation of realistic period-spacing patterns is thoroughly described in Appendix \ref{Appendix:detect}. This procedure is reproduced for values in the range $0.9<\alpha_{\rm rot}<1.05$ and $0.5 \mathrm{h}<\Gamma<60 \mathrm{h}$.
We then infer uncertainties on the parameters $\alpha_{\rm rot}, \Gamma$ by running a Markov-Chain Monte Carlo analysis on these simulated period-spacing patterns. 
\par
An open question is then of how many parameters would the fit to data best depend on. In the case of the dips found in \citet{Saio2021RotationModes}, their clear appearance in the period-spacing pattern suggests that a first estimation of $(\Omega_{\rm env}, \Pi_{0})$ can be performed excluding the modes belonging to the inertial dip. This was originally done in \citet{VanReeth2016InteriorSpacings} for the star KIC12066947, where the estimation of $\Omega_{\rm env}$ and $\Pi_{0}$ was done on the low-period part of the spectrum, while the region containing the dip was excluded when fitting the baseline. The global parameters can then be used to move to the co-rotating frame and detect the parameters which the dip shape and location depend on, $\epsilon$ (through $\Gamma$), and $\alpha_{\rm rot}$. This analysis can \textit{a priori} not be done for all dips, as we saw in subsection~\ref{subsec:effect_diff_rot} that in the regime of high coupling, or $\alpha_{\rm rot} > 1$, the dips can get shallow, hence uneasy to clearly distinguish from non-coupling gravito-inertial modes, and affect the measure of $\Omega_{\rm env}$ and $\Pi_{0}$. We then decide to investigate the detectability of core rotation effects when performing a multi-dimensionnal fit, taking into account simultaneously the previously inferred quantities ($\Omega_{\rm env}$, $\Pi_{0}$) and the ones characterizing the dip ($\alpha_{\rm rot}$, $\Gamma$) in the vector $\boldsymbol{\Theta} = (\Omega_{\rm env}, \Pi_{0}, \alpha_{\rm rot}, \Gamma, \mathrm{P_{0}})$, $\mathrm{P_{0}}$ being the period in the co-rotating frame of the first mode in the period-spacing pattern considered. We consider the period-spacing pattern in the inertial frame, as we aim to fit as well the internal rotation rate.
\par

To probe the parameter space, we build custom period-spacing patterns consisting of a series of periods $[\mathrm{P}_{k,\rm in }^{\mathrm{mod}}]$ in the inertial frame, depending on the five aforementioned parameters, following the differentially-rotating model (hereafter "5D model"), in the framework of a continuous near core Brunt-Väisälä profile:
\begin{equation}
    \mathcal{F}_{\rm5D, cont}: (\Omega_{\rm env},\Pi_{0},\mathrm{P_{0}}, \alpha_{\rm rot}, \Gamma) \rightarrow [\mathrm{P}_{k,\mathrm{mod}}] \, .
\end{equation}
     We compare it to the period-spacing pattern $[\mathrm{P}_{k,\rm in}^{\mathrm{pert}}]$ that was originated from a collection of "true" parameters, $\mathbf{\Theta}_{\rm true} = (\Omega_{\rm env} = 2\pi \times 2.16 ~\mathrm{d}^{-1}, \Pi_{0} = 4175 ~\mathrm{s}, \mathrm{P}_{0, \rm true}, \alpha_{\rm rot}, \Gamma))$ the three former ones being fixed to the values found in KIC12066947, and further perturbed to accurately represent realistic data following a procedure described in Appendix~\ref{App:mod}. We use the following loglikelihood:
\begin{equation}
    \log\mathcal{L}(\mathcal{F}) = -\frac{1}{2}\sum_{k}{\frac{(\mathrm{P}_{k,\rm in}^{\mathrm{mod}} - \mathrm{P}_{k,\rm in}^{\mathrm{pert}})^{2}}{\sigma_{k}^{2}}} \, .
    \label{eq:likelihood}
\end{equation}
The standard deviation $\sigma_{k}^{2}$ is the observational error we hypothesize, taking a fixed signal-over-noise ratio for all modes contained in the period-spacing pattern. This approach is similar to the one taken in e.g. \citet{Moravveji2015TightKIC10526294} and adapted to our case in which we model perturbations from the observational noise. We emphasize that even though this error represents modes with a fixed S/N ratio, the values of $\sigma_{k}$ can be derived for a realistic case of varying S/N ratio over modes. The form of this likelihood is discussed in Appendix \ref{App:mod}, as well as an extension to account for baseline variations that are not comprised in this analytical model.\\
We aim to compare the accuracy of the differentially-rotating model with the solid-body rotating one to fit the simulated period-spacing patterns containing an inertial dip. For this, we perform another analysis taking into account the following nested model (hereafter described as "4D model"), fitting it on the same simulated data as the 5D model:
\begin{equation}
    \mathcal{F}_{\rm4D, cont}: (\Omega_{\rm env},\Pi_{0}, \mathrm{P_{0}}, \Gamma) \rightarrow [\mathrm{P}_{k,\rm in}^{\mathrm{mod}}] \, .
\end{equation}
The MCMC is then computed using the \texttt{emcee} package \citep{Foreman-Mackey2013ttemcee/ttHammer}. Examples of cornerplots resulting from the 5D and 4D models are shown in Appendix~\ref{App:Bayes}.  We first perform a burn-in step of 100 iterations, taking walkers from a uniform prior described in table \ref{tab:prior}. The positions of the walkers are then reinitialized, sampling them from a normal distribution around the set of parameters found to maximize the loglikelihood in the burn-in phase. The MCMC is then run until 100 times the auto-correlation time is reached, with a limitation of 20000 iterations to get the posterior distributions for the parameters. If this limit on the number of iterations has been reached and is inferior to 50 times the auto-correlation time, the detection of the dip is considered "challenging": the parameters are insufficiently constrained by the fit.
\begin{table}[]
    \centering
    \begin{tabular}{|c|c|}
    \hline
    Parameter & range \\
    \hline
    & \\
    $\Omega_{\rm env}/2\pi$ & [2.14,2.18] $\mathrm{d}^{-1}$ \\
    $\Pi_{0}$ & [4070, 4270] s \\
    $\mathrm{P}_{0}$ & [$\mathrm{P}_{0,\rm true} - 4175/2 ; \mathrm{P}_{0,\rm true} +  4175/2 $] s \\
    $\alpha_{\rm rot}$ & [0.88;1.10]  \\
    $\Gamma$ & [0.5;60] h\\
    \hline
    \end{tabular}
        \vspace{10pt}
    \caption{Bounds of the uniform prior used in the MCMC analysis. $\mathrm{P}_{0,\mathrm{true}}$ designates the input value for the period of the first mode in the period-spacing pattern, varying with the number $n$ of modes in it.}
    \label{tab:prior}
\end{table}

From the posterior distribution, when available, we are interested in the marginal distributions of the parameters $\alpha_{\rm rot}$ and $\Gamma$, which give us uncertainties on the input parameters of our model. Then remains to evaluate for which set of parameters $(\alpha_{\rm rot},\Gamma)$ we can consider the 5D model presented in this work to significantly better fit the data than the 4D solid-body rotating model with only $\Gamma$ as a parameter.
We thus aim to test the following null hypothesis: "\textsl{The differentially-rotating and solid-body rotating model fit the data equally well}". If this hypothesis is true, the detection of differential rotation is not considered significant. To test the hypothesis, we calculate the p-value for the null model using a $\chi^{2}_{df =1}$ cumulative distribution function:
\begin{equation}
\mathrm{p} = P[\chi^{2}_{df =1} > \lambda] \, .
\end{equation}
With $\lambda$ the difference of the loglikelihoods:
\begin{equation}
    \lambda = 2[\log \mathcal{L}(\mathcal{F}_{\rm 5D, cont}) - \log \mathcal{L}(\mathcal{F}_{\rm 4D, cont})] \, .
\end{equation}
The analysis described here is extended to the case of models derived taking a discontinuous near-core Brunt-Väisälä profile, the 5D and 4D models being, in that case, $\mathcal{F}_{\rm 5D, disc}$ and $\mathcal{F}_{\rm 4D, disc}$. We detail hereafter our analysis of the results obtained in the framework of continuous Brunt-Väisälä profile, as the overall conclusions remain qualitatively the same in the discontinuous case, and bridge the two results in Subsection~\ref{subsec:degen}.
\par
\begin{figure}[h]
    \centering
    \includegraphics[width=\linewidth]{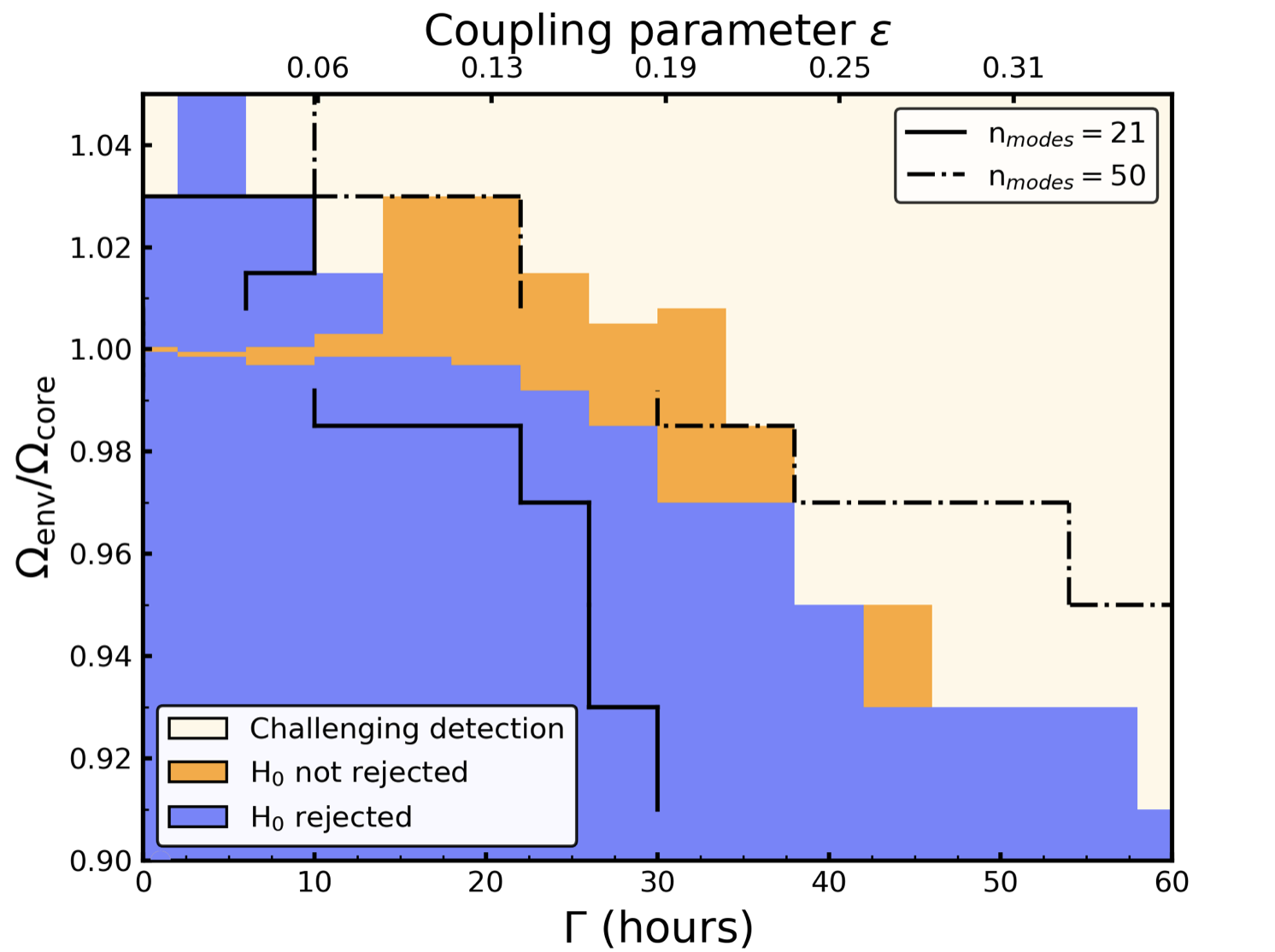}
    \caption{Detectability of differential rotation following the MCMC analysis, for a number of modes $n = 33$ in the period-spacing pattern. Zones in the $(\alpha_{\rm rot},\Gamma)$ plane for which we were able to reject the null hypothesis with a level of confidence of 95\% are coloured in blue, while zones for which the null hypothesis cannot be rejected are coloured in orange, and zones for which any fitting of an inertial dip is considered challenging (no convergence of our fits) are coloured in beige. The bin size is reduced towards $\alpha_{\rm rot} = 1$ to better see the regime of near solid-body rotation. Limits between rejection of $\mathrm{H}_{0}$ and no-rejection or challenging detection are shown in full and dotted-dashed lines for respectively $n = 21$ and $n=50$ modes. For these numbers of modes, limits in regimes for which $|\alpha_{\rm rot} - 1|<0.01$ are not shown for clarity.}
    \label{fig:Detect}
\end{figure}

Results are found in Fig.~\ref{fig:Detect}: the zones for which p < 0.05 are marked in blue, indicating that we can reject the null hypothesis with a level of confidence of $95 \%$. This treatment ensures that the detection is not subject to the noise. In other cases, the detection of differential rotation is not significant enough to stand out from the solid-body rotating case, and the region is marked in orange. These zones are then inscribed in Fig.~\ref{fig:two_cases}. This analysis is done both for the models with continuous or discontinuous Brunt-Väisälä near the boundary.

In the regime close to solid body rotation ($0.99<\alpha_{\rm rot}<1.01$), we notice the differential rotation to be detectable even for values of $\alpha_{\rm rot}$ different from 1 by less than 1 \%, depending on the considered $\Gamma$ parameter. The dip being more extended and shallower for high $\Gamma$ values, this threshold of detection is reduced with increasing values of $\Gamma$. For values inferior to 15 hours, this translates to a zone of 0.1 days around the minimum period of the dip in the corotating frame for which differential rotation would be impossible to disentangle from the solid-body rotating case. We however point out that a precision of less than a percent in this regime shows the great potential of taking the dip structure as a probe for core-to-envelope differential rotation.\\

When the core rotates slower than the envelope ($\alpha_{\rm rot}~>~1$), the detection is limited by the shallowness of the dip, as well as by the important observational noise at high periods. For values of $\Gamma > 12$ hours (in the case of n =33), the null hypothesis cannot be rejected at a 95\% level of confidence, indicating an impossibility of reliable detection of differential rotation from the analysis of a single inertial dip in data. In our particular case, above $\alpha_{\rm rot} = 1.02$ and $\Gamma > 4$ hours, the fitting shows no satisfactory results with dip parameters unconstrained in a reasonable time due to the high observational noise in the high-period part of the pattern in which the dip is situated. We thus consider the fit impossible in realistic data in the regime of a core rotating 2\% slower than the envelope, all the more if we consider slower rotators. This analysis provides the upper limit of the detectability region at large periods reported on the left panel of Fig.~\ref{fig:two_cases}, by choosing the case where $n=33$.


In the regime in which the core rotates faster than the envelope ($\alpha_{\rm rot} <1$), the detection is eased by the depth of the dip and the low impact of the noise in the region of the pattern in which the dip appears. We find the dip to be detectable for high values of the coupling parameter in this regime.


To give further insights, we consider the only study of the distribution of coupling parameter on a sample of $\gamma$ Doradus:
37 stars analysed in \citet{Aerts2023ModeStars}. In their sample of Kelvin g-mode series ($k=0,m=-1$), the mean value of radial order separation is 33, and the mean number of modes 29, not far from the number of modes found in the sample of \citet{Saio2021RotationModes} (33). The maximum value of the coupling parameter found is 0.25. The majority of the stars in the sample do not exhibit clear inertial dips, with only two common to the analysis of \citet{Saio2021RotationModes}. The authors state that those stars do not exhibit a value of coupling parameter different from the rest of the sample. We thus take this distribution of coupling parameters among "regular" $\gamma$ Doradus stars to be representative of a distribution of coupling parameters among $\gamma$ Doradus stars on which inertial dips can be fitted and a measurement of differential rotation made. According to Fig. ~\ref{fig:Detect}, if all of the stars had a detected core to near-core differential rotation of more than 6 \% (or $\alpha_{\rm rot} \leq 0.94)$ and $n=33$ modes in their period-spacing pattern, the differentially rotating model would significantly better fit their period-spacing patterns than the solid-body rotating model. The detection of differential rotation would be considered firm, with a confidence level of 95 \%. If the stars had a detected differential rotation of 1 \% ($\alpha_{\rm rot} = 0.99$), still with $n=33$ modes in their period-spacing patterns, only stars with $\epsilon < 0.16$ would fall in the region. In the sample of \citet{Aerts2023ModeStars}, this would translate to 25 stars having firm detection of differential rotation.

Interestingly, the limit of detectability of the dip in terms of $\Gamma$ parameter increases with increasing core-to-envelope rotation. Indeed, at high coupling parameter, the dip is shallow and an increase of the rotation of the core with fixed envelope rotation deepens the inertial dip, as seen in the bottom panel of Fig. \ref{fig:dips}: the dip in the regime of core rotating faster than the envelope (in black for $\alpha_{\rm rot} = 0.8$) is deepened compared to the solid-body rotating case (in red). This is an interesting property, as \citet{Aerts2023ModeStars} stated that among the fast rotators analyzed sample, $\gamma$ Doradus stars exhibiting dips do not stand out in term of their coupling parameter. We see here that in the regime of moderate to high coupling parameters, the presence of core-near-core differential rotation is one of the potential mechanisms rendering the dip clearer in the period-spacing pattern. 

Detectability limits taking hypotheses of lower ($n = 21$) or higher ($n = 50$) number of modes in the period-spacing pattern are also shown in Fig.~\ref{fig:Detect}, as respectively solid and dashed-dotted lines. As expected, a higher number of modes in the period-spacing pattern allows for a better constraint on the relevant parameters of the fit, related to the properties of the radiative envelope ($\Omega_{\rm env},\Pi_{0}$) or the inertial dip ($\alpha_{\rm rot}, \Gamma$). This pushes the detectability limit towards high coupling values, and conversely low coupling values for fewer modes.

We emphasize that even if the dip is detectable, one should be careful when studying dips in the regime $\epsilon^3/s_{\rm env}^2$>0.1, as explained in Section~\ref{sec:validity}. Lines under which $\epsilon^3/s_{\rm env}^2$<0.1 and $\epsilon^3/s_{\rm env}^2$<0.01 are shown on the left panel of Fig.~\ref{fig:comp}. This should encourage to be cautious using the analytical model described in this study for dips situated at low period values in the co-rotating frame related to the envelope, in the framework of a continuous Brunt-Väisälä frequency near the boundary.

We emphasize that this analysis assumes a constant value for the period spacing outside of the inertial dip. However, chemical and thermal structure features leave modulations in the pattern \citep{Miglio2008ProbingStars,Cunha2015StructuralModels,Degroote2010DeviationsStar}. The detectable zones described hereby thus correspond to the ideal case in which those modulations are accurately described from stellar modelling. We discuss the impact of chemical stratification on the detectability of the dip in Appendix \ref{App:mod}.

Additionally, due to the extended propagation region in the convective core, modes appearing in the dip are likely to have a low signal-over-noise ratio and thus can be unidentified in the period-spacing pattern. Such a phenomenon, with the extreme case of having dips appearing as gaps in the period-spacing pattern, will reduce the detectability region of differential rotation from the case studied here. This effect will be most prominent in the regions in which the dip is composed of a few modes, i.e. at low $\alpha_{\rm rot}$ and coupling parameters. We expect on the contrary dips in the $\alpha_{\rm rot} > 1$ detectability region to be still detectable with a few modes missing, as the dip extends over tens of modes in this regime even for low coupling parameter values.

For hypothetically strong core-to-envelope differential rotation below $\alpha_{\rm rot} = 0.9$, we chose not to pursue the MCMC analysis otherwise performed. In our routine of building perturbed period-spacing patterns, the detectability in this regime is not limited by the shallowness of the dip or the intrinsic observational noise hypothesized but by the presence or absence of the dip in the period-spacing pattern. Indeed the spin parameter value at which the dip appears, $s_{\rm env}^*$ is shifted towards values uncomprised in the period-spacing pattern analyzed. This is readily found in the case of $\alpha_{\rm rot} = 0.90$ for which the MCMC fitting does not give constraints on $\alpha_{\rm rot}$ and $\Gamma$, as the period of the first mode in the period-spacing pattern is too high compared to periods at which the effect of the interaction is sizeable.
Moreover, in this regime, the dip is formed of a few modes and the results would be vastly different if the signal-to-noise ratio of the few modes detected in the dip would significantly depart from the hypothesis we made for this analysis. Thus as both the extent of the period-spacing pattern in terms of spin parameter (see Appendix \ref{Appendix:detect}) and the Signal-to-Noise ratio of the modes in the dip are varying in each particular case, we find our analysis not to be applicable for this regime. Conversely, we point out that high core-to-envelope differential rotation is only to be found in stars displaying a low spin parameter for their mode of lowest radial order displayed in the period-spacing pattern.

We propose in the next sections to investigate the influences of other physical processes or uncertainties that could hinder the detection of differential rotation.

\subsection{Degeneracy with the treatment of the stratification gradient near the core: evolution constraints}\label{subsec:degen}

\begin{figure}
    \centering
    \includegraphics[width=\linewidth]{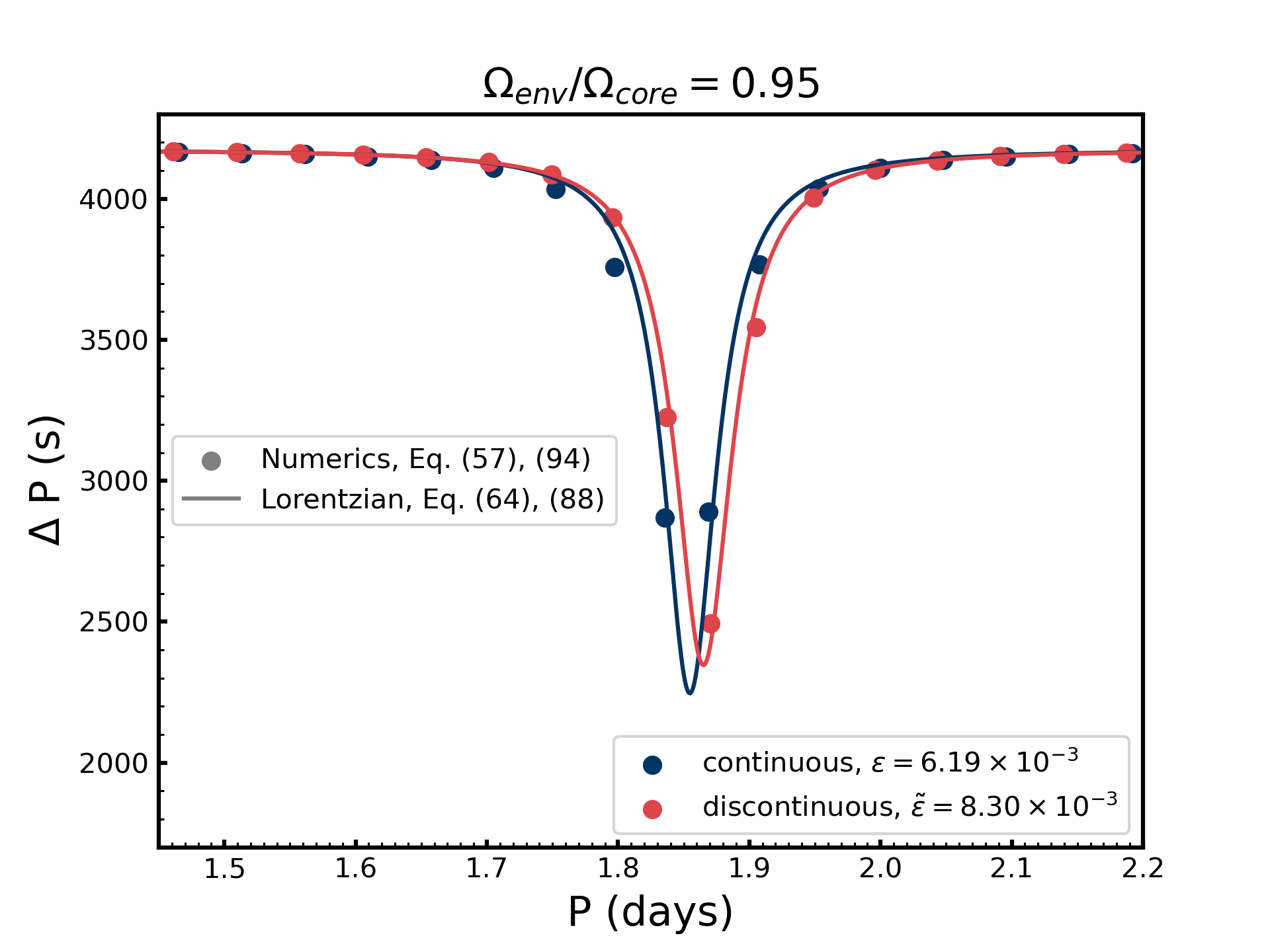}
    \caption{Comparison of inertial dips obtained in the case of a continuous (blue) or discontinuous (red) Brunt-Väisälä profile near-core, for the same value of differential rotation $\Omega_{\text{env}}/\Omega_{\text{core}} = 0.95$ and parameters $\Gamma$ = $\tilde{\Gamma} = 1.0 \text{h}$, corresponding to $\epsilon = 6.19\times10^{-3}$ and $\tilde{\epsilon} = 8.30\times 10^{-3}$. No discontinuity of the density in the near-core region is taken into account in the discontinuous case.}
    \label{fig:comp}
\end{figure}

We saw in subsection \ref{subsec:freq_dep} that by taking a continuous Brunt-Väisälä profile between the core and the envelope, the Lorentzian is shifted from the location of the pure inertial mode in the co-rotating frame related to the envelope and gets its mimimum at $\mathrm{P}_{\rm min} = \mathrm{P}_{*} - \Gamma_{\rm diff}/\sqrt{3}(\frac{\mathrm{d}G^{-1}}{\mathrm{d}s}\big{|}_{\frac{s^{*}+\bar{s}}{2}})^{-1}$, whereas the minimum is reached at $\mathrm{P}_{\rm min} = \mathrm{P}_{*}$ in the discontinuous case. The two continuous and discontinuous cases, represented in Figure 2 of \citet{Tokuno2022AsteroseismologyOscillations} can be considered as two extreme cases and thus represent a source of uncertainty on differential rotation, which also shifts the period at which the pure inertial mode forms a dip in the period-spacing pattern. 
Such a dependency is represented in Fig.~\ref{fig:comp}, in the case of the model star exhibiting a dip with parameters $\Gamma = \tilde{\Gamma} = 1.0$h, and a parameter $\alpha_{\rm rot} = 0.95$. We see that compared to the continuous treatment (in blue), the dip arising from the discontinuous treatment (in red) is shifted to high periods, as hinted from the theoretical model. This effect is due to the presence of an intermediate region between $R_{\rm core}$ and $r_{a}$ in the case of a continuous Brunt-Väisälä frequency, and is thus the same as a decrease in core rotation compared to the envelope rotation.
The shift in terms of envelope spin parameter, from the expression of $\mathrm{P}_{\rm min}$ in the continuous Brunt-Väisälä frequency framework, can be estimated as: 
\begin{align}
    \delta s_{\rm cont \rightarrow disc} = &6.7\times10^{-4}\left(\frac{\Omega_{\text{env}}}{2\pi\times 2.16 \mathrm{d}^{-1}}\right)\left(\frac{\Gamma}{1\mathrm{h}}\right) \nonumber \\ & \times \alpha_{\rm rot}\left(1+m(1/\alpha_{\rm rot}-1)s_{\rm core}^{*}\right)^{2} \, .
\end{align}
In the regime of typical fast rotators, with deep dips and core rotating faster than the envelope, the uncertainty coming from the discontinuity of the Brunt-Väisälä profile is thus under-dominant compared to the shift of the spin parameter due to differential rotation. In the precise case of Fig.~\ref{fig:comp}, the shift in the spin parameter of the location of the dip that is due to the discontinuous treatment of the Brunt-Väisälä is equivalent to a decreased value of $\alpha_{\rm rot}$ by $3\times 10^{-6}$ from $\alpha_{\rm rot} = 0.95$. However, the influence of this source of uncertainty becomes appreciable for high values of $\Gamma$ (hence high coupling). The right panel of Fig.~\ref{fig:two_cases} illustrates the different values of $\alpha_{\rm rot}$ forming an equivalent dip (same location of the minimum in period and half-width in the co-rotating frame related to the envelope): the colormap shows the value $\alpha_{\rm rot}$ in the discontinuous case in these coordinates, whereas the contour overplotted shows the value of $\alpha_{\rm rot}$ in the continuous case as fully described in the left panel. For instance, for the model star exhibiting a dip at $\mathrm{P}_{\rm min} = 1.93 ~\mathrm{d}$, with a half-width of $1.38 ~\mathrm{d}$, the continuous formalism gives a differential rotation of $\alpha_{\rm rot} = 0.94$, whereas the discontinuous formalism gives $\alpha_{\rm rot} = 0.98$, hence a relative error of $4\%$. 

In the right panel of Fig.~\ref{fig:two_cases}, we superimpose detectability zones inferred from the same treatment as the one described in Appendix \ref{Appendix:detect}, in the framework of a discontinuous near-core Brunt-Väisälä frequency, and $n=33$ modes visible in the period-spacing pattern. Results are consistent with the continuous case, in the $\alpha_{\rm rot} < 1$ regime, with differential rotation being detectable for higher values of $\Gamma$ as the dip increases in amplitude and is shifted to low periods. The situation is slightly different in the regime of $\alpha_{\rm rot} > 1$, with a differential rotation only found to be detectable for $\alpha_{\rm rot} \le 1.02$ and a low value of $\Gamma$. We interpret it as an incidence of the previously highlighted shift of the period of the minimum of the dip with increasing $\Gamma$ in the framework of the continuous near-core Brunt-Väisälä frequency, resulting in a detection of dip less attained at fixed $\alpha_{\rm rot}$ between the two regimes.

The detectability zone, considering the degeneracy of the treatment of the near-core Brunt-Väisälä profile, can thus be considered as the intersection between the detectability zones inferred in the two cases of continuous and discontinuous Brunt-Väisälä profile. This effectively strengthens the limits in terms of accessible coupling parameter values for which differential rotation is considered detectable.

 It is worth noting that the building of chemical gradient near-core during the evolution would hint towards a better accuracy of the discontinuous case for evolving main-sequence stars after the ZAMS. However, given the complex dependence of the Brunt-Väisälä profile on age or mixing parameters, one cannot state the most realistic case and reduce the uncertainty related to the Brunt-Väisälä profile near-core, unless a thorough forward-modelling study is led, taking into account the most up-to-date stellar modelling prescriptions. As it is not the aim of this present paper, we do not consider any bias towards one case or the other in this study.

\subsection{From differentiallity to the measure of convective core rotation}

We tackled two distinct internal structures: a continuous treatment of the near-core Brunt-Väisälä profile and a discontinuous one. In the discontinuous case, the period of the inertial dip is directly related to the eigenmode condition for the inertial dip ($s_{\rm core}^{*} = 11.32$ in the case we used). The sole measure of the period of the dip would then directly give an estimate of the convective core rotation rate, independently of a measure of the near-core rotation rate.
However, when the Brunt-Väisälä profile is continuous at the core-near-core interface, the period of the inertial dip not only depends on $s_{\rm core}^{*}$ but also on the $\Gamma$ parameter controlled by the sharpness of the Brunt-Väisälä gradient, and on level of the core to near-core differentiality through the function $G$. Therefore, an accurate measure of the convective core rotation depends on a reliable estimate of the stellar stratification and the near-core rotation rate.
Those two treatments can be considered as two extreme cases, and stars to be studied will be attributed one or the other formalism depending on the considered mode: the JWKB treatment made for the continuous case in subsection \ref{subsec:freq_dep} is only valid if the radial wavelength of the gravito-inertial modes is much shorter than the scale height related to the Brunt-Väisälä frequency, and the discontinuous Brunt-Väisälä profile can be attributed to modes having wavelength much longer than this scale-height.\\

For the shape of the Brunt-Väisälä profile to be constrained simultaneously as the differential rotation between the core and the envelope, we saw that one has to characterize the shape of the inertial dip in the co-rotating frame. The near-core rotation has first to be constrained to perform such a measurement. The method currently used to constrain near-core rotation is based on the study of the slope of period-spacing patterns in the inertial frame. When a clear dip is seen, it can be artificially removed from the period-spacing pattern to get a slope measurement. However, the impact of the dip spans several radial orders outside of what observers consider a dip (see for instance Fig~\ref{fig:perturbed_dip}). When no clear dip is detected, for instance in the regime of high coupling parameters, or core rotation inferior to the one of the envelope, the fitting of the slope in the period-spacing pattern does not include any information on the dip structure, even though it can be present but faint. Both removing coupled modes and ignoring the presence of a dip can lead to an error in estimating the near-core rotation rate, and therefore to errors in the convective core rotation rate estimate. In any case, especially when mechanisms of potential dip structure suppression will be known, it is of interest to integrate a dip model into the period-spacing pattern fitting routine to increase the precision of near-core rotation measurement.

\subsection{Effect of core density stratification along the evolution and need for forward modeling}\label{subsec:dens_strat}
Along the evolution, the density profile gets farther from the approximation of uniform density in the core \citep[see Fig. 5  of][]{Ouazzani2020FirstRevealed}. As highlighted by \cite{Wu2005ORIGINMODES}, this shifts the eigenfrequency of the pure inertial mode. Their work showed that with a density profile modeled by a power-law:
\begin{equation}
    \rho \propto \left(1-\left(\frac{r}{R_{\text{core}}}\right)^{2}\right)^{\beta} \, ,
\end{equation}
with $r$ the radial coordinate within the core, an increasing parameter $\beta$ would lower the spin parameter $s_{\text{core}}^{*}$, shifting the dip towards low periods. This effect is backed numerically by both \cite{Ouazzani2020FirstRevealed} and \cite{Saio2021RotationModes}: the mode studied in this work $(l = 3;m = -1)$, for which the spin parameter in a core of uniform density is at $s^{*}_{\text{core}} = 11.32$ will exhibit a lower spin parameter with density stratification. As \cite{Ouazzani2020FirstRevealed} previously pointed out, this effect is strongest for models close to the TAMS, reaching a value of $8.83$ for their most evolved model (Table 6), explaining the range of $[8,11]$ for dips found in \cite{Saio2021RotationModes} with limited differential rotation.

 To approach this problem semi-analytically, one has to solve the eigenfrequency problem of pure inertial modes in the presence of density stratification with the power-law profile or approximate the profile as described in section 2.3 of \cite{Wu2005ORIGINMODES}. An updated function $F_{l}^{m}$ can be derived and the full impact of density stratification estimated. In the case of more complex profiles, or different core-to-envelope mixing prescriptions such as convective penetration or overshooting, a numerical study is necessary \citep{Galoy2024PropertiesStars}. It is worth noting that as highlighted by \cite{Tokuno2022AsteroseismologyOscillations} and seen in \cite{Saio2021RotationModes}, an overshooting zone will result in a multiplicity of dips, which confirms the high potential of inertial dips to probe core to envelope processes happening in deep stellar interiors.

Yet, especially in the regime of high coupling parameters, thus shallow dips, forward modelling is necessary to be able to disentangle the effect of differential rotation from the effect of density stratification. But this study, as well as a semi-analytical one on density stratification, holds promising information on the prediction of the location of the dips and thus an easier detection of this feature from gravity-mode period-spacings.


\subsection{Limitations of the scope}\label{subsec:limit}

In this section, we discuss the impact of choices and assumptions we made on the results and evaluate the accuracy of the assumptions taken.

\subsubsection{Choice of the modes}
We took into account only one pure inertial mode in the analysis, mainly analyzing the interaction of the $(l=3, m=-1)$ pure inertial mode with the $(k = 0, m= -1)$ Kelvin g-modes. Another significant interaction exists with a non-negligible geometric factor: $(l = 5, m = -1)$. Yet, this interaction occurs at the spin parameter $s_{\rm core}^* = 29.33$. Even with an effect of density stratification, and differential rotation, it is very unlikely that this interaction results in a dip measurable in the range of observed period-spacing patterns. This interaction at a high spin parameter in turn justifies the assumption of the sparsity of the pure inertial modes spectra in period, which was used to derive the coupling equation. 

\subsubsection{Impact of the LTE eigenvalue}
\label{subsec:eigen}
    We have neglected the variation of the LTE horizontal eigenvalue $\Lambda_{k}^{m}$ with the spin parameter $s_{\text{env}}$ in the analysis in subsection \ref{subsec:effect_diff_rot}, which would lead to a non-constant buoyancy travel time $\Pi_0$ and an additional curvature in the period-spacing pattern otherwise. This is motivated by the fact that asymptotically, the eigenvalue of the LTE for Kelvin modes is constant. In the range that we analyzed, from 4.25 ($\alpha_{\rm rot} = 0.8$) to 16.59, ($\alpha_{\rm rot} = 1.05$), the eigenvalue $\Lambda_{0}^{-1}$ of the LTE varies for the mode $(k=0,m=-1)$ from $1.12$ to $1.03$, leading to a variation of the baseline of $4\%$. If the variation of the eigenvalue has to be taken into account, for instance, if the differential rotation considered is more important, or modes of different azimuthal orders have to be compared, the formalism here developed holds if we take a \textit{stretched} period $\sqrt{\Lambda_{k}^{m}}\mathrm{P}$ in the co-rotating frame. Indeed, modes in the asymptotic theory are equally spaced in this stretched period: $\sqrt{\Lambda_{k}^{m}}\mathrm{P}(n) = \Pi_0(n+\alpha_{g})$, with $n$ the radial order \citep[e.g.][]{Christophe2018DecipheringStars}. On the other hand, more dependencies have to be taken into account in the derivation of the period spacing, such as the decreased radial coordinate of the upper turning point $r_{b}$ with increasing frequency, thus resulting in a variation of $\Pi_0$, or modulations due to glitches in the Brunt-Väisälä frequency. These sources of potential deviation from constant baselines in period-spacing patterns are listed in Appendix A of \cite{Tokuno2022AsteroseismologyOscillations} and would require a systematic study on age, and mixing prescriptions on top of the sole influence of differential rotation described in this work. We nevertheless highlight that due to the very low frequency of the analysed sub-inertial Kelvin g-mode series, the latter effect of variation of $r_{b}$ with frequency will likely be unimportant as for this regime $N < L$ and the decrease of $N$ close to the upper boundary is very sharp.
    
\subsubsection{Underlying assumptions in the coupling mechanism}
    The reasoning held in subsection \ref{subsec:determinant} allowing us to simplify an infinite matrix problem into a simple coupling equation involving only one interacting mode relied on both the low value of $\epsilon$ and the non-negligible value of the geometric parameter $c_{k,l}$. As for the coupling parameter $\epsilon$, it is yet unclear why stars with dips do not show particularly low values of coupling parameter compared to stars with no detected inertial dips, as highlighted by \citet{Aerts2023ModeStars}. As already argued in subsection \ref{subsec:effect_diff_rot}, differential rotation leads to the formation of a clearer dip with high values of $\epsilon$, compared to the solid-body rotating case. However, a clear and definite picture will only be obtained with more realistic situations on the rotation profile or the density stratification, accessible with numerical calculations. The geometric factor, regarded as the main parameter controlling the interaction in \citet{Ouazzani2020FirstRevealed},  was not found to vary appreciably for the mode interaction considered (see Appendix \ref{App:geom}). Still, its dependence on differential rotation holds valuable information for newer theoretical developments, such as the one developed in \citet{Galoy2024PropertiesStars}. This work derived an improvement of the \cite{Tokuno2022AsteroseismologyOscillations} model, taking into account more modes from both sides of the boundary, rather than only considering the dominant ones, as done in this present work and \citet{Tokuno2022AsteroseismologyOscillations}'s one. The improved model shows that both the $\Gamma$ parameter and the dip's central spin parameter depend on the geometrical factors related to the dominant interactions from both sides of the boundary. As the projection of Hough functions on Legendre polynomials also varies with the differential rotation rate considered, it is also yet unclear if the number of interactions considered for a solid-body rotating star in \cite{Galoy2024PropertiesStars} will stay identical when considering differential rotation. A combination of the model derived in this work, accounting for differential rotation, and the model used in \citet{Galoy2024PropertiesStars}, taking into account the variation of the geometrical factors with differential rotation holds promising potential for an even more accurate modelling of the dip, and thus a more accurate detection of core and near-core properties by the dip study.
    
\subsubsection{Validity of the Traditional Approximation of Rotation}
The traditional approximation is used in this work in the radiative zone, allowing us to stay analytical throughout this study. This solution of the gravito-inertial modes under this approximation, as described in \citet{Prat2016AsymptoticDynamics}, can be seen as the limit of the full system of calculations in the regime $N/\sigma_{\rm env} \gg 1$, or $N/(2\Omega_{\rm env}) \gg 1$ in the sub-inertial regime we are interested in. As highlighted by \cite{Gerkema2008GeophysicalApproximation}, even for values such as $N=2\Omega$, the solutions under the TAR match reasonably with the full calculations. In the case of the radiative zone of the $\gamma$ Doradus stars, the high stratification of molecular weight ensures a high value of the Brunt-Väisälä frequency near the core. We thus expect calculations outside of the TAR not to diverge significantly from the results of the present analysis. However, the picture is blurrier with different core-envelope mixing prescriptions. If one is to consider a core-to-envelope mixing mechanism taking into account a non-radiative near-core gradient, the step overshoot with an adiabatic gradient being the most extreme case, the traditional approximation in the envelope will no longer hold in this case, and the matching of the solutions will have to be revisited with (1) an extended propagating zone of pure inertial modes and (2) an intermediate zone in which the gravito-inertial modes structure will significantly diverge from the one given by the TAR. We hereby point out that the study of the morphology of the inertial dip can be considered also as a probe of the core-to-envelope mixing mechanisms, processes that are otherwise difficult to constrain accurately. Analyses of the potentiality of gravity modes to probe the convective core boundary and different mixing prescriptions were made in \citet{Pedersen2018TheSpacings} and \citet{Michielsen2019ProbingAsteroseismology}. In this regard, complete numerical calculations of gravito-inertial modes \citep[such as the ones pursued in][]{Dintrans2000OscillationsTheory,Ballot20122DStars,Galoy2024PropertiesStars} are of a prime importance.

\section{Conclusion \& Perspectives}

We studied the coupling between gravito-inertial modes in the radiative envelope and pure inertial modes in the convective core of $\gamma$ Doradus star in the sub-inertial regime, allowing for a two-zones differential rotation from the two sides of the core to envelope boundary. We provide an analytical model for the interaction, and state on the possibility of constraining core to near-core differential rotation from the study of inertial dips in the period-spacing pattern of fast-rotating $\gamma$ Doradus stars.

We demonstrate that core to near-core differential rotation has an effect on the location in period of the inertial dip, as well as an incidence on the depth and the width of this characteristic feature. With envelope rotation taken constant, an increasing core rotation will shift the inertial dip to low periods, deepening and tightening it. On the contrary, a decreasing core rotation will shift the inertial dip to high periods, shallowing and widening it. The wider the dip, the higher the number of modes affected by the coupling. We retrieve as well the influence of the coupling parameter described in \cite{Tokuno2022AsteroseismologyOscillations}, making the dip shallower for decreasing Brunt-Väisälä near-core gradient, and dependant on the continuous or discontinuous treatment of such profile.

Taking realistic values of the coupling parameter from the recent study of \cite{Aerts2023ModeStars}, and differential rotation rates from \cite{Saio2021RotationModes}, we then state on the detectability of core to near-core differential rotation from the study of the inertial dip. We isolate two dip parameters regions in which we find differential rotation to be detectable and demonstrate that both core-to-envelope positive and negative differential rotation can be detectable in asteroseismic data from the $\textsl{Kepler}$ mission, even for a core rotation differing from the envelope one at a 1 \% level.

As a result of our study, the rotation rate of the convective core alone can be inferred from the dip location, with an increasing uncertainty for a low gradient of near-core stratification. In turn, we foresee that taking into account a dip model in the period-spacing pattern will also lower the uncertainty on the measurement of the near-core rotation rate. Both measurements can be enriched by an accurate modelling of the inertial dip.

We isolate two main caveats preventing detections from this model in real stars: the unconstrained shape of the Brunt-Väisälä near-core profile and the existence of a density gradient in the convective core. For the former, we find that the impact of the choice of one of the two formalisms for the Brunt-Väisälä profile has an appreciable impact on relatively high coupling parameters, introducing an additional uncertainty on the core-to-near-core differential rotation. For the latter, the existence of a density gradient is known to lower the period of the inertial dip from the constant density picture, as shown in \cite{Ouazzani2020FirstRevealed}. The effect of a core rotating faster than the envelope and a gradient of the density of the core is thus degenerated. Further efforts must be made to lift those degeneracies, with an investigation on the stratification gradient in the core and near-it, for the most up-to-date stellar models, taking into account the impact of age and core-to-envelope mixing prescriptions. Such a study will lower the uncertainties related to the measure of the differential rotation along evolution.

We took the side of remaining analytical for this study, by taking a bi-layer profile for the stellar rotation. This allowed us to isolate the effect of differential rotation and demonstrate its detectability. Going further, it would be of great interest to investigate a smoother rotation gradient in both the convective core and the radiative zone. Properties of inertial modes propagating in a differentially rotating shell, either with a shellular or cylindrical profile, have been proven to differ from the solid-body rotating case: the spectrum, ray paths, and damping are altered and waves tend to form attractors, both in the case of pure inertial \citep{Baruteau2013InertialShell,Guenel2016TidalModes} and gravito-inertial modes \citep{Mirouh2016Gravito-inertialShell}. Given the increased technical complexity of the mode computation, the use of a spectral code to investigate the coupling problem would be required for such a refinement \citep{Ouazzani2020FirstRevealed, Galoy2024PropertiesStars}.

The model hereby presented remains in the purely hydrodynamic framework. Yet, internal stellar magnetism is ubiquitous in stars and can potentially be intense in convective zones and near-core regions in MS stars, due to the dynamo processes happening in the core, which could be strengthened by a potential fossil field in the radiative zone arising from the past evolution of the stars \citep{Brun2005SimulationsAction,Featherstone2009EffectsStars, Augustson2016THESTARS}. Moreover, stars displaying no evidence of differential rotation as derived from the formalism developed in this work could be potential candidates for strong internal magnetism, as the dynamo field can contribute to a locked rotation between the convective core and the radiative zone. It would be thus of prime interest to revisit the coupling problem in a magneto-hydrodynamical framework, derive the potential effects on the inertial dip formation picture and degeneracies between differential rotation, or the effect of density stratification in the core. This will be the scope of a follow-up paper.

\begin{acknowledgements}\
We thank the referee for very constructive and detailed comments that led to an improvement of the quality of our study. L.B. and L.B. gratefully acknowledge support from the European Research Council (ERC) under the Horizon Europe programme (Calcifer; Starting Grant agreement N$^\circ$101165631). S. Mathis acknowledges support from the PLATO CNES grant at CEA/DAp and from the European Research Council through HORIZON ERC SyG Grant 4D-STAR 101071505. While partially funded by the European Union, views and opinions expressed
are however those of the author only and do not necessarily reflect those of the European Union or the European Research Council. Neither the European Union nor the granting authority can be held responsible for them. L. Barrault thanks the members of the asteroseismology group of the Institute of Astronomy (IvS) of KU Leuven, in particular T. Van Reeth, M. Vanrespaille, Z. Guo and C. Aerts, for their warm welcome during a work visit in Spring 2024, and very insightful input on the present study. The authors thank also the members of the Asteroseismology and Stellar Dynamics group of the Institute of Science and Technology Austria (ISTA) for very useful discussion: K. M. Smith, L. Einramhof, S. Torres and A. Cristea.
\end{acknowledgements}



\bibliographystyle{aa} 
\bibliography{references}
\begin{appendix} 
\section{Derivation of gravito-inertial modes structure in the radiative envelope and expansion near the core}
\label{Appendix:g-i}

This Appendix describes the structure of the gravito-inertial modes propagating in the radiative envelope of the star, along with an expansion of the Lagrangian displacement and pressure perturbation at the inner boundary of the radiative zone. 

\par Under the Traditional Approximation of Rotation (TAR) and the Cowling approximation, the oscillation variables are separable in the spherical coordinates ($r$,$\theta$,$\varphi$), and can be expended as:

\begin{equation}
\xi'_{r}(r,\theta) = \sum_{k} \xi'_{r;k,m}(r)\Theta_{k}^{m}(\mu;s_{\text{env}}) 
\end{equation}

\noindent and

\begin{equation}
p'(r,\theta) = \sum_{k} p'_{k,m}(r)\Theta_{k}^{m}(\mu;s_{\text{env}})   \, ,
\end{equation}

\noindent 
where $\Theta_{k}^{m}(\mu;s_{\text{env}})$ are the Hough functions, defined as the eigenfunctions of the Laplace Tidal Equation (LTE). Hough functions depend both on $\theta$ through $\mu = \cos\theta$ and $\sigma_{\text{env}}$ through $s_{\text{env}} = 2\Omega_{\text{env}}/\sigma_{\text{env}}$ the spin parameter in the envelope. The LTE describes the horizontal variations of physical quantities within the TAR \citep{Lee1997Low-frequencyDependence}, and reads:
\begin{equation}
    \mathcal{L}_{s_{\text{env}}}[\Theta_{k}^{m}(\mu;s_{\text{env}})] = -\Lambda_{k}^{m}(s_{\text{env}})\Theta_{k}^{m}(\mu;s_{\text{env}}) \, ,
\end{equation}
with the Laplace tidal operator:
\begin{align}
    \mathcal{L}_{s_{\text{env}}} = &\frac{\mathrm{d}}{\mathrm{d}\mu}\left(\frac{1-\mu^{2}}{1-s_{\text{env}}^{2}\mu^{2}}\frac{\mathrm{d}}{\mathrm{d}\mu}\right)\nonumber\\ &- \frac{1}{1-s_{\text{env}}^2\mu^2}\left(\frac{m^{2}}{1-\mu^{2}}+ms_{\text{env}}\frac{1+s_{\text{env}}^2\mu^2}{1-s_{\text{env}}^2\mu^2}\right) \, .
\end{align}
The radial functions of $\xi_{r}'$ and $p'$ satisfy the following system:
\begin{align}
    \frac{\mathrm{d}}{\mathrm{d}r}(\xi'_{r;k,m}) & = -\left(\frac{2}{r} - \frac{1}{\Gamma_{1}H_{p}}\right)\xi'_{r;k,m} \nonumber \\ & +\frac{1}{\bar{\rho}c_{S}^{2}}\left(\frac{\Lambda_{k}^{m}(s_{\text{env}})c_{S}^{2}}{r^{2}\sigma_{S, \text{env}}^{2}}-1\right)p'_{k,m}(r) \, ,
    \label{eq:displacement}
\end{align}

\begin{equation}
    \frac{\mathrm{d}p'_{k,m}(r)}{\mathrm{d}r} = \bar{\rho}\left(\sigma_{\text{env}}^{2} - N^{2}\right)\xi'_{r;k,m} - \frac{p'_{k,m}(r)}{\Gamma_{1}H_{p}} \, ,
    \label{eq:pressure}
\end{equation}

\noindent with $H_{p}$, $N$, $\Gamma_{1}$, $\bar{\rho}$, and $c_{S}$ being the pressure scale height, the Brunt-Väisälä frequency, the first adiabatic index, the mean density, and the sound speed respectively. $\Lambda_{k}^{m}(s_{\text{env}})$ is the eigenvalue of the LTE,  quantized by $k$ \citep[introduced by][]{Lee1997Low-frequencyDependence} and $m$, depending on the spin parameter of the envelope $s_{\text{env}}$. For Kelvin modes, the eigenvalue reaches a constant value in the regime of high spin parameters. The asymptotic development has been derived by \cite{Townsend2003AsymptoticStars}, later extended in \cite{Townsend2020ImprovedEquations}. \cite{Tokuno2022AsteroseismologyOscillations} derive, using a Jeffreys-Wentzel-Kramers-Brillouin (JWKB) analysis, the radial wavenumber \citep[see also e.g.][]{Unno1989NonradialStars}:
\begin{equation}
    k_{r}^{2}= \frac{\sigma_{\text{env}}^{2} - N^{2}}{c_{S}^{2}}\left(1-\frac{\Lambda_{k}^{m}(s_{\text{env}}) c_{S}^{2}}{r^{2}\sigma_{\text{env}}^{2}}\right) \, .
\end{equation}
The integral over the whole radiative zone reads, in the low-frequency regime:
\begin{equation}
      \int_{r_a}^{r_b}k_{r}\mathrm{d}r \simeq \int_{r_a}^{r_b}\frac{N}{\sigma_{\rm env}} \frac{\sqrt{\Lambda_{k}^{m}(s_{\text{env}})}}{r}\mathrm{d}r= \frac{\pi^{2}s_{\text{env}}}{\Omega_{\text{env}} \Pi_{0}}  \, ,
\end{equation}
$r_{a}$ and $r_{b}$ being respectively the lower and the upper turning points of the cavity (see Fig.~\ref{fig:sketch_gamma}), and $\Pi_{0}$ the asymptotic period spacing:
\begin{equation}
    \Pi_{0} \simeq \frac{2\pi^{2}}{\sqrt{\Lambda_{k}^{m}(s_{\text{env}})}}\left(\int_{r_a}^{r_b}\frac{N}{r}\mathrm{d}r\right)^{-1} \, .
\end{equation}
This period spacing is approximately constant for Kelvin modes, neglecting the variation of the eigenvalue of the LTE. A discussion of this approximation can be found in Appendix A of \cite{Tokuno2022AsteroseismologyOscillations}, complemented by a discussion taking into account the differentially rotating situation in our subsection \ref{subsec:eigen}. \\ 

\subsection{The continuous case}
The expressions derived using this JWKB analysis for the Eulerian pressure and the Lagrangian displacement remain valid in the range $[r_{\text{a}}, r_{\text{b}}]$. Yet, the matching of the inertial waves in the core and the gravito-inertial waves in the envelope occurs at the edge of the convective core, at the radial coordinate $R_{\text{core}}$. In the case of a continuous Brunt-Väisälä frequency near-core, the two radial coordinates $r_{a}$ and $R_{core}$ differ.
The next key step of the analysis is thus to derive an approximate expression for the radial displacement and the pressure at the edge of the convective core, from the expressions of the Eulerian pressure perturbation and the Lagrangian displacement at $r_{\text{a}}$. We perform a Taylor expansion near $r=r_{\text{a}}$, given that the distance between $R_{\text{core}}$ and $r_{\text{a}}$ is small. \\
The vertical wave number can be written as
\begin{equation}
    k_{r}^{2} \simeq \frac{\Lambda_{k}^{m}(s_{\text{env}})s_{\text{env}}^{2}}{\epsilon^{3} r_{a}^{3}}(r-r_{a}) \, ,
\end{equation}
\noindent where we have introduced the small parameter
\begin{equation}
    \epsilon = \left(\frac{r_{a}}{4\Omega_{\text{env}}^{2}}\frac{\text{d}N^{2}}{\text{d}r}\Big|_{r_a}\right)^{-1/3}
    \label{eq:eps}
\end{equation}

\noindent assuming $\epsilon \ll 1$ because of the steepness of the gradient of the Brunt-Väisälä frequency at the lower turning point. Expanding all quantities, \citet{Tokuno2022AsteroseismologyOscillations} show that the radial Lagrangian displacement and the Eulerian pressure perturbation can be rewritten synthetically as such:

\begin{equation}
    \frac{\xi'_{r;k,m}}{r}\Big{|}_{r_{a}} \simeq Q \epsilon X_{k}^{m}(s_{\text{env}}) 
\end{equation}
and
\begin{equation}
    \frac{p'_{k,m}}{\bar{p}\Gamma_{1}}\Big|_{r_{a}} \simeq Q \frac{r_{a}^{2} \sigma_{\text{env}}^{2}}{c_{S}^{2}} Y_{k}^{m}(s_{\text{env}}) \, ,
\end{equation}
where we have introduced, very similarly to \cite{Tokuno2022AsteroseismologyOscillations}, the following functions:
\begin{equation}
    X_{k}^{m}(s_{\text{env}}) = \Lambda_{k}^{m}(s_{\text{env}})^{1/6}s_{\text{env}}^{2/3} \sin \left(\frac{\pi^{2}s_{\text{env}}}{\Omega_{\rm env}\Pi_{0}}-\frac{\pi}{6}\right)
\end{equation}
and
\begin{equation}
    Y_{k}^{m}(s_{\text{env}}) = \alpha \Lambda_{k}^{m}(s_{\text{env}})^{-1/2}s_{\text{env}}^{4/3} \sin \left(\frac{\pi^{2}s_{\text{env}}}{\Omega_{\text{env}}\Pi_{0}}-\frac{5\pi}{6}\right) \, .
\end{equation}
In those equations, the dimensionless parameters $\alpha$ and $Q$ are defined as:
\begin{equation}
    \alpha \equiv \frac{3^{1/3}\Gamma(2/3)}{\Gamma(1/3)} \simeq 0.73\, ,
\end{equation}

\begin{equation}
    Q \equiv -\frac{c}{3^{2/3} \Gamma(2/3)}\frac{\Lambda_{k}^{m}(s_{\text{env}})^{1/6}}{\epsilon^{1/2}\Omega_{\text{env}}r_{a}^{5/2}}\frac{1}{\bar{\rho}^{1/2}}\Bigg{|}_{r_{a}}\, ,
\end{equation}
c being an arbitrary constant and $\Gamma(x)$ the Gamma function.\\
For this expansion to be accurate, and to match the approximate expressions at the first order with the respective quantities at the edge of the convective core, we need to assert that the values of $\xi_{r}$ and $p'$ at $R_{\text{core}}$ are close to the ones at $r_{\text{a}}$. From a first order expansion of $N^{2}$ from $r_{\text{a}}$ to $R_{\text{core}}$, we obtain taking Eq.~ \eqref{eq:eps}:
\begin{equation}
    \frac{R_{\text{core}}}{r_{\text{a}}} \sim 1-\frac{\epsilon^{3}}{s_{\text{env}}^{2}} \, .
    \label{eq:dev}
\end{equation}

\noindent We thus highlight here that the formalism of \cite{Tokuno2022AsteroseismologyOscillations} only holds for small values of $\epsilon$ and high values of the spin parameter. This limitation will be further discussed in subsection 
\ref{subsec:limit}. The expressions of the pressure perturbation and the radial Lagrangian displacement are, in the inertial frame:

\begin{equation}
\frac{\xi_{r}'}{r}\Bigg|_{R_{\text{core}}} = \sum_{k}a_{k}\epsilon X_{k}^{m}(s_{\text{env}})\Theta_{k}^{m}(\mu; s_{\text{env}})
\end{equation}

\noindent and
\begin{equation}
\frac{p'}{\bar{p}\Gamma_{1}}\Bigg|_{R_{\text{core}}} = 
\frac{\sigma_{\text{env}}^{2} R_{\text{core}}^{2}}{c_{S}^{2}}\sum_{k}a_{k}Y_{k}^{m}(s_{\text{env}})\Theta_{k}^{m}(\mu ; s_{\text{env}}) \, .
\end{equation}

\subsection{The discontinous case}
The vertical wave number can be written at the upper edge of the boundary as
\begin{equation}
    k_{r}^{2} = \frac{\Lambda_{k}^{m}(s_{\rm env})s_{\rm env}^{2}}{4\tilde{\epsilon}^{2}R_{\rm core}^{2}} \, ,
\end{equation}
with $\tilde{\epsilon} = \frac{\Omega_{\rm env}}{N_{0}}$.
Very similarly to the continuous case, the terms appearing in the decomposition of the Lagrangian displacement and Eulerian pressure perturbation are:
\begin{equation}
    \frac{\xi'_{r;k,m}}{r}\Bigg|_{R_{\text{core}}^{+}} \simeq Q \epsilon \tilde{X}_{k}^{m}(s_{\text{env}}) 
\end{equation}
and
\begin{equation}
    \frac{p'_{k,m}}{\bar{p}\Gamma_{1}}\Bigg|_{R_{\text{core}}^{+}} \simeq Q \frac{R_{\rm core}^{2} \sigma_{\text{env}}^{2}}{c_{S}^{2}} Y_{k}^{m}(s_{\text{env}}) \, .
\end{equation}

The JWKB analysis described in \cite{Tokuno2022AsteroseismologyOscillations} is providing:
\begin{equation}
    \tilde{X}_{k}^{m}(s_{\text{env}}) = 2\Lambda_{k}^{m}(s_{\text{env}})^{1/4}s_{\text{env}}^{1/2}\sin\left(\frac{\pi^{2}s_{\text{env}}}{\Omega_{\text{env}}\Pi_0} - \frac{\pi}{4} \right)
\end{equation}
and:
\begin{equation}
    \tilde{Y}_{k}^{m}(s_{\text{env}}) = -\Lambda_{k}^{m}(s_{\text{env}})^{-1/4}s_{\text{env}}^{3/2}\cos\left(\frac{\pi^{2}s_{\text{env}}}{\Omega_{\text{env}}\Pi_0} - \frac{\pi}{4} \right) \, .
\end{equation}
We end up with the following expression for the Lagrangian displacement and the Eulerian pressure perturbation:

\begin{equation}
    \frac{\xi'_{r}}{r}\Bigg|_{R_{\text{core}}^{+}} = \sum_{k}a_{k}\tilde{\epsilon}\tilde{X}_{k}^{m}(s_{\text{env}})\Theta_{k}^{m}(\mu;s_{\text{env}})
\end{equation}
and:
\begin{align}
    \frac{p'}{\Gamma_{1}\bar{p}}\Bigg|_{R_{\text{core}}^{+}} & = \frac{\sigma_{\text{env}}^{2}R_{\text{core}}^{2}}{c_{S}^{2}} \\ \nonumber
    & \times \sum_{k}a_{k}\tilde{Y}_{k}^{m}(s_{\text{env}})\Theta_{k}^{m}(\mu;s_{\text{env}})
\end{align}

\section{Deriving Bryan solutions for pure inertial modes in the convective core}\label{Appendix:pure_inertial}

We hereby derive the structure of pure inertial modes propagating in the convective core of the star and give an expression for the Lagrangian displacement and pressure perturbation at the outer boundary of the convective core.\\
In the framework of solid-body rotation and uniform averaged density in the core, under the Cowling approximation, using the scalar variable $\Psi$ defined as:
\begin{equation}
    \Psi = \frac{1}{\sigma_{\text{core}}^{2}}\frac{p'}{\bar{\rho}}
\label{eq:def_psi} \, ,
\end{equation}
the momentum and mass conservation equations reduce to the Poincaré equation:
\begin{equation}
    \nabla^{2}\Psi - s_{\text{core}}^{2}\frac{\partial^{2} \Psi}{\partial z^{2}} = 0 \, ,
\end{equation}
z being the axial coordinate in the cylindrical coordinate system, the cylindrical axis being along the rotation vector.

The radial Lagrangian displacement vector projected on $\widehat{\mathbf{e}}_{r}$ in the formalism of prograde modes having m<0 is: 
\begin{equation}
    \xi'_{r} = \frac{1}{1-s_{\text{core}}^{2}}\left[\frac{\partial}{\partial r} + \frac{m s_{\text{core}}}{r} - \mu s_{\text{core}}^{2}\frac{\partial}{\partial z}\right]\Psi
\end{equation}
Eq.~ \eqref{eq:def_psi} can also be written as:
\begin{equation}
    p'=\bar{p}\Gamma_{1}\frac{\sigma_{\text{core}}^{2}}{c_{S}^{2}}\Psi \, , 
\end{equation}
\noindent where we have used the sound speed velocity definition: $c_{S}^{2} = \Gamma_{1}\bar{p}/\bar{\rho}$.
Using the ellipsoidal coordinates, \citet{Wu2005ORIGINMODES} shows that the expressions of the relevant quantities near the core are:

\begin{equation}
    \frac{\xi'_{r;l,m}(r)}{r}\Big{|}_{R_{\text{core}}^{-}} \propto C_{l}^{m}(1/s_{\text{core}})P_{l}^{m}(\mu)\, ,
\end{equation}
\begin{equation}
    \frac{p'_{l,m}(r)}{\bar{p}\Gamma_{1}}\Big{|}_{R_{\text{core}}^{-}} \propto \frac{R_{\text{core}}^{2}\sigma_{\text{core}}^{2}}{c_{S}^{2}}P_{l}^{m}(1/s_{\text{core}})P_{l}^{m}(\mu)
\end{equation}

\noindent with $C_{l}^{m}$ being as defined in \cite{Ouazzani2020FirstRevealed}:
\begin{equation}
 C_{l}^{m}(x) = x \left(\frac{\mathrm{d}P_{l}^{m}(x)}{\mathrm{d}x} - \frac{m}{1-x^{2}}P_{l}^{m}(x)\right) \, .
 \label{eq:def_C}
\end{equation}
These analytical solutions are well-known as Bryan solutions, having been first described by \citet{Bryan1889TheEllipticity}. 

The fixed boundary condition $\xi'(r) = 0$ at the edge of the core translates into $C_{l}^{m}(1/s_{\text{core}}) = 0$, which is an eigenvalue problem setting the spin parameters of the pure inertial modes in the core. Table 1 of \cite{Ouazzani2020FirstRevealed} reports
the spin parameters of the prograde $m=-1$ modes in the core.
Finally, the solution of the variables of interest near the core is the linear combination of such solutions, in the inertial frame taking into account the Doppler shift:

\begin{equation}
\frac{\xi_{r}'}{r}\Big{|}_{R_{\text{core}}^{-}}= \sum_{l}b_{l}C_{l}^{m}(1/s_{\text{core}})\tilde{P}_{l}^{m}(\mu) \, ,
\end{equation}
and

\begin{equation}
\frac{p'}{\bar{p}\Gamma_{1}} \Big{|}_{R_{\text{core}}^{-}} 
=\frac{\sigma_{\text{core}}^{2}R_{\text{core}}^{2}}{c_{S}^{2}}\sum_{l}b_{l}P_{l}^{m}(1/s_{\text{core}})\tilde{P}_{l}^{m}(\mu) \, .
\end{equation}

\noindent where $b_{l}$ are constant factors, and $\tilde{P}_{l}^{m}$ are the normalized Legendre polynomials:
\begin{equation}
    \tilde{P}_{l}^{m} \equiv \sqrt{\frac{(2l+1)(l-m)!}{2(l+m)!}}P_{l}^{m}(x)\, .
\end{equation}

\section{Derivation of the modified lorentzian profile}\label{App:Mod_lorentz}

To compute the period spacing expression, we write Eq.~\eqref{eq:coupling_simplified} for two neighbouring solutions $s_{1}$ and $s_{2}$ with $G^{-1}(s_{2}) > G^{-1}(s_{1})$. We assume that the frequency spectrum of the gravito-inertial modes is dense. The two solutions being on two different branches of the cotangent, and the function $1/(G^{-1}(s) - G^{-1}(s)^*)$ not varying appreciably due to the density of the spectrum, we have $\frac{\pi^2(s_{2}-s_{1})}{\Omega_{\text{env}}\Pi_0}-\pi = 2x \ll 1$.
In the case of $G^{-1}(s_{1})<s_{\text{core}}^{*}<G^{-1}(s_{2})$ the solutions belong to the same branch, as pointed out by \citet{Tokuno2022AsteroseismologyOscillations}. We thus rewrite Eq.~\eqref{eq:coupling_simplified} as:
\begin{equation}
    \tan\left(\frac{\pi^{2}s}{\Omega_{\text{env}}\Pi_0}\right) \simeq -\frac{1}{\frac{1}{\sqrt{3}} + \frac{\epsilon/V_{\text{diff}}}{G^{-1}(s) - G^{-1}(s^*)}}
    \label{eq:tangent}
\end{equation}
and the solutions are on two branches of the tangent.
Thus, in both cases:
\begin{equation}
    \frac{\pi^2 s_{1,2}}{\Omega_{\text{env}}\Pi_0} - \frac{\pi}{6} = \frac{\pi^2 \bar{s}}{\Omega_{\text{env}}\Pi_0} - \frac{\pi}{6} \pm \frac{\pi}{2} \mp x
\end{equation}
with $\bar{s} = \frac{s_{1}+s_{2}}{2}$.
The resulting system is thus, noting $\Bar{S} = \frac{\pi^2 \bar{s}}{\Omega_{\text{env}}\Pi_0} - \frac{\pi}{6}$ and using a first order Taylor expansion:
\begin{align}
    \begin{cases}
        \cot(\bar{S}) - x(1+\cot^{2}(\bar{S})) +1/\sqrt{3}&\simeq -\frac{\epsilon/V_{\text{diff}}}{G^{-1}(s_{1})-G^{-1}(s)^{*}}\\
        \cot(\bar{S}) + x(1+\cot^{2}(\bar{S})) +1/\sqrt{3}&\simeq -\frac{\epsilon/V_{\text{diff}}}{G^{-1}(s_{2})-G^{-1}(s)^{*}}
    \end{cases}
    \, .
\end{align}
We then take the sum and the difference. Considering the right-hand side terms, with $\overline{G^{-1}(s)} = (G^{-1}(s_{1}) + G^{-1}(s_{2}))/2$:
\begin{align}
    &\frac{1}{G^{-1}(s_{2})-G^{-1}(s)^{*}} + \frac{1}{G^{-1}(s_{1})-G^{-1}(s)^{*}} \nonumber \\ &= \frac{2(\overline{G^{-1}(s)}-G^{-1}(s)^{*})}{(\overline{G^{-1}(s)}-G^{-1}(s)^{*})^{2} - (\frac{G^{-1}(s_{2})-G^{-1}(s_{1})}{2})^{2}}
\end{align}
and
\begin{align}
    &\frac{1}{G^{-1}(s_{2})-G^{-1}(s)^{*}} - \frac{1}{G^{-1}(s_{1})-G^{-1}(s)^{*}} \nonumber \\ &= \frac{G^{-1}(s_{2})-G^{-1}(s_{1})}{(\overline{G^{-1}(s)}-G^{-1}(s)^{*})^{2} - (\frac{G^{-1}(s_{2})-G^{-1}(s_{1})}{2})^{2}} \, .
\end{align}
We approximate the denominator considering: \\ $\overline{G^{-1}(s)}-G^{-1}(s)^{*} \gg \frac{G^{-1}(s_{2})-G^{-1}(s_{1})}{2}$. This approximation is accurate moving away from the location of the pure inertial mode.
With such approximation, we retrieve:

\begin{align}
    &\left[1+\Big(\frac{\epsilon /V_{\text{diff}}}{\overline{G^{-1}(s)}-G^{-1}(s)^{*}}+\frac{1}{\sqrt{3}}\Big)^{2}\right] \nonumber \\
    & \times \Big[\frac{\pi^{2}(s_{1}-s_{2})}{\Omega_{\text{env}} \Pi_{0}}-\pi\Big]\simeq -\frac{\epsilon}{V_{\text{diff}}}\frac{G^{-1}(s_{1})-G^{-1}(s_{2})}{(\overline{G^{-1}(s)})-G^{-1}(s)^{*})^{2}} \, .
\end{align}
We further manipulate this equation to give an analytical insight. Provided that $s_{1}-s_{2} \ll \bar{s} $, and defining $s^{*}_{\text{env}} = G(s_{\text{core}}^{*}) = G(G^{-1}(s)^{*})$, we rewrite the equation as such:
\begin{align}
    &\left[1+\Big(\frac{\epsilon /V_{\text{diff}}}{G^{-1}(\bar{s})-G^{-1}(s^{*})}+\frac{1}{\sqrt{3}}\Big)^{2}\right] \nonumber \\
    & \times \Big[\frac{\pi^{2}(s_{1}-s_{2})}{\Omega_{\text{env}} \Pi_{0}}-\pi\Big]\simeq -\frac{\epsilon}{V_{\text{diff}}}\frac{G^{-1}(s_{1})-G^{-1}(s_{2})}{(G^{-1}(\bar{s})-G^{-1}(s^{*}))^{2}} \, .
\end{align}
We then perform a first order expansion of $G^{-1}$, both around $\bar{s}$ and $\frac{\bar{s}+s^{*}}{2}$.The smoothness of $G^{-1}$ ensures that the latter expansion remains accurate near the dip. This gives Eq.~\eqref{eq:Taylor_exp}.
Reasoning on Eq.~\eqref{eq:tangent}, the expansion holds the same result.

\section{Building perturbed period-spacing patterns}\label{Appendix:detect}
We investigate the detectability of the differential rotation from the inertial dip analysis, in real data, by considering a typical noise from the observations.

\begin{figure}
    \centering
    \includegraphics[width = \linewidth]{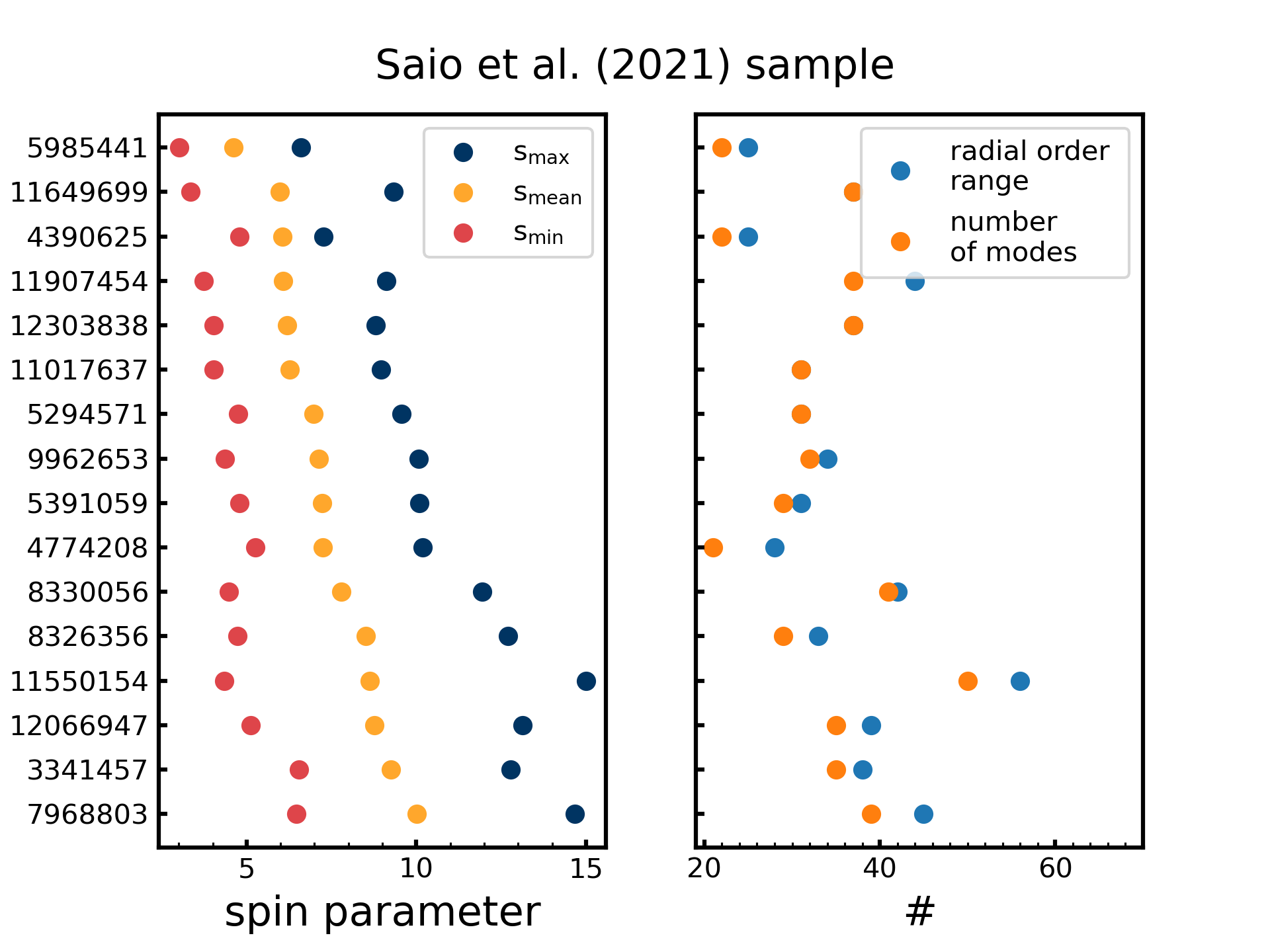}
    \caption{Analysis of the \cite{Saio2021RotationModes} sample, complemented by \cite{Li2020Gravity-modeKepler} analysis. Left panel: minimum, mean, and maximum period-spacing pattern for the 16 stars analyzed by \cite{Saio2021RotationModes} (KIC numbers as the y-axis). Right panel: number of modes in the related period-spacing patterns, and extent in radial order. Note that all period-spacing patterns are taken from \cite{Li2020Gravity-modeKepler} related catalog, although \cite{Saio2021RotationModes} used the period-spacing pattern of \cite{VanReeth2016InteriorSpacings} for KIC12066947.}
    \label{fig:Saios}
\end{figure}

The question of the uncertainty to attach to the period-spacing patterns has been extensively analyzed by \citet{Bowman2021TowardsPulsators} in the case of a well-studied star, KIC 7760680, a Slowly-Pulsating B star. It is highly non-trivial and has potential sources of uncertainty that can come from (1) the light-curve used, (2) the iterative pre-whitening process, (3) the building of the period-spacing pattern, which has still a part of subjectivity from the scientist. The excitation and damping mechanisms of g-modes are still unsufficiently known and the lowest the signal of a peak, and the highest the uncertainty on the period obtained. Thus, no accurate model of noise can be derived from first principles, such as the finite observation times of space missions. We take our noise model from \citet{Kjeldsen2003SpaceAsteroseismology} and \citet{Montgomery1999AData}, which inferred the following structure for the uncertainty on the frequency:
\begin{equation}
    \delta f = 0.44\frac{\langle a \rangle}{a}\frac{1}{\Delta T} \, ,
    \label{eq:S/N}
\end{equation}
where $\langle a \rangle$ is the noise level in the amplitude spectrum, a is the amplitude of peak, so that $a/\langle a \rangle$ is the signal over noise ratio. $\Delta T$ is the observing time of the considered space mission. In our case, we take $\Delta T$ to be of 4 years. Without prior information on the signal-over-noise ratio for each peak, the ratio $\langle a \rangle/a$ is not possible to compute. We then take a hypothesis of $\rm{S/N} = 4.6$ for every signal composing the period-spacing pattern. This value of signal over noise ratio was used in the study of \citet{Bowman2021TowardsPulsators} as a stopping criterion for their analysis. It is worth noting that as mentioned by \cite{VanReeth2015DetectingStudies}, the value of S/N to be taken is controversial and a priori distinct for every star. \cite{Li2019PeriodStars} for instance used a S/N stopping criterion of 3. Converting this relation in the period domain, with the adopted S/N:
\begin{equation}
    \delta \mathrm{P}_{\rm in} = \frac{\mathrm{P}_{\rm in}^{2}}{10.45 \Delta T} \, .
\end{equation}
Then remains the issue of the extent of the generated period-spacing pattern in terms of accessible spin parameters. For this, we analyze the period-spacing patterns with a confirmed analysis by \cite{Saio2021RotationModes}. The results are shown in Fig.~\ref{fig:Saios}. First, the minimum, mean and maximum amount of modes in the analysis of \cite{Saio2021RotationModes} are respectively 21, 33, and 50. We thus take these values as three distinct cases for our detectability analysis.

\begin{figure}[h]
    \centering
    \includegraphics[width=\linewidth]{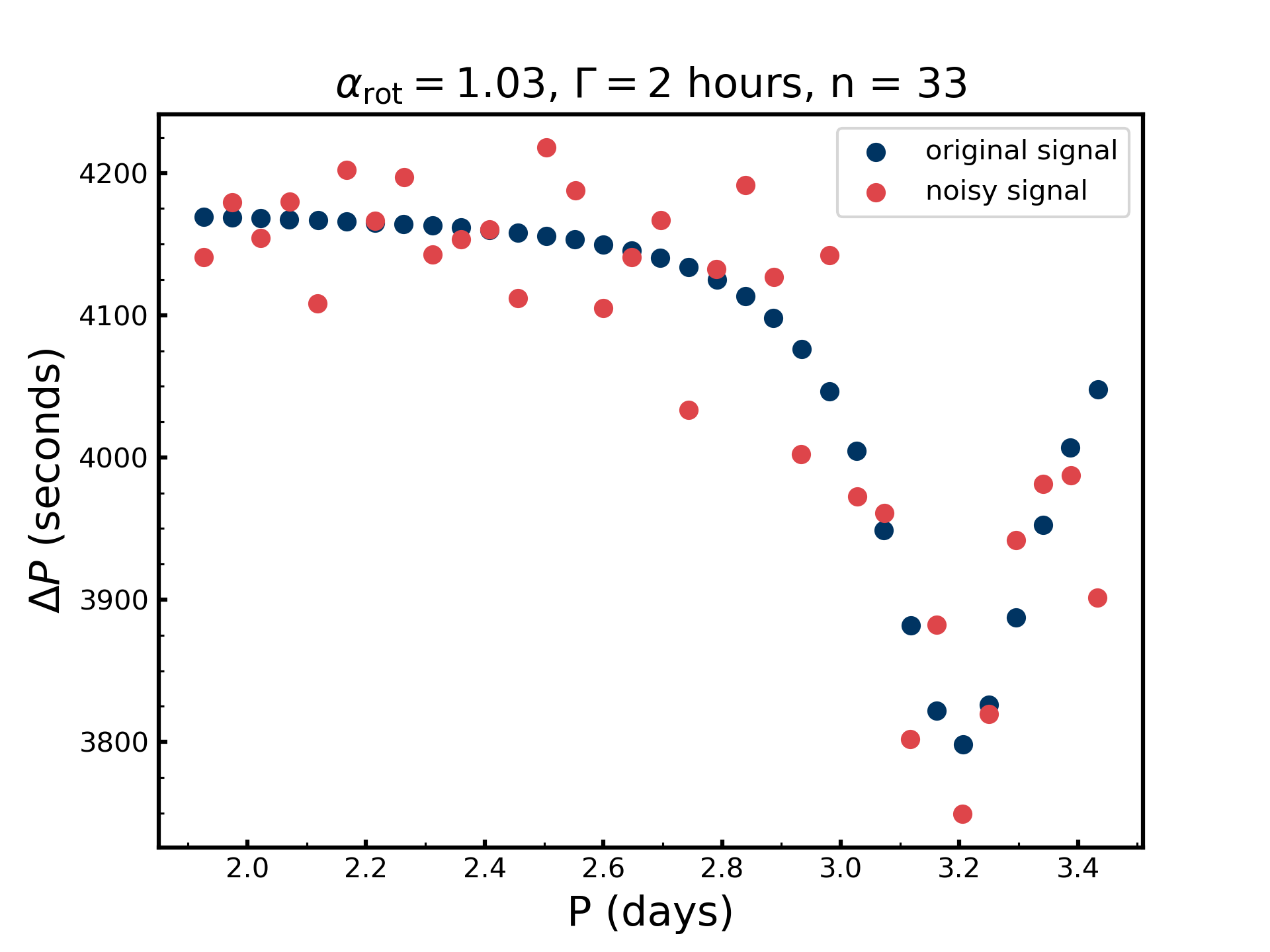}
    \includegraphics[width=\linewidth]{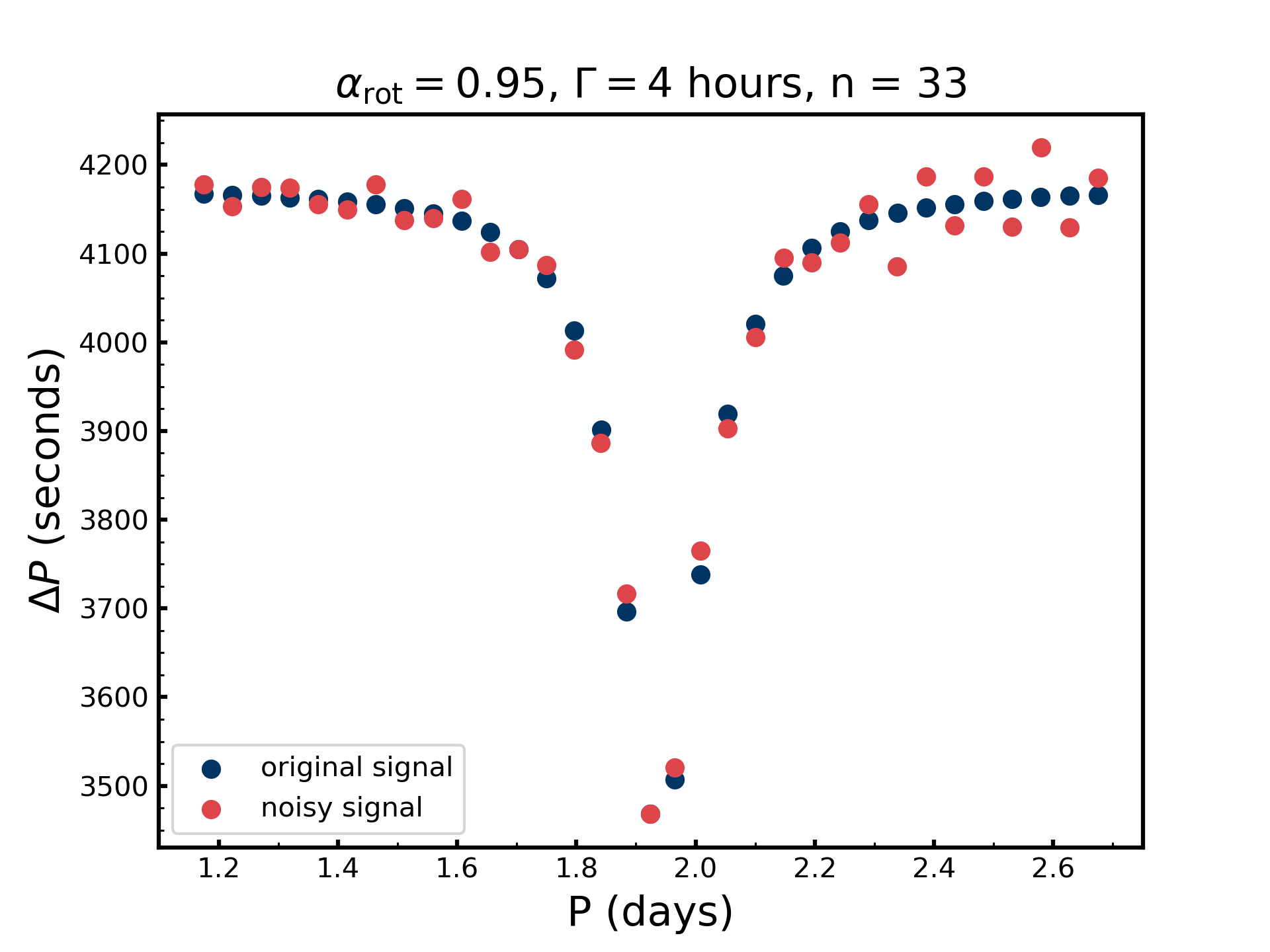}
    \caption{Examples of perturbed signals obtained from the described routine: top: $(\alpha_{\rm rot};\Gamma) = $ (1.03 ; 2 hours), bottom: $(\alpha_{\rm rot};\Gamma) = $ (0.95 ; 4 hours)}
    \label{fig:perturbed_dip}
\end{figure}

Second, we aim to build a representative period-spacing pattern in term of the range of periods of the gravito-inertial modes. Sticking to the case of KIC12066947 for which the extremal values of periods in the inertial frame are $\mathrm{P_{in,min}} = 0.33 ~\mathrm{d}$ and  $\mathrm{P_{in,max}} = 0.405 ~\mathrm{d}$, corresponding to values $\mathrm{P_{co,min}}$ and $\mathrm{P_{co,max}}$ in the co-rotating frame, we take $n/2$ modes around the mean value $(\mathrm{P_{co,min}} + \mathrm{P_{co,max}})/2$ to build the period-spacing pattern, then convert it in the inertial frame for the MCMC analysis described in the main text to fit simultaneously the internal rotation rate $\Omega_{\rm env}$ and the buoyancy travel time $\Pi_{0}$ as well as the parameters $(\alpha_{\rm rot}$ and $\Gamma$ specific to the inertial dip. Two examples of mock period-spacing patterns are shown in Fig. \ref{fig:perturbed_dip} in the co-rotating frame for a better vizualisation of the dip structure.

\section{Bayesian fitting}\label{App:Bayes}

This appendix presents the cornerplots issued following the procedure described in subsection \ref{subsubsec:detect}, using the the \texttt{emcee} package \citep{Foreman-Mackey2013ttemcee/ttHammer}, for the particular situation of input parameters ($\Omega_{\rm env}/2\pi$, $\Pi_{0}$, $\mathrm{P_{0}}$, $\alpha_{\rm rot}$,$\Gamma$) = $(2.16 \, \mathrm{c.d^{-1}}, 4175 \, \mathrm{s},  1.419 \, \mathrm{d}, 0.98, 12 \, \mathrm{hours})$.

\begin{figure}
    \centering
    \includegraphics[width=\linewidth]{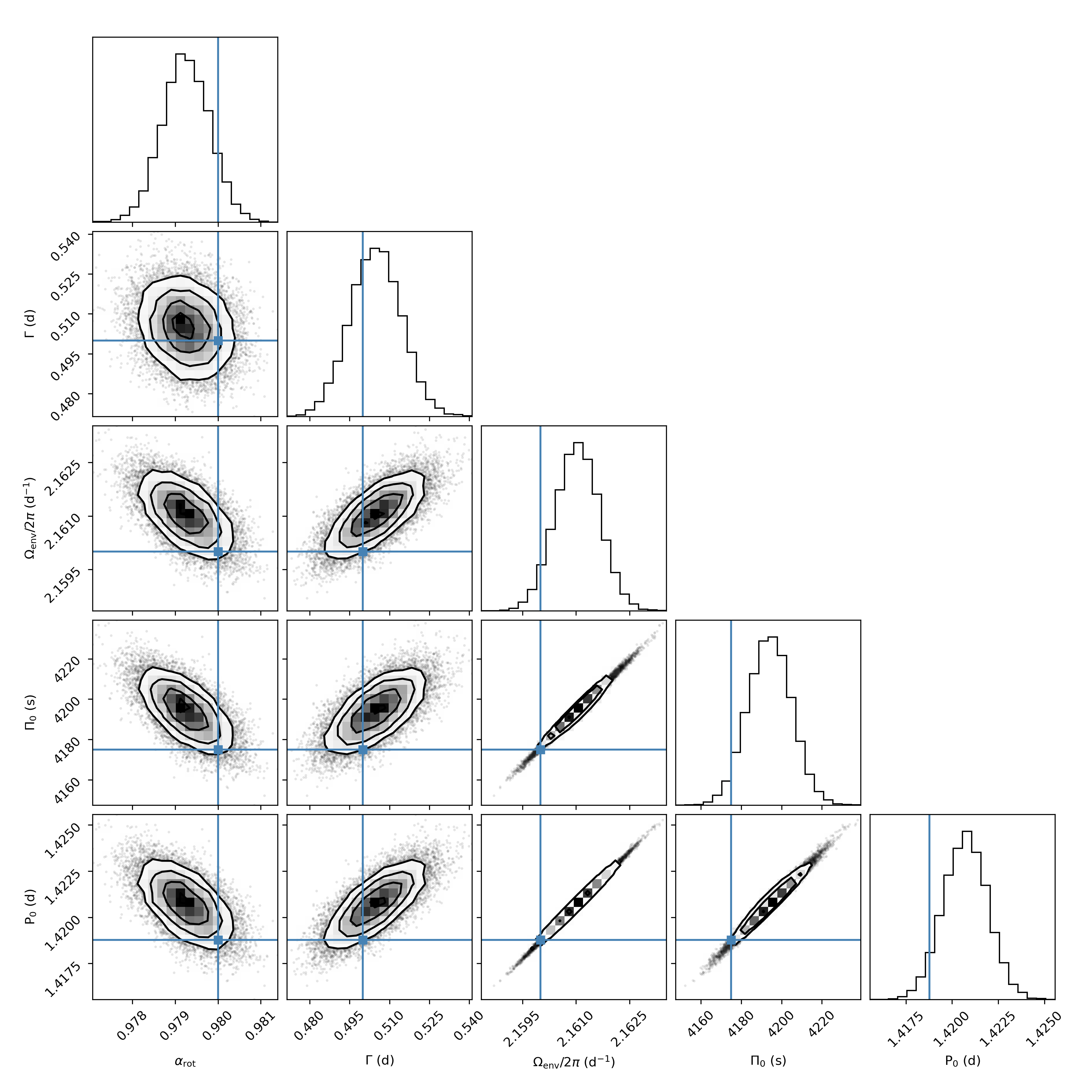}
    \caption{Corner plot showing the posterior distribution of $\Omega_{\rm env}/2\pi$, $\Pi_{0}$, $\mathrm{P}_{0}$, $\alpha_{\rm rot}$, $\Gamma$ for input values of respectively $(2.16 \, \mathrm{d^{-1}}, 4175 \, \mathrm{s}, \mathrm{P}_{0} = 1.419 \, \mathrm{d}, 0.98, 12 \, \mathrm{hours})$, shown in blue lines on the plot.}
    \label{fig:5D_plot}
\end{figure}

\begin{figure}
    \centering
    \includegraphics[width=\linewidth]{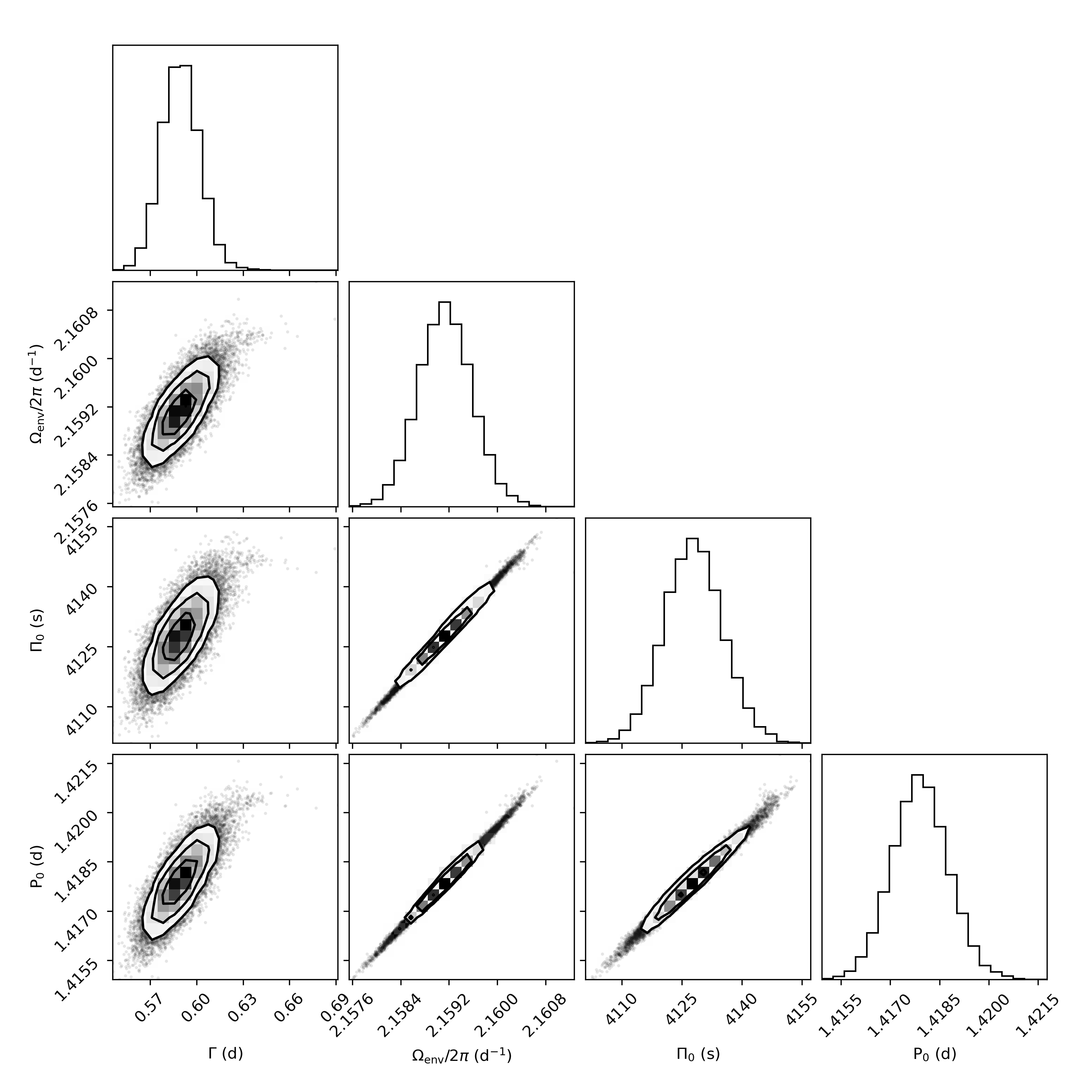}
    \caption{Corner plot showing the posterior distribution of $\Omega_{\rm env}/2\pi$, $\Pi_{0}$, $\mathrm{P_{0}}$, $\Gamma$ for the period-spacing pattern generated with the five parameters previously described.}
    \label{fig:4D_plot}
\end{figure}

\section{Variation of the geometric factor with the amount of differential rotation}\label{App:geom}

The geometric factor $c_{k,l}$ measures the geometrical overlap between Hough functions $\Theta_{k}^{m}(\mu,s_{\text{env}})$ and Bryan solutions, represented by a Legendre polynomial $P_{l}^{m}(\mu)$, as defined in Eq.~\eqref{eq:def_c}. The two definitions given both in \citet{Tokuno2022AsteroseismologyOscillations} and in this present work on the one hand, and in \citet{Ouazzani2020FirstRevealed} on the other hand, are consistent if both the Hough functions and the Legendre polynomials are duly normalized. For the reasoning held in subsection \ref{subsec:determinant} to hold, this geometric factor needs not to be negligibly small. We thus compute the geometric factors corresponding to the different $s_{\text{env}}^*$ at which the interaction occurs with varying differential rotation rate $\alpha_{\rm rot}$. Hough functions are computed using a code described in Appendix A of \citet{Prat2019PeriodComponents}, using a Chebyshev collocation scheme described in \citet{Wang2016OnFunctions}.

\begin{figure}[h!]
    \centering
    \includegraphics[width=1\linewidth]{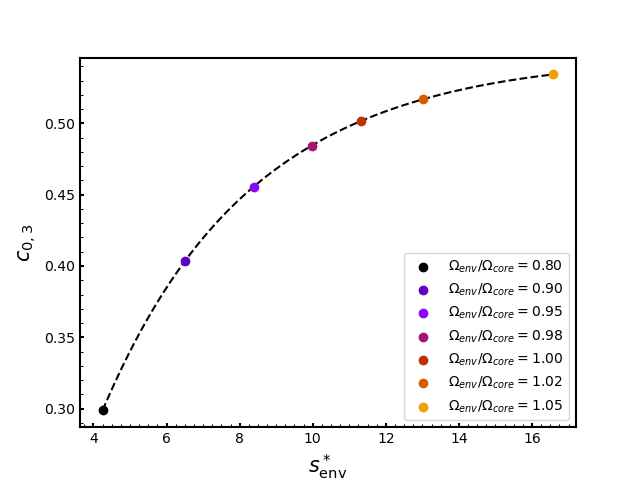}
    \caption{Variation of the geometric factor with the considered spin parameter, in the case of $(k=0,m=-1)$ Kelvin g-modes interacting with a $(l=3,m=-1)$ pure inertial mode. Specific cases corresponding to relevant values of $\alpha_{\rm rot}$ are overplotted.}
    \label{fig:var_geom}
\end{figure}

We find that at the differential rotation rates taken into account in this work, the geometric factor does not reach negligibly small values. With increasing differential rotation rates, the geometric factor is gradually increasing. This comes from the fact that Hough functions, which reduce to a simple Legendre polynomial in the non-rotating case, progressively diverge from their non-rotating counterparts, being composed of further harmonics in their Legendre polynomial decomposition. In the case of the considered interaction, the spin parameter at which the interaction occurs is progressively shifted towards low values with increasing core rotation over constant envelope rotation. The geometric factor thus decreases in this regime.

\section{Impact of modulations due to chemical or thermal stratification on the inertial dip detectability}\label{App:mod}

\begin{figure}
    \centering
    \includegraphics[width=\linewidth]{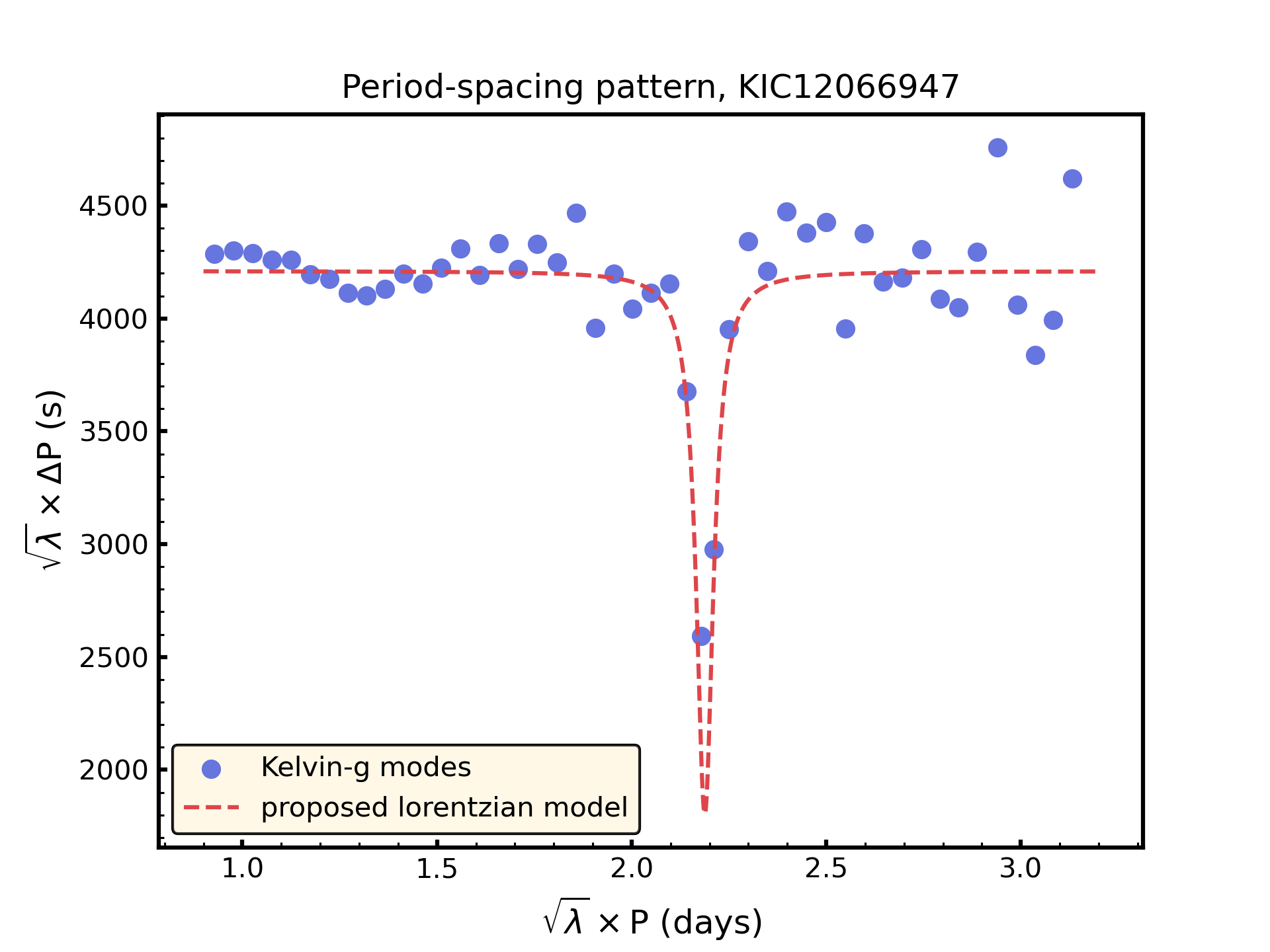}
    \caption{Period-spacing pattern extracted using the algorithm described in \citet{VanReeth2015DetectingStudies}, overplotting the Kelvin-g mode series (k=0,m=-1) corrected by the eigenvalue of the LTE, with an inverse Lorentzian (red, dashed line) to show the location of the inertial dip.}
    \label{fig:data}
\end{figure}

The impact of buoyancy glitches is enscribed in period-spacing patterns through modulations, whose shape depend on the particular hypothesized glitch profile \citep{Miglio2008ProbingStars, Cunha2015StructuralModels, Cunha2019AnalyticalDiagram, Cunha2024BuoyancyRevisited}. The location along the stellar radius, number and shape of those glitches depends on the age and particular choices of stellar modelling parameters, which renders the sole analysis of chemical or thermal stratification challenging in itself, and often relying on forward modelling \citep[see e.g.][]{Mombarg2022PredictionsLevitation}. We thus chose not to consider these modulations in our fitting analysis, and to assume a knowledge of the baseline only limited by the intrinsic observational errors. We aim to discuss this approach in this appendix. Rather than taking an arbitrary glitch location and shape in this analysis, we aim to base ourselves on observed modulations in real data. We take the case of KIC12066947, for which a periodic modulation is observed at low periods in the period-spacing pattern (Fig.~\ref{fig:data}), as well as a clear inertial dip (highlighted with a red dashed inverse Lorentzian profile).

\begin{figure}
    \centering
    \includegraphics[width=\linewidth]{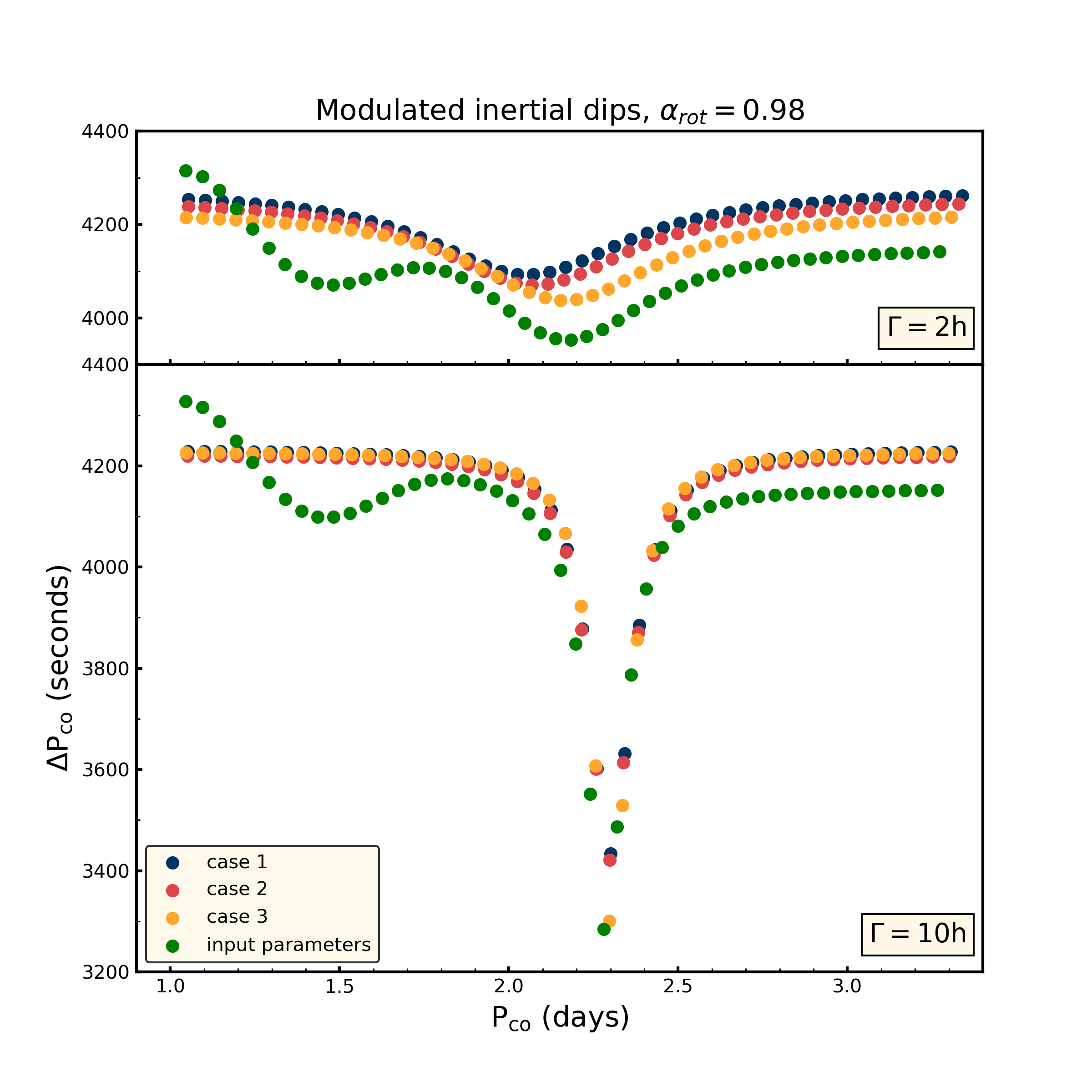}
    \caption{Inertial dips retrieved from an input parameter of $\alpha_{\rm rot} = 0.98$ with different values of $\Gamma$, introducing a modulation of the period-spacing pattern, and the intrinsic perturbation due to the observationnal noise (not represented here for clarity). Period-spacing patterns contain 48 modes, as in the observed case of KIC12066947. Green dots designate the modulated period-spacing pattern without this latter perturbation. The blue dip is retrieved fitting the periods as in the main text (case 1), red with an added free parameter in the likelihood (case 2) and orange fitting the period-spacings with the added free parameter (case 3).}
    \label{fig:mod}
\end{figure}

We introduce this periodic modulation in custom period-spacing patterns reproducing the behaviour of the baseline shown in Fig.~\ref{fig:data}, with a same number of modes, while introducing an inertial dip whose amplitude is made varying, allowing us to study the impact of oscillatory features on the dip fitting. We apply our MCMC analysis on two different period-spacing patterns exhibiting (1) an inertial dip of a significant amplitude compared to the chemical modulation (Bottom panel of Fig.~\ref{fig:mod}, green points) (2) a shallow inertial dip of a similar amplitude compared to the modulation (Top panel of Fig.~\ref{fig:mod}, green points).

By its treatment of the sole errors related to the intrinsic finite observing time, the likelihood described in Eq.~\eqref{eq:likelihood} would not be fitted for this analysis, as it gives increasing weight for low-period modes, that are primarily perturbed in this present analysis. We thus investigate the results given by three different implementations of the likelihood. Alongside with the likelihood used in the main text (case 1), we adapt our likelihood to take into account a free parameter in our error estimations (case 2). This implementation is related to the one of \citet{Aerts2018ForwardSelection}, without taking the Mahalanobis distance in our simplified case. The log-likelihood takes the following form:
\begin{equation}
    \log\mathcal{L}(\mathcal{F}) = -\frac{1}{2}\sum_{k}{\left[\frac{(\mathrm{P}_{k,\rm in}^{\mathrm{mod}} - \mathrm{P}_{k,\rm in}^{\mathrm{pert}})^{2}}{s_{k}^{2}} + \ln(2\pi s_{k}^{2})\right]} \, ,
\end{equation}
with $s_{k}^{2} = \sigma_{k}^{2} + f^2(\mathrm{P}_{k,\rm in}^{\mathrm{mod}})^{2}$, $\sigma_{k}^{2}$ being the error previously taken for the likelihood related to intrinsic observationnal uncertainty and $f$ an additional free parameter. This choice makes the model flexible to modulations not comprised in the observationnal error. We also consider fitting period-spacings instead of period with a similar treatment of the error (case 3) for completeness, as results can differ taking one or the other formulation \citep[see e.g.][]{Cunha2024BuoyancyRevisited}.

Results are shown in Fig.~\ref{fig:mod}, for each of the considered cases. The precise estimated distribution of the parameters is given in Table~\ref{tab:val_10} for the shallowest dip.  We find that in the first configuration, the parameters related to the dip $(\alpha_{\rm rot},\Gamma)$ are retrieved accurately irrespective of the likelihood used. This in turn justifies our use of the first likelihood for the dips studied in the main text, without modulation. The rotation rate is constrained as well, but $\Pi_{0}$, the baseline is overestimated. In the second configuration, a difference emerges between the values retrieved using the different likelihoods. We notice our method used so far (case 1, blue points) to be biased towards high core-to-envelope differential rotation in this configuration, as we were only considering the intrinsic observationnal error in the calculations of the likelihood. Modes of low periods being less affected by this noise, they take a significant weight in the obtained result. We find that fitting period-spacings (case 3) provides the closest estimation of dip parameters ($\alpha_{\rm rot},\Gamma$), all cases introducing a bias on the fitted baseline as well in this case. We nevertheless highlight first that this case shows a relative robustness of the differential rotation ($\alpha_{\rm rot}$) and near-core sratification ($\Gamma$) detection even in the hypothesis of a shallow dip relative to the chemical modulations. Second, this fitting routine does not comprise any information on the chemical modulations and a more educated fitting method would show better constraints on both the inertial dips and the chemical modulations. A dedicated data-oriented study would be necessary in this case but is outside of the scope of our present proof-of-concept study.

\begin{table}[
]
    \centering
    \begin{tabular}{c|c|c|c|c|}
    &$\alpha_{\rm rot}$ & $\Gamma ( \rm h)$ & $\Omega_{\rm env}/2\pi$ & $\Pi_{0} (\rm s)$\\
    \hline & & & & \\
    Input & 0.98 & 10 & 2.16 & 4175 \\[0.1cm] 
    Case 1 & $0.974^{0.001}_{-0.001}$ & $12.96^{0.12}_{-0.12}$ & $2.167^{0.13}_{-0.29}$ & $4232^{2}_{-2}$ \\[0.1cm] 
    Case 2 & $0.981^{0.003}_{-0.003}$ & $12.50^{1.13}_{-1.13}$ & $2.166^{0.001}_{-0.001}$ & $4260^{10}_{-17}$ \\[0.1cm] 
    Case 3 & $0.981^{0.004}_{-0.004}$ & $10.66^{1.80}_{-1.49}$ & $2.168^{0.008}_{-0.011}$ & $4235^{28}_{-37}$ \\

    \end{tabular}
    \vspace{0.3cm}\caption{Values of parameters retrieved in the different implementations of the MCMC, and their uncertainty, from the listed input parameters. Uncertainties are given in term of the 16th and 84th percentiles around the 50th percentile of the samples in the marginalized distributions.}
    \label{tab:val_10}
\end{table}

\end{appendix}

\end{document}